\documentclass[aps,pre,10pt,showpacs,floatfix,onecolumn,nofootinbib,a4paper]{revtex4-2}
\usepackage{amssymb, amsmath, amsthm, mathtools}
\usepackage{bm}
\usepackage{mathrsfs}
\usepackage{hyperref,bookmark}
\usepackage{graphicx}
\usepackage{epsfig,color}
\usepackage{mathrsfs}
\usepackage{verbatim}
\usepackage{url}
\usepackage{subcaption}
\usepackage{float}
\usepackage[greek,english]{babel}
\usepackage{dcolumn}
\usepackage{accents}

\usepackage{cases}

\usepackage{geometry}
\newcommand{\tr}{\operatorname{tr}}
\newcommand{\dd}{\operatorname{d}\!}

\newcommand{\diver}{\operatorname{div}}
\newcommand{\curl}{\operatorname{curl}}

\newcommand{\n}{\bm{n}}
\newcommand{\e}{\bm{e}}
\newcommand{\esp}{\mathrm{e}}
\newcommand{\nper}{\bm{n}_\perp}
\newcommand{\normal}{\bm{\nu}}
\newcommand{\body}{\mathscr{B}}

\newcommand{\arcsinh}{\operatorname{arcsinh}}

\newcommand{\WOF}{W_\mathrm{OF}}
\newcommand{\WQT}{W_\mathrm{QT}}

\newcommand{\bend}{\bm{b}}

\newcommand{\Wn}{\mathbf{W}(\n)}
\newcommand{\Pn}{\mathbf{P}(\n)}

\newcommand{\vv}{\bm{v}}
\newcommand{\frameca}{(\e_x,\e_y,\e_z)}

\newcommand{\nigh}[1]{{\color{black}{#1}}}
\newcommand{\twist}{w}
\newcommand{\sgn}{\operatorname{sgn}}

\newtheorem{theorem}{Theorem}
\newtheorem{definition}{Definition}
\newtheorem{lemma}{Lemma}
\newtheorem{corollary}{Corollary}

\newtheorem{proposition}{Proposition}

\theoremstyle{definition}
\newtheorem{remark}{Remark}

\begin{document}
\latintext

\title{Singular Twist Waves in Chromonic Liquid Crystals}
\author{Silvia Paparini\footnote{\emph{Present address}: Dipartimento di Matematica ``Tullio Levi-Civita'', Universit\`a degli Studi di Padova, Via Trieste 63, I-35121 Padova, Italy.}}
	\email{silvia.paparini@unipv.it}	
\affiliation{Dipartimento di Matematica, Universit\`a di Pavia, via Ferrata 5, I-27100 Pavia, Italy}
\author{Epifanio G. Virga}
\email{eg.virga@unipv.it}
\affiliation{Dipartimento di Matematica, Universit\`a di Pavia, via Ferrata 5, I-27100 Pavia, Italy}

\begin{abstract}
Chromonic liquid crystals are lyotropic nematic phases whose applications span from food to drug industries. It has recently been suggested that the elastic energy density governing the equilibrium distortions of these materials may be \emph{quartic} in the measure of \emph{twist}. Here we show that the non-linear twist-wave equation associated with such an energy has smooth solutions that break down in a finite time, giving rise to the formation of a shock wave, under rather generic assumptions on the initial profile. The critical time at which smooth solutions become singular is estimated analytically with an accuracy that numerical calculations for a number of exemplary cases prove to be satisfactory.
\end{abstract}
\date{\today}

\pagenumbering{arabic}

\maketitle

\section{Introduction}\label{sec:intro}
\emph{Lyotropic} liquid crystals phases arise in colloidal solutions (mostly aqueous) when the concentration of the solute is sufficiently high or the temperature is sufficiently low. Chromonic liquid crystals (CLCs) are special lyotropic phases with potential applications in life sciences \cite{shiyanovskii:real-time,mushenheim:dynamic,mushenheim:using,zhou:living}. These materials are constituted by plank-shaped molecules that arrange themselves in stacks when dissolved in water. For sufficiently high concentrations or low temperatures, the constituting stacks give rise to an ordered phase, either \emph{nematic} or \emph{columnar} \cite{lydon:chromonic_1998,lydon:handbook,lydon:chromonic_2010,lydon:chromonic,dierking:novel}. Here, we shall only be concerned with the nematic phase, in which the elongated  microscopic constituents of the material display  a certain degree of \emph{orientational} order, while their centres of mass remain disordered. Numerous substances have a CLC phase; these include organic dyes (especially those common in food industry), drugs, and oligonucleotides. 

The classical quadratic elastic  theory of Oseen~\cite{oseen:theory} and Frank~\cite{frank:theory} was proved to have potentially paradoxical consequences when applied to free-boundary problems, such as those involving the equilibrium shape of CLC droplets surrounded by their isotropic phase \cite{paparini:paradoxes}. To remedy this state of affairs, a \emph{quartic} elastic theory was proposed for CLCs in \cite{paparini:elastic}, which alters the Oseen-Frank energy density by the addition of a \emph{single} quartic term in the \emph{twist} measure of nematic distortion. Preliminary experimental confirmations of the validity of this theory are presented in \cite{paparini:spiralling,ciuchi:inversion,paparini:what}.

All these applications of the quartic elastic theory fall in the realm of statics. In this paper, we move a step forward and explore the dynamical consequences of this theory. To this end, in Sect.~\ref{sec:EL_inertial} we present a general dynamical setting, indeed far more general than it would strictly be fit for our purposes, as we also contemplate the possibility that the scalar degree of order $s$ be variable in space and time like the nematic director $\n$.

In Sect.~\ref{sec:energetics}, we recall the basic features of the elastic quartic twist theory and in Sect.~\ref{sec:twist_waves} we write the general dynamical equations in the special case where the fluid remains quiescent, while $\n$  exhibits a single distortion mode, the \emph{twist}, which is described by a single scalar function $w$ in a single spatial variable $x$ and time $t$. In Sect.~\ref{sec:method}, for the conservative case, which is the only one considered in this paper, $\twist$ is found to obey a non-linear wave equation with propagation velocity that depends on the spatial derivative $\twist_{,x}$. Even if the initial profile $\twist_0$ is smooth, in its  evolution $\twist$ is bound to brood singularities that erupt in a finite time $t^\ast$, at which second derivatives become unbounded and first derivatives develop discontinuities. This is when a smooth wave breaks down and a \emph{shock} wave emerges from it, a typical non-linear phenomenon that the Oseen-Frank quadratic theory could not embrace.

As also shown in Sect.~\ref{sec:method}, the classical theory of hyperbolic equations does not suffice to our needs, for the wave velocity is not in a standard form. Building on more recent analytical results, we construct the mathematical framework that serves our purposes. Thus, in Sect.~\ref{sec:critical_time}, we prove that under mild assumptions on the initial twist profile $\twist_0$, all smooth solutions $\twist$ to the wave equation do indeed break down, and  we give an upper estimate for the critical time $t^\ast$ that numerical solutions illustrated in Sect.~\ref{sec:applications} prove to be quite accurate.

Finally, in Sect.~\ref{sec:conclusions}, we collect the conclusions of our study and comment on possible ways to extend it. The paper is closed by two appendices, where we present mathematical details that are needed to make our development self-contained, but which would disrupt the reader's attention, if placed in the main body of the paper.

\section{Generalized Ericksen-Leslie Theory}\label{sec:EL_inertial}
Here, mainly following \cite{ericksen:liquid} (and Chapt.~3 of \cite{sonnet:dissipative}), we present the general dynamical theory of nematic liquid crystals that extends the early formulation of the foundation papers \cite{ericksen:anisotropic,ericksen:conservation,leslie:some_1966,leslie:some,leslie:thermal,leslie:continuum}, also summarized in \cite{ericksen:continuum}. Although in the rest of the paper this theory will not be employed in its full-fledged version, it provides the general framework within which our study is developed; it would serve as the natural environment for possible future extensions. As in the pioneering work of Ericksen~\cite{ericksen:twist}, both macro- and micro-\emph{inertia} of the  motion will systematically be included in the picture.

The local molecular organization of nematic liquid crystals is described by a 
director field $\n$, representing the average orientation of the (elongated) molecular aggregates that constitute the material, and a scalar field  $s$, the \emph{degree of orientation}, which vanishes where the orientational order is lost.  Nematic liquid crystals are commonly described as incompressible, dissipative, ordered fluids:\footnote{A nematic fluid is considered to be incompressible insofar as the processes connected with the reorientation of the director are slow compared with the frequency of sound waves.} a continuum theory capable of describing their dynamics must primarily model the coupled evolution of the orientation of the microscopic  constituents, their degree of order, and the macroscopic flow, described by $\n$, $s$, and the velocity field $\vv$, respectively. 

In the general variational approach proposed in \cite{sonnet:dissipative}, which we also adopt here, the dynamical equations are derived from a Lagrange-Rayleigh \emph{dissipation principle}.  
Letting $\body_t$ be the smooth region in three-dimensional space occupied at the time $t$ by a generic \emph{sub-body} during its motion,\footnote{In continuum mechanics, a sub-body is a generic part of a larger body, for which balance laws are written in integral form.} we denote by $(\bm{t}, \bm{c}, K)$ the system of generalized tractions and by $(\bm{b}, \bm{k}, L)$ the system of generalized body forces  expending power against the generalized velocities  $(\vv, \dot\n, \dot{s})$, on the boundary $\partial\body_t$ and in the interior $\mathring{\body}_t$, respectively. A standard localization argument implies the following evolution equations (see Sect.\,3.2 of \cite{sonnet:dissipative} for more details about the general method), 
\begin{subequations}\label{eq:balance_equations}
\begin{gather}
\rho\dot\vv=\bm b+\diver \bm{\mathrm{T}},\label{eq:momentum_balance}\\
\sigma\ddot{\n}+\frac{\partial R}{\partial \accentset{\circ}{\n}}+\frac{\partial W}{\partial\n}-\diver\left(\frac{\partial W}{\partial\nabla \n}\right)-\bm{k}=\mu\n,\label{eq:balance_torques}\\
\frac{\partial W}{\partial s}-\diver\frac{\partial W}{\partial\nabla s}+\frac{\partial R}{\partial\dot{s}}=L,\label{eq:evolution_s_bulk}
\end{gather}
\end{subequations}
in $\body_t$, and
\begin{subequations}
\begin{gather}
\bm{t}=\bm{\mathrm{T}}\normal,\label{eq:traction}\\
\bm{c}=\frac{\partial W}{\partial\nabla \n}\normal,\label{eq:second_boundary_1}\\
\frac{\partial W}{\partial\nabla s}\cdot\normal= K,\label{eq:second_boundary_2}
\end{gather}
\end{subequations}
on $\partial\body_t$, where a superimposed dot denotes the material time derivative, $\rho$ is the mass density, $\sigma$ is the density of (microscopic) moment of inertia, $\mu$ is a  Lagrange multiplier  which ensures that $\n$ obeys the constraint $\n\cdot\n\equiv1$, $W$ is the elastic free-energy density, $R$ is the Rayleigh dissipation function, and $\normal$ is the outer unit normal to $\partial\body_t$.

Typically, for liquid crystals of small molecular weight, $\sigma$ is set equal to zero in the equations that govern the evolution of $\n$, indicating that no inertial torque acts on the director. In these cases, the microkinetic energy associated with the motion of $\n$ is systematically neglected in favor of the predominant macroscopic kinetic energy $\frac12\rho\vv^2$ of the fluid.
However, for lyotropic liquid crystals, particularly for CLCs, this assumption may be inaccurate, as in these fluids large molecular complexes act as elementary constituents of the material, making inertia important.

For nematic materials, the elastic free-energy density $W$ is taken to be a frame-indifferent function $W=W(s,\nabla s,\n,\nabla\n)$ that is positive
definite,
\begin{equation}
	\label{eq:W_positive_definite}
	W(s,\nabla s,\n,\nabla\n)\ge W(s,\bm{0},\n,\bm{0})=W_0(s)\ge0,
\end{equation}
and reflects the nematic symmetry,
\begin{equation}
	\label{eq:W_nematic_symmetry}
	W(s,\nabla s,\n,\nabla\n)=W(s,\nabla s,-\n,-\nabla\n).
\end{equation}

The Rayleigh dissipation function $R$ plays a central role in the dynamics of dissipative fluids. Here it is taken as a quadratic function in the (indifferent)  measures of dissipation. These latter are  the \emph{stretching} tensor $\mathbf{D}$, that is, the symmetric part of the velocity gradient $\nabla\vv$, the \emph{corotational} time derivative $\mathring{\n}$ of the director,
\begin{equation}
	\label{eq:corotational_n_derivatve}
	\mathring{\n}:=\dot{\n}-\mathbf{W}\n,
\end{equation}
where $\mathbf{W}$ is the \emph{vorticity} tensor, that is, the skew part of $\nabla\vv$, and finally the material time derivative of the scalar order parameter, $\dot{s}$. We thus write $R$ as the following function,
 \begin{equation}
\label{eq:dissipation_function}
R(s,\n;\bm{\mathrm{D}},\accentset{\circ}{\n},\dot s)=\beta_1\dot{s}\n\cdot\bm{\mathrm{D}}\n+\frac{\beta_2}{2}\dot s^2+\frac{\gamma_1}{2}\accentset{\circ}{\n}^2+\gamma_2\accentset{\circ}{\n}\cdot\bm{\mathrm{D}}\n
+\frac{\gamma_3}{2}(\bm{\mathrm{D}}\n)^2+\frac{\gamma_4}{2}(\n\cdot\bm{\mathrm{D}}\n)^2+\frac{\gamma_5}{2}\tr{\bm{\mathrm{D}}}^2,
\end{equation}
where the coefficients $\beta$'s and $\gamma$'s are the generalized \emph{viscosities}, functions of $s$ subject to the requirement that $R$ in \eqref{eq:dissipation_function} be positive semidefinite. In particular, the \emph{rotational} (or twist) viscosity $\gamma_1$ must satisfy the inequality $\gamma_1\geq0$.

Equation \eqref{eq:momentum_balance} expresses the balance of linear momentum, while \eqref{eq:balance_torques} and \eqref{eq:evolution_s_bulk} are additional equations  governing the evolution of $\n$ and $s$. It can be shown reasoning as in \cite[p.\,198]{sonnet:dissipative} that the balance equation of rotational momentum is  a consequence of the frame-indifference of the function $W$ and equation \eqref{eq:balance_torques}.\footnote{Equation \eqref{eq:evolution_s_bulk} is a genuinely additional evolution equation, with no bearing on the balance of torques; accordingly, $L$ gives no contribution to the body couple.}

The Cauchy stress tensor $\mathbf{T}$ comprises both elastic and viscous components, as is clear from the following formula (see \cite[p.\,197]{sonnet:dissipative}),
\begin{equation}
\label{eq:stress}
\mathbf{T}=-p\mathbf{I}\underbrace{-\left(\nabla\n\right)^{\mathsf{T}}\frac{\partial W}{\partial\nabla\n}-\nabla s \otimes \frac{\partial W}{\partial\nabla s}}_{\text{elastic stress}}+\underbrace{\frac{1}{2}\left(\n\otimes\frac{\partial R}{\partial\mathring{\n}}-\frac{\partial R}{\partial\mathring{\n}}\otimes\n\right)
+\frac{\partial R}{\partial\mathbf{D}}}_{\text{viscous stress}},
\end{equation}
where $p$ is the pressure, an unknown function representing the Lagrange multiplier that enforces the incompressibility constraint, $\diver\vv\equiv0$. 

The generalized Ericksen-Leslie equations recalled above afford a conveniently simplified description of \emph{defects}, which can be identified with the regions in space (typically, points or lines) where $s$ vanishes. In the dynamical problem that we shall be concerned with, the director field is unlikely to develop defects; it is instead expected to develop another type of singularity, not in $\n$ itself, but both in its gradient $\nabla\n$ and time derivative $\dot{\n}$. For this reason, the assumption that
\begin{equation}
	\label{eq:s_0}
	s\equiv s_0 
\end{equation}
will be adopted in the rest of the paper. Under this assumption, which requires that both $K\equiv0$ and $L\equiv0$ for consistency, equations \eqref{eq:evolution_s_bulk} and \eqref{eq:second_boundary_2} will be void.

In the following section, we shall derive the form of the elastic free-energy density $W=W(\n,\nabla\n)$ appropriate for CLCs; we shall see then in Sect.~\ref{sec:twist_waves} how a type of non-linearity arises there that makes the twist waves generated in CLCs differ from the \emph{solitons} studied in ordinary nematics.\footnote{The reader interested in seeing how solitons of different types featured in both the physics and mathematics of condensed matter at the time when they had already become fashionable may consult \cite{bishop:solitonsl}.}

The solitons mainly involved in liquid crystals are solutions to an appropriate form of the (dissipative) sine-Gordon equation. They were also called \emph{walls} in the first, pioneering studies \cite{helfrich:alignment,degennes:mouvements,brochard:mouvements,leger:observation,leger:static}. These usually turned out to be of small amplitude (and so, difficult to observe) and sustained by an external magnetic field. Later works \cite{guozhen:experiments,lei:soliton,magyari:inertia,lei:generation} then proved the existence of solitons in the director field driven by a hydrodynamic flow; these were also governed by a sine-Gordon equation, but easier to observe experimentally. At variance with the case treated here, in that equation the non-linearity manifested itself in the forcing term, while the differential part remained linear.\footnote{On a similar note, the paper \cite{fergason:liquid} is often credited to present the first prediction of solitons in liquid crystals. However, as also noted in \cite{lei:soliton}, the only evolution equation for the director field  that can be retraced in there is the classical, \emph{linear} wave equation, which cannot sustain solitons.}

\section{Energetics of Chromonics}\label{sec:energetics}
Here, following closely \cite{paparini:what}, we recall the \emph{quartic} elastic theory for CLCs adopted in this paper. What makes chromonic nematics differ from ordinary ones is the \emph{ground state} of their distortion: a \emph{double} twist for the former, a uniform field (along any direction) for the latter. We now explore this difference in more detail.

Like Selinger~\cite{selinger:interpretation}, we write  the elastic energy density of the Oseen-Frank theory \cite{oseen:theory,frank:theory} in an equivalent form,
\begin{equation}
	\label{eq:Frank_equivalent}
	\WOF(\n,\nabla\n)=\frac12(K_{11}-K_{24})S^2+\frac12(K_{22}-K_{24})T^2+\frac12K_{33}B^2+2K_{24}q^2,
\end{equation}
where $S:=\diver\n$ is the \emph{splay}, $T:=\n\cdot\curl\n$ is the \emph{twist}, $B^2:=\bend\cdot\bend$ is the square modulus of the \emph{bend} vector $\bend:=\n\times\curl\n$, and $q>0$ is the \emph{octupolar splay}  \cite{pedrini:liquid} derived from the following equation
\begin{equation}
	\label{eq:identity}
	2q^2=\tr(\nabla\n)^2+\frac12T^2-\frac12S^2.
\end{equation}
Since $(S,T,B,q)$ are independent \emph{distortion measures}, it easily follows from \eqref{eq:Frank_equivalent} that $\WOF$ is positive semi-definite whenever
\begin{subequations}\label{eq:Ericksen_inequalities}
	\begin{eqnarray}
		&K_{11}\geqq K_{24}\geqq0,\label{eq:Ericksen_inequalities_1}\\
		&K_{22}\geqq K_{24}\geqq0, \label{eq:Ericksen_inequalities_2}\\
		&K_{33} \geqq 0,\label{eq:Ericksen_inequalities_3}
	\end{eqnarray}
\end{subequations}
which are the celebrated \emph{Ericksen's inequalities} \cite{ericksen:inequalities}. If these inequalities are satisfied in strict form, the global ground state of $\WOF$ is attained on any uniform director field, characterized by
\begin{equation}
	\label{eq:uniform_ground_state}
	S=T=B=q=0,
\end{equation}
which designates the ground state of ordinary nematics.

CLCs are characterized by a different ground state, which we call a \emph{double twist}, one where all distortion measures vanish, \emph{but} $T$. Here, we adopt the terminology proposed by Selinger~\cite{selinger:director} (see also \cite{long:explicit}) and distinguish between \emph{single} and \emph{double} twists. The former is characterized by
\begin{equation}
	\label{eq:single_twist}
	S=0,\quad B=0,\quad T=\pm2q,
\end{equation}
which designates a director distortion capable of filling \emph{uniformly} the whole space \cite{virga:uniform}. \nigh{Unlike this, a double twist \emph{cannot} fill space uniformly: it can possibly be realized locally, but not everywhere. The essential feature of the quartic twist theory proposed in \cite{paparini:elastic} is to envision a double twist (with two equivalent chiral variants) as ground state of CLCs in three-dimensional space,
\begin{equation}
	\label{eq:double_twist}
	S = 0, \quad T = \pm T_0, \quad B = 0, \quad q = 0.
\end{equation}
The degeneracy of the ground  double twist  in \eqref{eq:double_twist} arises from  the achiral nature of the molecular aggregates that constitute these materials, which is reflected in the lack of chirality of their condensed phases; the elastic stored energy must equally penalize both ground chiral variants in \eqref{eq:double_twist}.} For the Oseen-Frank theory to accommodate such a ground state, inequality \eqref{eq:Ericksen_inequalities_2} must be replaced by $K_{24}\geqq K_{22}\geqq0$, but this comes at the price of making $\WOF$ unbounded below \cite{paparini:stability}. Our minimalist proposal to overcome this difficulty was to add  a \emph{quartic twist} term to the Oseen-Frank stored-energy density,
\begin{equation}
	\label{eq:quartic_free_energy_density}
	\WQT(\n,\nabla\n):=\frac{1}{2}(K_{11}-K_{24})S^2+\frac{1}{2}(K_{22}-K_{24})T^2+ \frac{1}{2}K_{23}B^{2}+2K_{24}q^2 + \frac14K_{22}a^2T^4,
\end{equation}
where $a$ is a \emph{characteristic length}. $\WQT$ is bounded below whenever 
\begin{subequations}\label{eq:new_inequalities}
	\begin{eqnarray}
		&K_{11}\geqq K_{24}\geqq0,\label{eq:new_inequalities_1}\\
		&K_{24}\geqq K_{22}\geqq0, \label{eq:new_inequalities_2}\\
		&K_{33}\geqq0.\label{eq:new_inequalities_3}
	\end{eqnarray}
\end{subequations}
If these inequalities  hold, as we shall assume here, then $\WQT$ is minimum at the  degenerate double twist \eqref{eq:double_twist}
characterized by
\begin{equation}
	\label{eq:T_0min}
	T_0:=\frac{1}{a}\sqrt{\frac{K_{24}-K_{22}}{K_{22}}}.
\end{equation}
Here, we shall treat $a$ as a phenomenological parameter to be determined experimentally.

\section{Twist Waves in Chromonics}\label{sec:twist_waves}
\emph{Twist} waves in nematic liquid crystals were fist studied by Ericksen in \cite{ericksen:twist}. They are special solutions to the hydrodynamic equations under the assumption that  the flow velocity $\vv$ vanishes: this implies that the motion of the director induces \emph{no backflow}. The governing  one-dimensional wave equation derived  in \cite{ericksen:twist} presumes that no extrinsic body forces or couples act on the system, and that the material occupies the whole space, assumptions that will be retained in this paper.

Since $\vv\equiv\bm{0}$ and $\mathbf{D}\equiv\mathbf{0}$, both the material and corotational derivatives of $\n$ reduce to its partial time derivative $\partial_t\n$. Moreover, by combining \eqref{eq:dissipation_function}, \eqref{eq:stress}, and \eqref{eq:s_0}, we readily give the Cauchy stress tensor $\mathbf{T}$ the following form
\begin{equation}
\label{eq:momentum_balance_tw}
 \mathbf{T}=-p\mathbf{I}-\left(\nabla\n\right)^{\mathsf{T}}\frac{\partial W}{\partial\nabla\n}+\mu_2\partial_t\n\otimes\n+\mu_3\n\otimes\partial_t\n,
\end{equation}
where 
\begin{equation}
	\label{eq:mu_2_3}
	\mu_2:=\frac12(\gamma_2-\gamma_1)\quad\text{and}\quad\mu_3:=\frac12(\gamma_2+\gamma_1).
\end{equation}
In the absence of body forces and couples, the balance equations in \eqref{eq:balance_equations} thus reduce to 
\begin{subequations}
	\label{eq:balance_equations_reduced}
	\begin{gather}
		\diver\mathbf{T}=\bm{0},\label{eq:balance_momentum_reduced}\\
		\sigma\partial_{tt}\n+\gamma_1\partial_t\n+\frac{\partial W}{\partial\n}-\diver\left(\frac{\partial W}{\partial\nabla \n}\right)=\mu\n.
	\end{gather}
\end{subequations}
Letting $\n$ be represented in a Cartesian frame $\frameca$ as
\begin{equation}
\label{eq:director_tw}
\n = \cos \twist\e_y + \sin \twist\e_z, 
\end{equation}
where $\twist = \twist(t, x)$ denotes the \emph{twist} angle for  $(t, x) \in [0, \infty) \times \mathbb{R}$ and setting $W=\WQT$, we easily see that $T$ and $q$ are the only (related) distortion measures that do not vanish, 
\begin{equation}
	\label{eq:distortion_measures}
	S=0,\quad T=-\twist_{,x},\quad B=0,\quad 2q^2=\frac12T^2,
\end{equation}
the governing equations \eqref{eq:balance_equations_reduced} are equivalent to
\begin{equation}
	\label{eq:balance_torques_wave}
	\sigma \twist_{,tt}-K_{22}\left(1+3a^2\twist_{,x}^2\right)\twist_{,xx}=-\gamma_1\twist_{,t},
\end{equation}
\begin{equation}
	\label{eq:lagrange_multiplier}
	\mu=-\sigma \twist_{,t}^2+K_{22}\twist_{,x}^2\left(1+a^2\twist_{,x}^2\right),
\end{equation}
and
\begin{equation}
	\label{eq:pressure}
	p=-K_{22}\twist_{,x}^2(1+a^2\twist_{,x}^2)+p_0(t)=-T\frac{\partial\WQT}{\partial T}+p_0(t),
\end{equation}
where $p_0(t)$ is an arbitrary function of time. While equations \eqref{eq:pressure} and \eqref{eq:lagrange_multiplier} determine the Lagrange multipliers associated with the constraints enforced by the theory, \eqref{eq:balance_torques_wave} is the genuine evolution equation of the system, whose solutions thus provide a complete solution to the governing equations.

The following sections will be devoted to the analysis of a special instance of equation \eqref{eq:balance_torques_wave}. The molecular inertia $\sigma$ is responsible for its hyperbolic character: equation \eqref{eq:balance_torques_wave} becomes parabolic if $\sigma$ vanishes. 

\subsection{Non-dimensional form}
We find it convenient to rescale lengths to $\sqrt{3}a$ and times to the  characteristic time 
\begin{equation}
\label{eq:time_characteristic}
\tau:=\sqrt{3}a\sqrt{\frac{\sigma}{K_{22}}}.
\end{equation}
Keeping the original names for the rescaled variables $(t,x)$, we write \eqref{eq:balance_torques_wave} as 
\begin{equation}
\label{eq:wave_equation_characteristic}
\twist_{,tt}-Q^2(\twist_{,x})\twist_{,xx}=-\lambda \twist_{,t} \quad\text{for}\quad (t,x)\in[0,+\infty)\times\mathbb{R},
\end{equation}
where $Q(\xi)$, the positive root of
\begin{equation}
	\label{eq:Q_definition}
	Q^2(\xi):=1+\xi^2,
\end{equation}
is the dimensionless  \emph{wave velocity} and $\lambda$ is a dimensionless \emph{damping} parameter defined as
\begin{equation}
\label{eq:lambda_def}
\lambda:=a\gamma_1\sqrt{\frac{3}{\sigma K_{22}}}.
\end{equation}

In our scaling, the molecular inertia $\sigma$ affects both $\lambda$ and $\tau$, making the former larger and the latter smaller when it is decreased, so that the director evolution becomes overdamped and (correspondingly) its hyperbolic character applies to an ever shrinking time scale.

\nigh{\subsection{Non-dissipative limit}\label{sec:non-dissipative}
In this paper, we shall only be concerned with the non-dissipative limit, where $\lambda\ll1$. Although this limit is precisely identified by the definition of the dimensionless damping parameter in \eqref{eq:lambda_def}, 
singling out a definite class of physical conditions that ensure its validity could be questionable due to the substantial variability of the physical parameters involved in the model. For lyotropic phases, such as CLCs, this variability arises from the intrinsic polydispersity of molecular aggregates and the dependence of their geometry on concentration and temperature, a complication not present in thermotropic phases. Specifically, the average length of molecular aggregates $L$ varies significantly with concentration and temperature, decreasing as the nematic-to-isotropic transition is approached \cite{collings:nature}.

Such a state of affairs is responsible for the considerable variability in $\gamma_1$, $K_{22}$, and $\sigma$, all of which are expected to be proportional to $L^2$, \cite{yu:rotational,zhou:elasticity_2014}. For example, the experimental evidence collected in \cite{zhou:elasticity_2014} and \cite{zhou:ionic} indicates that the rotational viscosity $\gamma_1$ of SSY and DSCG can easily span a few orders of magnitude: for the latter, $\gamma_1$ ranges between  $1$ and $300\, \mathrm{Pa}\,\mathrm{s}$, and for the former  between  $0.1$ and $10 \, \mathrm{Pa}\,\mathrm{s}$, depending on ionic content, temperature, and concentration.

The estimate of the inertial density $\sigma$, which is also affected by the geometric structure of molecular aggregates, is even more uncertain for CLCs than for ordinary thermotropic materials, where molecular lengths are fixed and $\sigma$ can be assumed to be independent of the scalar order parameter (as remarked in \cite{golo:chaos,golo:new}), an assumption which is not valid here.

In previous studies \cite{paparini:elastic,paparini:spiralling,paparini:what,ciuchi:inversion}, we estimated the phenomenological length $a$ from experimental data for CLCs under different spatial confinements; we found that  $a$ is an increasing function of $L$ and we estimated $a\sim10\mu\mathrm{m}$ in ordinary physical conditions. Assuming that $\gamma_1$, $K_{22}$, and $\sigma$ scale with the same power of $L$, we may expect that the decrease of $\lambda$ with $L$ is driven by $a$. This would suggest that $\lambda$ should become small as the nematic-to-isotropic transition is approached, although this might not be the only regime where our approximation is valid.}

In the rest of the paper, we shall set $\lambda=0$ and study systematically the non-dissipative limit of equation \eqref{eq:wave_equation_characteristic}. The physical relevance of our conclusions for actual CLCs will be higher as closer these materials are to \nigh{physical conditions for which $\lambda\ll1$.}  

\section{Mathematical Methodology}\label{sec:method}
In this section, we study the following global Cauchy problem for the function $\twist(t,x)$,
\begin{subequations}
\label{eq:wave_system} 
\begin{numcases}{}
\twist_{,tt}-Q^2(\twist_{,x})\twist_{,xx}=0 &for $(t,x)\in[0,+\infty)\times\mathbb{R}$,\label{eq:wave_eq}\\
\twist(0,x)=\twist_0(x) &for $x\in\mathbb{R}$,\label{eq:wave_initial_w} \\
\twist_{,t}(0,x)=0 &for $x\in\mathbb{R},$\label{eq:wave_initial_w_t}\label{eq:wave_system_c} 
\end{numcases}
\end{subequations}
where $w_0$ is a function of class $\mathcal{C}^2$ such that $w_0'$ is not constant, but bounded, and $Q$ is the function defined in \eqref{eq:Q_definition}.
\begin{remark}\label{rmk:genuinely_nonlinear}
Since $Q(\xi)>0$ for all $\xi\in\mathbb{R}$, we may say that  equation \eqref{eq:wave_eq} is \emph{strictly} hyperbolic, but since $Q'(0)=0$, it is not  \emph{genuinely} nonlinear (according to the definition given in \cite[p.\,15]{majda:compressible}).
\end{remark}

We are mainly interested in providing  conditions sufficient to guarantee that the solution $\twist$ to \eqref{eq:wave_system} breaks down in a finite time, meaning that some second  derivatives of $\twist$ become infinite. Such a breakdown ushers the formation of a twist \emph{shock} wave, where discontinuities in the first derivatives $\twist_{,x}$ and $\twist_{,t}$ arise across a plane traveling in time (with law $x=x_0(t)$), while $\twist$ remains continuous. We shall not study these waves here; we shall be contented to determine initial conditions that necessarily lead to their formation.

The study of twist shock waves in quiescent liquid crystals has a long and interesting history starting with the works of Ericksen~\cite{ericksen:propagation,ericksen:continuum}, but it is limited to \emph{weak} shocks, for which the traveling discontinuities occur either in the second derivatives of the twist angle $\twist$ (acceleration shock waves) or even in higher derivatives (weaker shock waves).

Acceleration and weaker shock waves behave quite differently in liquid crystals if the elastic energy density $W$ grows faster than $T^2$, as is the case for $\WQT$ in \eqref{eq:quartic_free_energy_density}. As shown in \cite{shahinpoor:finite,shahinpoor:effect}, whenever $W$ is more than quadratic in $T$, weaker shock waves decay in a short time, whereas acceleration shock waves may survive for longer times and possibly evolve into ordinary shock waves. Conversely, when $W$ is at most quadratic in $T$, both acceleration and weaker shock waves decay in a finite time. 

Here, we are considering a different approach to twist shock waves: we describe how they can arise in a finite time from a regular (smooth) solution of system \eqref{eq:wave_system}. To this end, the quartic growth of $\WQT$ in $T$ suffices to make \eqref{eq:wave_eq} non-linear.

The occurrence of ordinary shock waves in one-dimensional director distortions in a quiescent liquid crystal has also been studied in both dissipative \cite{chen:singularity} and conservative \cite{glassey:singularities,chen:energy} settings.\footnote{Here we cite just a few relevant works from a vast literature, an accurate account of which is given in \cite{chen:singularity}.} These waves, however, are governed by an equation where the wave velocity $Q$ is a (nonlinear) function of the angle $w$, instead of $w_{,x}$: they are \emph{splay-bend} waves instead of \emph{twist} waves, and so they fall outside the scope of this paper.

Equation \eqref{eq:wave_eq} that governs conservative twist waves has some antecedents in the literature, which we now briefly recall. It was studied in \cite{zabusky:exact} by applying a general method earlier developed in \cite{ludford:extension}, in the case where
\begin{equation}
	\label{eq:Q_fermi}
	Q^2(\xi)=(1+\varepsilon\xi)^\alpha,
\end{equation}
with both $\varepsilon$ and $\alpha$ positive parameters. This special form of the wave velocity was suggested by the pioneering numerical study of a discretized non-linear string \cite{fermi:studies}. While it was proved in \cite{zabusky:exact} that the equation for the continuum would predict a breakdown of the solution after a time $t^\ast=O(1/\varepsilon\alpha)$, the discretized version studied in \cite{fermi:studies} remained smooth at all times. With yet another method, the breakdown result of \cite{zabusky:exact} was extended in \cite{lax:development} to a general class of positive functions $Q$ such that
\begin{equation}
	\label{eq:Q_lax}
	|Q'(\xi)|>0\quad\forall\ \xi\in\mathbb{R}.
\end{equation}
Moreover, a deep analysis of solution breakdown was performed in \cite{maccamy:existence} in a complementary case, where $Q$ obeys the following assumptions.
\begin{equation}
	\label{eq:Q_mizel}
	Q(\xi)>0,\quad Q(0)=1,\quad\text{and}\quad\sgn(Q'(\xi)\xi)=-1\quad\forall\ \xi\neq0.
\end{equation}

Clearly, neither \eqref{eq:Q_lax} nor \eqref{eq:Q_mizel} apply to the function $Q$ in \eqref{eq:Q_definition} that occurs in our system \eqref{eq:wave_system}, for which we thus need newer analytical methods. We found them in the work \cite{manfrin:note}; they will be recalled and adapted to our needs in the rest of this section.

\subsection{Problem Reformulation}
\begin{figure}[h] 
\begin{subfigure}[c]{0.48\linewidth}
	\centering
	\includegraphics[width=.65\linewidth]{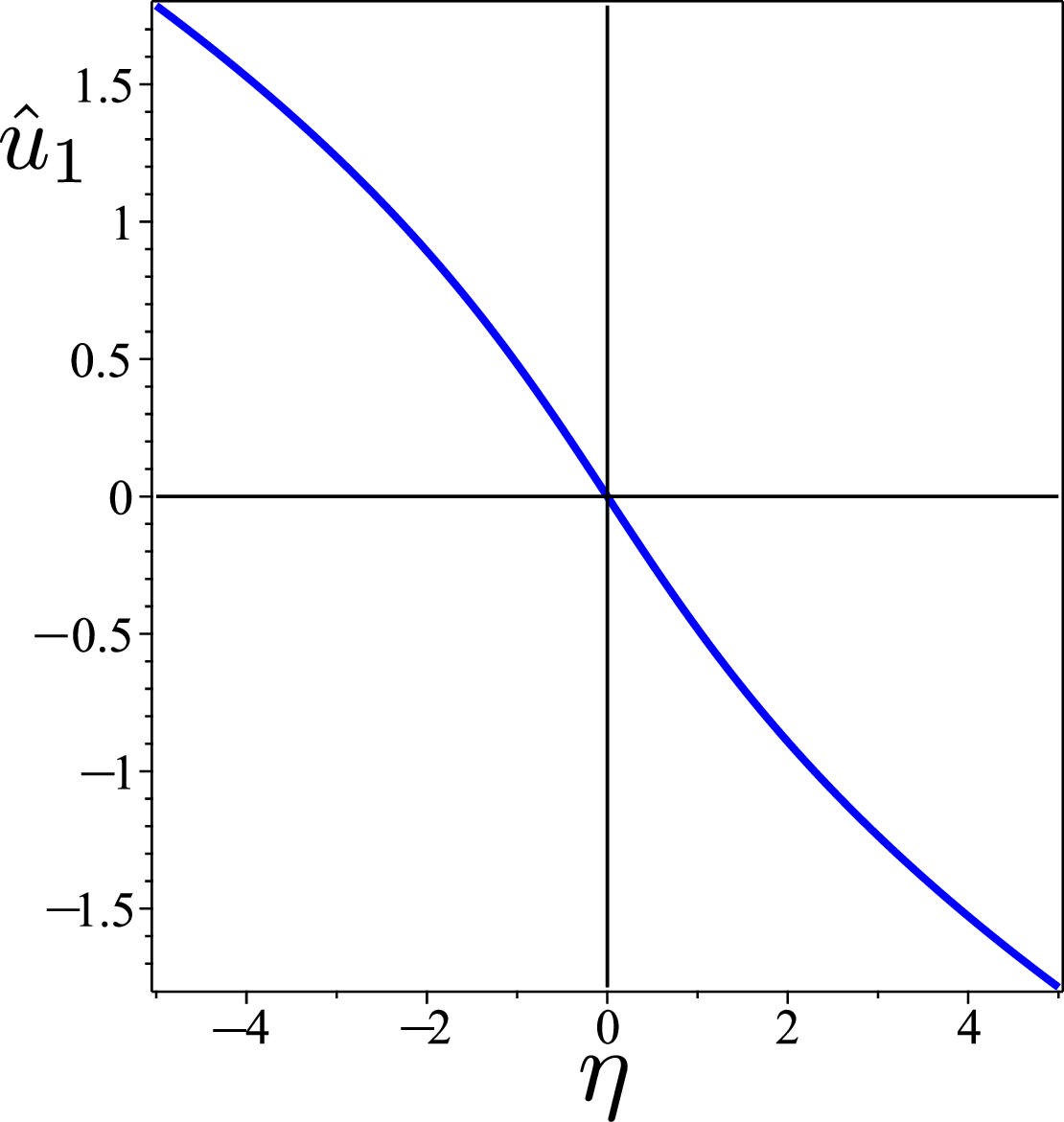}
	\caption{Graphical solution $u_1$ of \eqref{eq:L_u_1} depicted as a function of $\eta$.}
	\label{fig:u_1}
	\end{subfigure}
\quad
	\begin{subfigure}[c]{0.45\linewidth}
	\centering
	\includegraphics[width=0.6\linewidth]{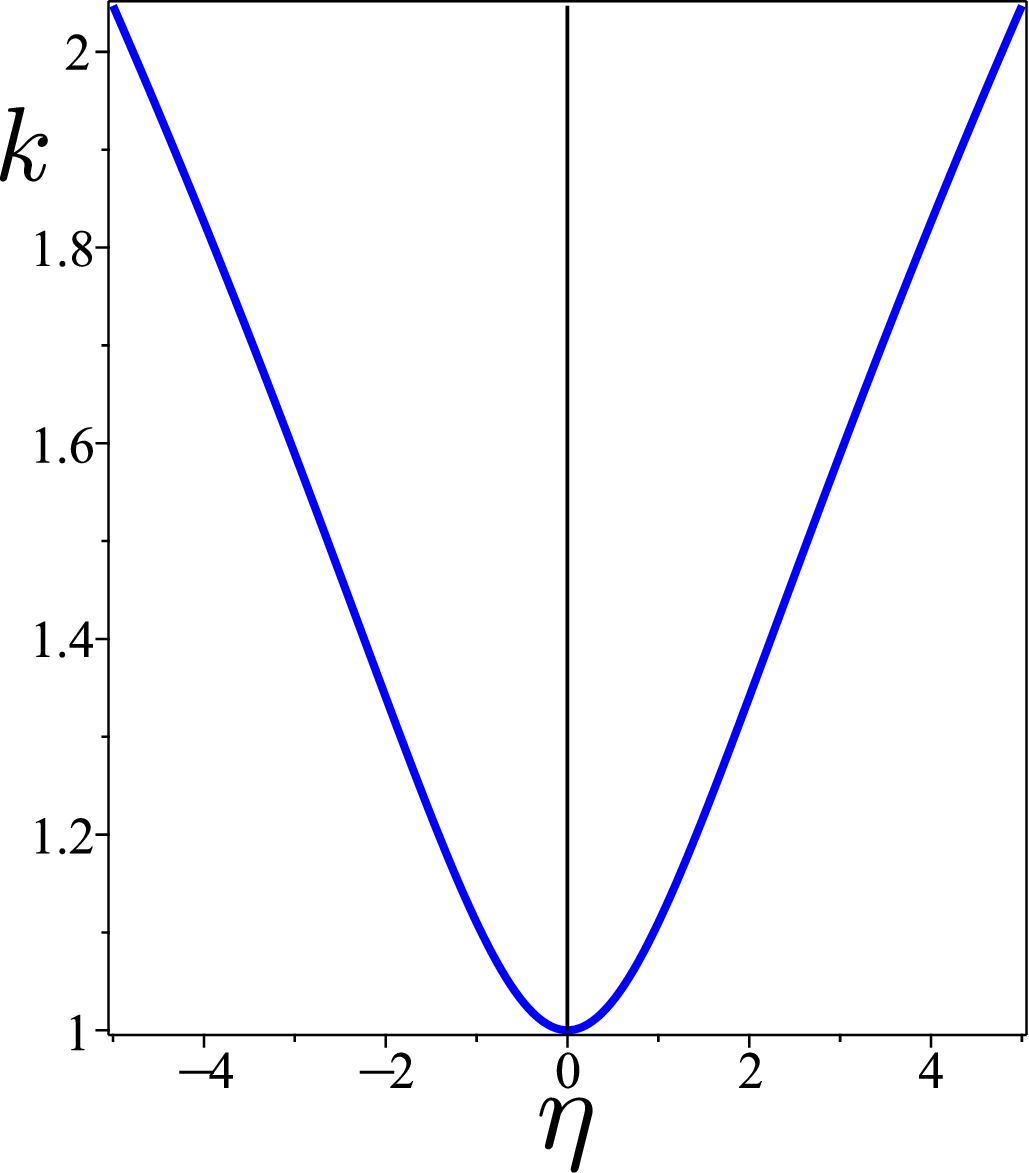}
	\caption{Graph of the function $k$ in \eqref{eq:k_l_r}, illustrating its dependence on $\eta$ through $\widehat{u}_1$ defined in \eqref{eq:u_1_rl}.}
	\label{fig:k_rl}
\end{subfigure}
\caption{Functions $\widehat{u}_1$ and $k$ defined in \eqref{eq:u_1_rl} and \eqref{eq:k_l_r}, respectively, expressed in terms  of $\eta:=r-\ell$.}
\label{fig:k_prime_special}
\end{figure}

We start by considering as independent unknowns the following first-order derivatives,
\begin{equation}
\label{eq:first_order_unknowns}
u_1(t,x):=\twist_{,x}(t,x), \quad u_2(t,x):=\twist_{,t}(t,x)
\end{equation}
By their use, we transform \eqref{eq:wave_system} into a first-order system, 
\begin{subequations}
\label{eq:first_order_system} 
\begin{numcases} {}
u_{1,t}=u_{2,x},\\
u_{2,t}=Q^2(u_1)u_{1,x}.
\end{numcases}
\end{subequations}
Applying classical methods (see, for example, \cite{chang:existence,maccamy:existence,klainerman:formation}), we diagonalize
system \eqref{eq:first_order_system} with the aid of 
Riemann's \emph{invariants} $r$ and $\ell$ defined by
\begin{subequations}
\label{eq:riemann_system} 
\begin{numcases} {}
r(t,x):=u_2(t,x)-L(u_1(t,x)),\label{eq:riemann_system_a} \\
\ell(t,x):=u_2(t,x)+L(u_1(t,x))\label{eq:riemann_system_b} ,
\end{numcases}
\end{subequations}
where $L$ is an appropriate mapping.
It is a simple matter to show that by setting 
\begin{equation}
\label{eq:L_prime_u1}
L'(u_1)=Q(u_1),
\end{equation}
the system \eqref{eq:first_order_system} can be written as
\begin{subequations}
	\label{eq:pre_diagonal_system} 
	\begin{numcases} {}
		r_{,t}+Qr_{,x}=0 &for $(t,x)\in[0,t_\ast)\times\mathbb{R}$, \label{eq:pre_r_syst}\\
		\ell_{,t}-Q\ell_{,x}=0 &for $(t,x)\in[0,t_\ast)\times\mathbb{R}$, \label{eq:pre_l_syst}
	\end{numcases}
\end{subequations}
where $[0,t_\ast)$ is  the \emph{maximal interval} of classical existence. For \eqref{eq:pre_diagonal_system} to acquire the desired diagonal form, we  need to express $Q$ as a function of $r$ and $\ell$ only. This can be achieved by use of \eqref{eq:Q_definition} and \eqref{eq:riemann_system}, which  lead us to  
\begin{equation}
\label{eq:L_u_1}
\ell(t,x)-r(t,x)=2L(u_1)=2\int_0^{u_1}Q(\xi)\dd\xi=u_1\sqrt{1+u_1^2}+\arcsinh u_1,
\end{equation}
where we have set $L(0)=0$, with no prejudice for the validity of \eqref{eq:L_prime_u1}. By inverting \eqref{eq:L_u_1}, we obtain that
\begin{equation}
\label{eq:u_1_rl}
u_1=\widehat{u}_1(\eta):=L^{-1}\left(-\frac12\eta\right),
\end{equation}
where
\begin{equation}
	\label{eq:eta_definition}
	\eta:=r-\ell.
\end{equation}
In \eqref{eq:pre_diagonal_system} we can thus formally replace the function $Q(u_1)$ with 
\begin{equation}
	\label{eq:k_l_r}
	k(\eta):=Q(\widehat{u}_1(\eta)),
\end{equation}
finally arriving at
\begin{subequations}
	\label{eq:diagonal_system} 
	\begin{numcases} {}
		r_{,t}+k(r-\ell)r_{,x}=0 &for $(t,x)\in[0,t_\ast)\times\mathbb{R}$, \label{eq:r_syst}\\
		\ell_{,t}-k(r-\ell)\ell_{,x}=0 &for $(t,x)\in[0,t_\ast)\times\mathbb{R}$, \label{eq:l_syst}
	\end{numcases}
subject to the initial conditions
\begin{equation}
	\label{eq:IC_riemann}
	r(0,x)=r_0(x)=-L(\twist_0'(x)), \quad \ell(0,x)=\ell_0(x)=L(\twist_0'(x))=-r_0(x).
\end{equation}
\end{subequations}
The graphs of both  functions $\widehat{u}_1(\eta)$ and $k(\eta)$ are illustrated in Fig.~\ref{fig:k_prime_special}.

The \emph{characteristics} of \eqref{eq:diagonal_system} are family of curves $x=x_1(t,\alpha)$ and $x=x_2(t,\beta)$, indexed in $\alpha, \, \beta\in\mathbb{R}$,  along which the Riemann invariants $r$ and $\ell$ remain constant. It readily follows from \eqref{eq:r_syst} that the curves along which $r$ is constant solve the differential problem
\begin{equation}
\label{eq:x_1}
\begin{cases}
\frac{\dd x_1}{\dd t}(t,\alpha)=k(\eta((t,x_1(t,\alpha))), \\
x_1(0,\alpha)=\alpha.
\end{cases}
\end{equation}
In view of \eqref{eq:IC_riemann}, we can then write that
\begin{equation}
\label{eq:r_constant}
r(t,x_1(t,\alpha))=r_0(\alpha) \quad \hbox{for every } t\in[0,t_\ast).
\end{equation}
Similarly, the characteristics along which $\ell$ is constant solve the differential problem
\begin{equation}
\label{eq:x_2}
\begin{cases}
\frac{\dd x_2}{\dd t}(t,\beta)=-k(\eta(t,x_2(t,\beta))), \\
x_2(0,\beta)=\beta,
\end{cases}
\end{equation}
and
\begin{equation}
\label{eq:l_constant}
\ell(t,x_2(t,\beta))=\ell_0(\beta) \quad \hbox{for every } t\in[0,t_\ast).
\end{equation}
\begin{definition}
	\label{def:forward_backward}
	Since $k\ge0$, we shall also say that $x_1$ is the \emph{forward} characteristic, whereas $x_2$ is the \emph{backward} characteristic.
\end{definition}
\begin{remark}
	\label{rmk:forward_backward}
The reader should be advised though that this definition is not universally accepted: in equations (3.52) of \cite{majda:compressible}, for example, the role of the two characteristics is interchanged.
\end{remark}
\begin{remark}\label{rmk:boundedness}
Again by \eqref{eq:IC_riemann}, since $\twist_0'$ is bounded by assumption, the initial values of the Riemann invariants, $r_0$ and $\ell_0$ are also  bounded continuous functions with bounded continuous derivatives. Thus, by the general theory presented in \cite{douglis:existence}, system \eqref{eq:diagonal_system} has a unique $\mathcal{C}^1$  solution $(r,\ell)$ locally in time.  
Furthermore, it follows from \eqref{eq:r_constant} that 
\begin{equation}
\label{eq:riemann_bounded}
|r(t,x)|\leq||r_0||_{\infty} \quad\text{and}\quad |\ell(t,x)|\leq||\ell_0||_{\infty} \quad \text{for all}\quad (t,x)\in[0,t_\ast)\times \mathbb{R},
\end{equation}
where $||\cdot||_{\infty}$ denotes the $L^\infty$-norm. Thus, both $r(t,x)$ and $\ell(t,x)$ are also  uniformly bounded in $[0,t_\ast)\times\mathbb{R}$, and so is also $\eta$.
\end{remark}
\begin{remark}
	\label{rmk:dichotomy}
	By the \emph{continuation principle} (see, for example, \cite[p.\,100]{majda:compressible} for this particular incarnation), the estimate \eqref{eq:riemann_bounded} implies the following \emph{dichotomy}: either there is a \emph{critical} time $t^\ast$ such that
	\begin{equation}
		\label{eq:dichotomy_estimate}
		||r_{,x}||_\infty+||\ell_{,x}||_\infty\to+\infty\quad\text{for}\quad t\to t^\ast,
	\end{equation}
or there is a \emph{global} smooth solution of \eqref{eq:riemann_system} for all $0\leq t<+\infty$. In the former case, which is the one we are interested in, a \emph{shock} wave is formed in a finite time. In the latter case, we conventionally set $t^\ast=+\infty$.
\end{remark}

\subsection{Conservation Laws}
Before analyzing in detail the characteristics of  \eqref{eq:diagonal_system}, we pause to study two conservation laws enjoyed by the regular solutions of the original system \eqref{eq:wave_system}. We may identify one conserved quantity with an effective \emph{mass} of the system, and the other with its \emph{energy}. We shall employ the following lemma, which asserts that the limits of the spatial and time derivatives of the twist angle $\twist(t,x)$ as $x$ approaches $\pm\infty$, coincide with the limits of the initial data's derivatives.
\begin{lemma}
	\label{lemma:w_x_t_infty}
	Let $\twist'_0(\pm\infty):=\lim_{x\to\pm\infty}\twist_0'(x)$ be finite. For a $\mathcal{C}^1$ solution $\twist(t,x)$ of the system \eqref{eq:wave_system} in $[0,t^\ast)\times\mathbb{R}$, the following limits holds for every $t\in[0,t^\ast)$:
	\begin{equation}
		\label{eq:lim_infinity_wx}
		\lim_{x\to\pm\infty}\twist_{,x}(t,x)=\twist_0'(\pm\infty).
	\end{equation}  
	Similarly, since $\twist_{,t}(0,x)=0$ for all $x\in\mathbb{R}$, then
	\begin{equation}
		\label{eq:lim_infinity_wt}
		\lim_{x\to\pm\infty}\twist_{,t}(t,x)=0,
	\end{equation}  
	for all $t\in[0,t^\ast)$.
\end{lemma}
\begin{proof}
	After rewriting  \eqref{eq:u_1_rl} with the aid of \eqref{eq:eta_definition} and \eqref{eq:first_order_unknowns} as
	\begin{equation}
		\label{eq:w_x_r_l}
		\twist_{,x}(t,x)=-L^{-1}\left(\frac{r(t,x)-\ell(t,x)}{2}\right),
	\end{equation}
	we consider the two characteristic curves $x=x_1(t,\alpha_0)$ and $x=x_2(t,\beta_0)$ with $\alpha_0$ and $\beta_0$ selected so that these curves meet at a given point  $(t_0,x_0)$; as long as the solution remains regular, they are uniquely identified. From \eqref{eq:r_constant} and \eqref{eq:l_constant}, $r$ and $\ell$ remain correspondingly constant along these curves. Therefore $\twist_{,x}(t_0,x_0)$ can be expressed as
	\begin{equation}
		\label{eq:w_x_characteristic}
		\twist_{,x}(t_0,x_0)=-L^{-1}\left(\frac{r_0(x_1(0,\alpha_0))-\ell_0(x_2(0,\beta_0)}{2}\right).
	\end{equation}
	Since, for any given $t$, $\lim_{\alpha\to\pm\infty}x_{1}(t,\alpha)=\lim_{\beta\to\pm\infty}x_{2}(t,\beta)=\pm \infty$, by the arbitrariness of $(t_0,x_0)$ it follows from \eqref{eq:w_x_characteristic} that
	\begin{equation}
		\label{eq:w_x_characteristic_infinity}
		\lim_{x\to\pm\infty}\twist_{,x}(t,x)=-\lim_{\alpha,\beta\to\pm\infty}L^{-1}\left(\frac{r_0(x_1(0,\alpha))-\ell_0(x_2(0,\beta)}{2}\right)=-L^{-1}\left(\frac{r_0(\pm\infty)-\ell_0(\pm\infty)}{2}\right)=\twist_0'(\pm\infty),
	\end{equation}
	where we have set $r_0(\pm\infty):=\lim_{\alpha\to\pm\infty}r_0(x_1(0,\alpha))$ and $\ell_0(\pm\infty):=\lim_{\beta\to\pm\infty}\ell_0(x_2(0,\beta))$. 
	Similarly, \eqref{eq:lim_infinity_wt} is obtained by treating in the same way the equation
	\begin{equation}\label{eq:w_t_characteristic}
		\twist_{,t}(t,x)=\frac{r(t,x)+\ell(t,x)}{2},
	\end{equation}
	which follows from \eqref{eq:riemann_system}, and by making use of \eqref{eq:wave_initial_w_t}.
\end{proof}

Building upon Lemma \ref{lemma:w_x_t_infty}, we now derive two conservation laws for the system \eqref{eq:wave_system}.

\begin{proposition}\label{prop:mass_conservation}
	Let a regular solution $w(t,x)$ of system \eqref{eq:wave_system} be integrable over $\mathbb{R}$ for all $t\in[0,t^\ast)$ and let $M$ be the function of $t$ defined by
	\begin{equation}
		\label{eq:M_defnition}
		M(t):=\int_{\mathbb{R}}w(t,x)\dd x.
	\end{equation}
	If the initial datum $w_0(x)$ in \eqref{eq:wave_initial_w} is such that $w_0'(+\infty)=w_0'(-\infty)$ then
	\begin{equation}
		\label{eq:M_conservation}
		M(t)=M(0)\quad\forall\ t\in[0,t^\ast).
	\end{equation}
\end{proposition}
\begin{proof}
	Equation \eqref{eq:wave_eq} can also be rewritten in the form
	\begin{equation}
		\label{eq:wave_eq_rewritten}
		\nigh{\twist_{,tt}}-\partial_x\left(\twist_{,x}+\frac13\twist_{,x}^3\right)=0.
	\end{equation}
	By integrating over $\mathbb{R}$ both sides of \eqref{eq:wave_eq_rewritten}, by Lemma~\ref{lemma:w_x_t_infty}, we readily obtain that $\ddot{M}(t)=0$. The desired conclusion then follows from $\dot{M}(0)=0$, which is a consequence of \eqref{eq:wave_initial_w_t} and \eqref{eq:M_defnition}.
\end{proof}
If $M$ can be seen as a conserved effective mass, the existence of a conserved energy $E$ is established by the following proposition.
\begin{proposition}
	\label{prop:conservation_law}
	For a regular solution of the system \eqref{eq:wave_system}, the following  conservation law holds,
	\begin{equation}
		\label{eq:energy_constant}
		E(t):=\frac{1}{2}\int_{\mathbb{R}}\left[\twist_{,t}^2+\twist_{,x}^2\left(1+\frac{1}{6}\twist_{,x}^2\right)\right]\dd x=\frac{1}{2}\int_{\mathbb{R}}\left[\twist_{0}'^2\left(1+\frac{1}{6}\twist_0'^2\right)\right]\dd x=E(0).
	\end{equation}
\end{proposition}
\begin{proof}
	We multiply both sides of equation \eqref{eq:wave_eq_rewritten} by $\twist_{,t}(t,x)$ and then integrate by parts with respect to $x$ over $\mathbb{R}$; thus we arrive at
	\begin{equation}
		\label{eq:energy_law0}
		\partial_{t}\int_{\mathbb{R}}\frac{1}{2}\left[\twist_{,t}^2+\twist_{,x}^2\left(1+\frac{1}{6}\twist_{,x}^2\right)\right]\dd x=\left.\left[\twist_{,t}\twist_{,x}\left(1+\frac{1}{3}\twist_{,x}^2\right)\right]\right|_{-\infty}^{+\infty}.
	\end{equation}
	By Lemma \ref{lemma:w_x_t_infty}, the boundary terms in \eqref{eq:energy_law0} vanish for every $t$, leading to
	\begin{equation}
		\label{eq:energy_law}
		\dot{E}=0,
	\end{equation}
	which shows that the total energy $E(t)$ is conserved.
\end{proof}
\begin{remark}\label{rmk:quartic}
	It should be noted that as a consequence of the quartic term in $T$ featuring in $\WQT$ the integrand of $E$ is also quartic in $w_{,x}$.
\end{remark}
\nigh{
	\begin{remark}\label{rmk:law_role}
These conservation laws will be instrumental in Section \ref{sec:applications} to assess the accuracy of our numerical solutions. Specifically, we will compute the conserved quantities 
$M(t)$ and $E(t)$ for a sequence of times in the interval $[0, t^*)$ and check whether they remain indeed constant.
\end{remark}
}

\subsection{Properties of Characteristics}
Here, to study the system \eqref{eq:diagonal_system}, we apply the general method proposed in \cite{keller:periodic} (see also Chapt.\,3 of \cite{majda:compressible}), which has in \cite{manfrin:note} one of its most recent extensions.  We will derive identities for smooth solutions of \eqref{eq:diagonal_system} through a geometric approach that focuses on the behavior of the characteristic curves belonging to the families \eqref{eq:x_1} and \eqref{eq:x_2}. This will prepare the ground for the analysis of the formation of shocks along characteristics performed in the following section. Specifically, we shall focus on solution breakdowns  associated with the degeneracy of characteristics.

Detailed proofs of the following preparatory results are deferred to Appendix~\ref{sec:manfrin}. Here, we concentrate on their statements, along with a brief discussion of their significance in our context.

A special role is played in our analysis by the wave \emph{infinitesimal compression ratios}, which are formally defined as follows.
\begin{definition}
	\label{def:infinitesimal_compression_ratios}
	For each characteristic curve, $x=x_1(t,\alpha)$ and $x=x_2(t,\beta)$, corresponding to a smooth solution of \eqref{eq:diagonal_system}, the wave \emph{infinitesimal compression ratios}, $c_1$ and $c_2$, are defined by
	\begin{equation}
		\label{eq:infinitesimal_compression_ratios}
		c_1(t,\alpha):=\frac{\partial x_1}{\partial\alpha}\quad\text{and}\quad c_2(t,\beta):=\frac{\partial x_2}{\partial\beta}.
	\end{equation}
\end{definition}
The following Proposition is an adaptation to our context of a result proved in \cite{majda:compressible} (see, in particular, their equations (3.74) and (3.76)). 
\begin{proposition}
\label{prop:infinitesimal_compression_ratios}
If the pair $(r,\ell)$ is a solution of class $\mathcal{C}^1$ of \eqref{eq:diagonal_system}, then  the wave infinitesimal compression ratios $c_1$ and  $c_2$ are given by
\begin{subequations}
\label{eq:infinitesimal_compression_ratios_formulae}
\begin{align}
c_1(t,\alpha)&=\sqrt{\frac{k(r_0(\alpha)-\ell(t,x_1(t,\alpha)))}{k(2r_0(\alpha))}}\left\{1+r_0'(\alpha)\sqrt{k(2r_0(\alpha))}\int_0^t f\left(r_0(\alpha)-\ell(\tau,x_1(\tau,\alpha))\right)\dd \tau\right\},\label{eq:infinitesimal_compression_ratio_x1}\\
c_2(t,\beta)&=\sqrt{\frac{k(r(t,x_2(t,\beta))-\ell_0(\beta))}{k(2r_0(\beta))}}\left\{1+\ell_0'(\beta)\sqrt{k(2r_0(\beta))}\int_0^t f\left(r(\tau,x_2(\tau,\beta))-\ell_0(\beta)\right)\dd \tau\right\},\label{eq:infinitesimal_compression_ratio_x2}
\end{align}
\end{subequations}
where $f$ is the function defined by
\begin{equation}
	\label{eq:f_definition}
	f(\eta):=\frac{k'(\eta)}{\sqrt{k(\eta)}}.
\end{equation}
\end{proposition}
\begin{remark}
\label{rmk:boundness}
To estimate $c_1$ and $c_2$
in \eqref{eq:infinitesimal_compression_ratios_formulae}, we can rely on the boundedness of the Riemann invariants. Specifically, from \eqref{eq:riemann_bounded} we have that
\begin{equation}
	\label{eq:invariant_boudness}
	|r(t,x)-\ell(t,x)|\leq ||r_0||_\infty+||\ell_0||_\infty\quad\text{for every}\quad (t,x)\in[0,t^\ast)\times\mathbb{R}.
\end{equation}
Consequently, from the definition of $k$ in \eqref{eq:k_l_r} and the monotonicity of $L$ as expressed by \eqref{eq:L_u_1}, we also have that under the assumptions of Proposition~\ref{prop:infinitesimal_compression_ratios} $k$ is subject to the following lower and upper bounds,
\begin{equation}
\label{eq:estimates}
1=k(0)\leq k(\eta)\leq\delta \quad\text{with}\quad \delta:=k\left(||r_0||_{\infty}+||\ell_0||_{\infty}\right)>1.
\end{equation}
Thus, since $\eta:=r-\ell$ is bounded, so is also $f$.
\end{remark}
\begin{remark}
	\label{rmk:f_parametric}
	Since $f$ is defined implicitly by \eqref{eq:k_l_r}, it is better represented in parametric form. By differentiating both side of \eqref{eq:k_l_r}, making use of \eqref{eq:Q_definition}, \eqref{eq:L_prime_u1}, and \eqref{eq:u_1_rl}, we easily arrive at 
	\begin{equation}
		\label{eq:f_parametric}
		\eta=-u\sqrt{1+u^2}-\arcsinh u,\quad f=-\frac12\frac{u}{(1+u^2)^{5/4}},
	\end{equation}
the former being simply \eqref{eq:L_u_1} rewritten. It is clear from \eqref{eq:f_parametric} that $f$ is an odd function of $\eta$. It exhibits an isolated minimum at $\eta=\eta_0$ and an isolated maximum at $\eta=-\eta_0$, whose values are $\mp f_0$, respectively, corresponding to the values $u_0=\pm\sqrt{2/3}$ of the parameter $u$ in \eqref{eq:f_parametric}.
\end{remark}

\begin{figure}[h] 
	\centering
	\includegraphics[width=0.3\linewidth]{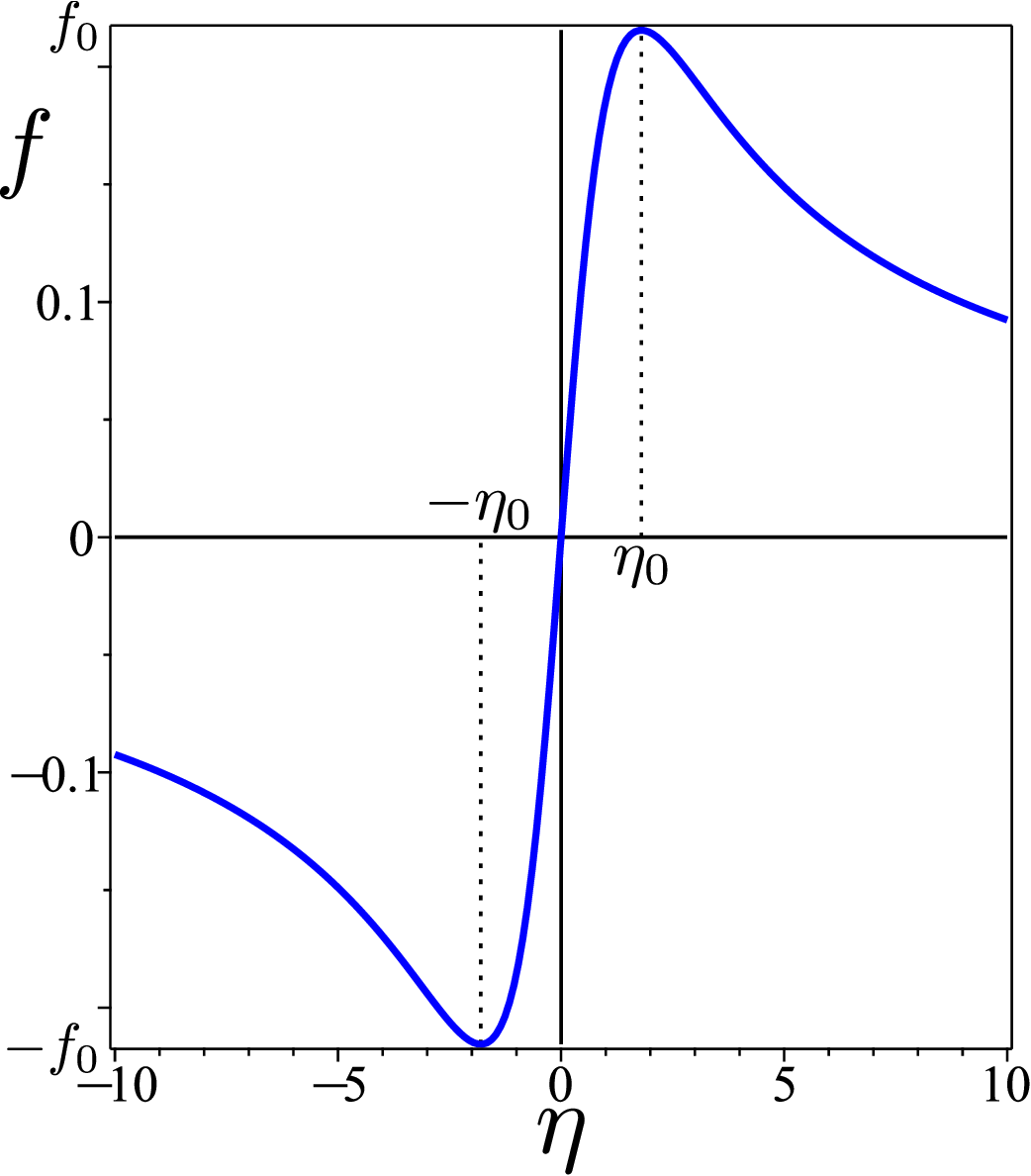}
	\caption{Graph of the odd function $f(\eta)$. It is bounded between $-f_0$ and $+f_0$, which are the values attained by $f$ at $\eta=\mp\eta_0$, respectively. The numerical values are $\eta_0\doteq1.80$ and $f_0\doteq0.22$, which correspond to $u_0=-\sqrt{2/3}$ in \eqref{eq:f_parametric}.}
	\label{fig:k_prime_rl}
\end{figure}

The graph of the function $f(\eta)$ is illustrated in Fig. \ref{fig:k_prime_rl}. 

The infinitesimal compression ratios computed in \eqref{eq:infinitesimal_compression_ratios_formulae} serve as \emph{sentinels} for shock formation.
\begin{remark}
	\label{rmk:sentinels}
	By differentiating both sides of \eqref{eq:r_constant} and \eqref{eq:l_constant} with respect to $\alpha$ and $\beta$, respectively, we obtain that 
	\begin{equation}
		\label{eq:r_x_characteristic}
		r_{,x}(t,x_1(t,\alpha))c_1(t,\alpha)=r_0'(\alpha), \quad \ell_{,x}(t_0,x_2(t,\beta))c_2(t,\beta)=\ell_0'(\beta).
	\end{equation}
Thus, by Remark~\ref{rmk:dichotomy}, if either $r_0'(\alpha)\neq0$ or $\ell_0'(\beta)\neq0$, then a shock wave is formed in a smooth solution whenever either $c_1$ or $c_2$ vanishes.
\end{remark}

\begin{figure}[h] 
	\centering
	\includegraphics[width=0.4\linewidth]{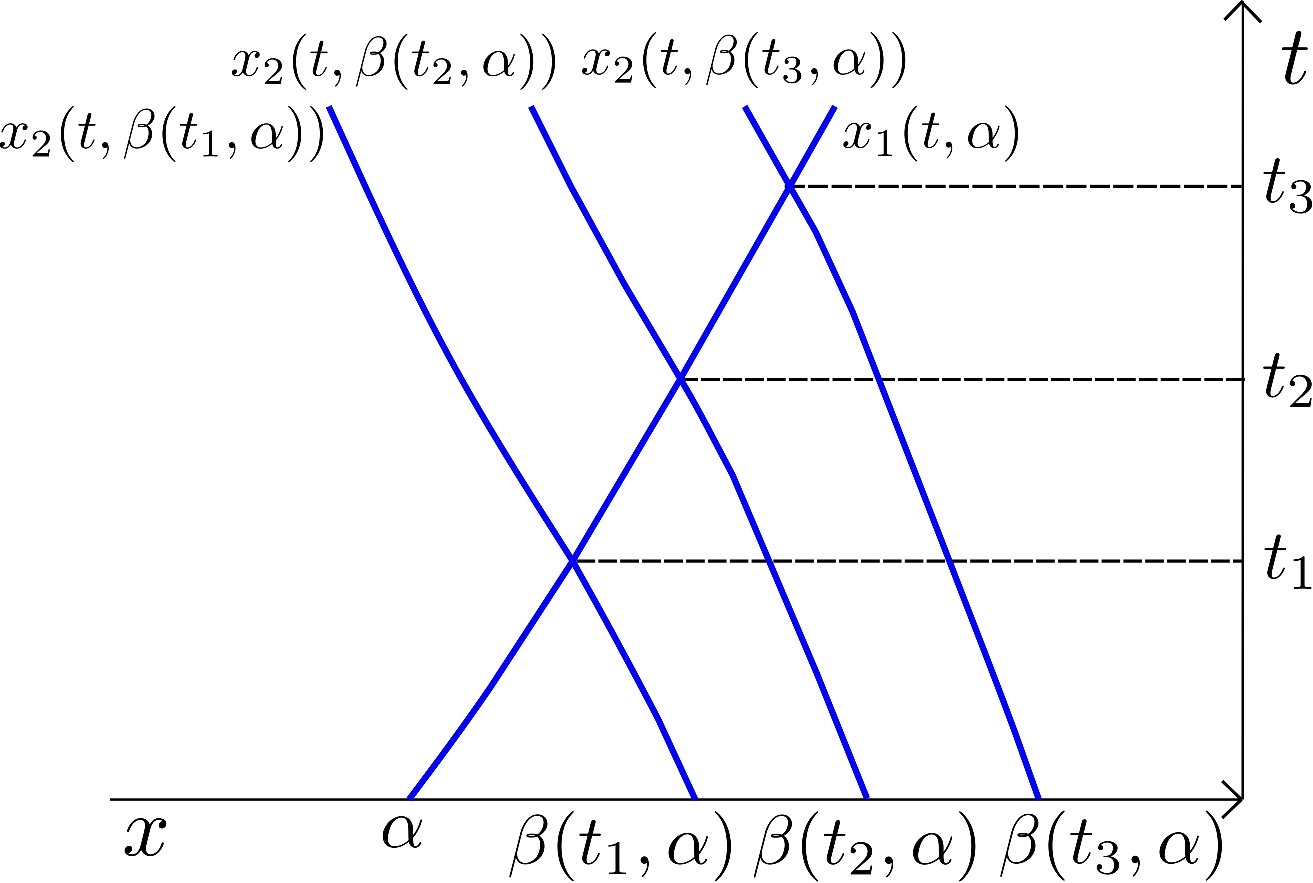}
	\caption{A forward characteristic $x=x_1(t,\alpha)$ starting  from $\alpha$ at $t=0$ and traversing different time levels, $0<t_1<t_2<t_3$, where it intersects different backward characteristics emanating from $\beta(t_i,\alpha)$, for $i=1, \, 2, \, 3$.}
	\label{fig:characteristics_beta}
\end{figure}

The following Proposition (a more general version of which is stated in \cite{manfrin:note}) concerns the sign of $c_1$ and $c_2$.
\begin{proposition}
If the pair $(r,\ell)$ is a local solution  of class $\mathcal{C}^1$ of the system \eqref{eq:diagonal_system} in $[0,t^\ast)\times\mathbb{R}$, then 
\begin{equation}
\label{eq:no_overlapping}
c_1(t,\alpha)>0, \quad c_2(t,\beta)>0, \quad \text{for}\ t\in[0,t^\ast), \, \text{and every}\ \alpha,\beta\in\mathbb{R}.
\end{equation}
\end{proposition}
\begin{remark}
	\label{rmk:no_overlapping}
	As long as both $r$ and $\ell$ are of class $\mathcal{C}^1$, inequalities \eqref{eq:no_overlapping} remain valid, and \emph{vice versa}, meaning that characteristics in the same family do \emph{not} crash on one another. If, on the other hand, there exist a value of $\alpha$  and a finite time $t^\ast(\alpha)$ or a value of $\beta$ and a finite time $t^\ast(\beta)$ such that the corresponding compression ratio vanishes, then by Remark~\ref{rmk:sentinels} either $r_{,x}$ or $\ell_{,x}$ diverges  along the corresponding characteristic, provided that $r_0'(\alpha)\neq0$ or $\ell_0'(\beta)\neq0$.
\end{remark}
\begin{remark}
	\label{rmk:compact_support}
	It is proven in \cite{majda:compressible} (see Theorem\,3.4, which builds upon the earlier work \cite{klainerman:formation}) that for all data $r_0$ and $\ell_0$ with \emph{compact support} and values  ranging in appropriate intervals, system \eqref{eq:diagonal_system} develops a shock wave in a finite time.
\end{remark}
\begin{remark}
	\label{rmk:2_times_2}
The analysis of hyperbolic systems of two scalar equations in a single spatial variable, such as \eqref{eq:diagonal_system}, is much richer in results than the analysis of more general hyperbolic systems. Notable among these are the precise breakdown estimates achieved with Lax's geometric method \cite{lax:hyperbolic} (recounted in Theorem\,3.5 of \cite{majda:compressible}). However, they are \emph{not} generally applicable to our setting, as they would require constraining the data so that $k'\neq0$, which would be rather restrictive an assumption, as by \eqref{eq:k_l_r} and \eqref{eq:IC_riemann} this would amount to require that $w_0'\neq0$.
\end{remark}
Here we build instead on more recent work \cite{manfrin:note} to establish similar breakdown estimates for more general data. To this end, we collect below a number of preliminary properties of the solutions of \eqref{eq:diagonal_system} that will be instrumental to a detailed analysis of the specific cases we are interested in.
\begin{remark}
	\label{rmk:characteristic_crossing}
	Every point  $(t,x)$ reached by a forward characteristic, so that $x=x_1(t,\alpha)$ for some $\alpha\in\mathbb{R}$, can be seen as the end-point at time $t$ of a backward characteristic within the family defined by \eqref{eq:x_2}. Such a  backward characteristic originates at $\beta=\beta(t,\alpha)$, a point in $\mathbb{R}$ depending on $t$, the time of intersection between the two characteristics, and $\alpha$, the starting point in $\mathbb{R}$ of the forward characteristic  (see Fig.~\ref{fig:characteristics_beta}). 
\end{remark}
Formally, the function $\beta(t,\alpha)$ is implicitly defined by the equation
\begin{subequations}\label{eq:alpha_beta_implicit_definitions}
	\begin{equation}
	\label{eq:beta_of_alpha_implicit_definition}
	x_1(t,\alpha)=x_2(t,\beta(t,\alpha).
\end{equation}
Conversely, by exchanging the roles of forward and backward characteristics, we may regard $\alpha$ as a function of $(t,\beta)$ defined implicitly by
\begin{equation}
	\label{eq:alpha_of_beta_implicit_definition}
	x_2(t,\beta)=x_1(t,\alpha(t,\beta)).
\end{equation}
\end{subequations}
\begin{proposition}
\label{prop:beta_t_alpha}
Equations \eqref{eq:alpha_beta_implicit_definitions} have a 
unique solution, $\beta=\beta(t,\alpha)$ and $\alpha=\alpha(t,\beta)$, respectively. The mappings $t\mapsto\beta(t,\alpha)$ and $t\mapsto\alpha(t,\beta)$ are both of class $\mathcal{C}^1$, strictly increasing and decreasing, respectively, and such that
\begin{equation}
\label{eq:d_t_beta}
\beta_{,t}(t,\alpha)=\frac{2k(r_0(\alpha)-\ell_0(\beta(t,\alpha)))}{c_2(t,\beta(t,\alpha))}>0 \quad \text{and}\quad \alpha_{,t}(t,\beta)=-\frac{2k(r_0(\alpha(t,\beta))-\ell_0(\beta))}{c_1(t,\alpha(t,\beta))}<0.
\end{equation}
Moreover they obey the following bounds,
\begin{equation}
\label{eq:beta_confined}
\alpha+\frac{2}{f_0||\ell_0'||_\infty}\ln\left(1+f_0||\ell_0'||_\infty t\right)\leq\beta(t,\alpha)\leq\alpha+2\delta t \quad \text{and}\quad\beta-2\delta t\leq\alpha(t,\beta)\leq\beta-\frac{2}{f_0||r_0'||_\infty}\ln\left(1+f_0||r_0'||_\infty t\right),
\end{equation}
where $\delta$ is defined in \eqref{eq:estimates} and $f_0$ is the maximum of  $f$.
\end{proposition}
\begin{remark}
	\label{rmk:alpha_beta_nesting}
	It readily follows from \eqref{eq:beta_confined} that $\beta(t,\alpha)>\alpha$  and  $\alpha(t,\beta)<\beta$, for every $t>0$, and that $\beta(0,\alpha)=\alpha$ and  $\alpha(0,\beta)=\beta$.
\end{remark}
\begin{definition}
	\label{def:degeneracy}
	According to the geometrical approach adopted here, a   shock wave is formed when a characteristic curve becomes \emph{degenerate}. This occurs when either family becomes \emph{infinitely compressive}, meaning that there either $c_1$ or $c_2$ vanishes.
\end{definition}
\begin{remark}
	\label{rmk:merging_forward_characteristics}
	If a forward characteristic becomes degenerate, then there is $\alpha\in\mathbb{R}$ and a finite time $t^\ast(\alpha)$ such that 
	\begin{equation}
		\label{eq:compressive_x1}
		c_1(t,\alpha)\to 0 \quad\text{as}\quad t\to t^\ast(\alpha).
	\end{equation}
Thus, by \eqref{eq:r_x_characteristic}, $r_{,x}$ must accordingly diverge, provided that $r_0'(\alpha)\neq0$, namely
\begin{equation}
	\label{eq:r_x_characteristic_special}
	r_{,x}(t,x_1(t,\alpha))=\frac{r_0'(\alpha)}{c_1(t,\alpha)}\to\sgn(r_0'(\alpha))\infty \quad \text{as}\quad t\to t^\ast(\alpha).
\end{equation}
By \eqref{eq:riemann_system_a} and \eqref{eq:first_order_unknowns}, this in turn implies that  the second derivatives of $w$   diverge, while the first derivatives develop a discontinuity.
\end{remark}
\begin{remark}
	\label{rmk:economy}
	In a completely parallel way, the degeneracy of a backward characteristic would lead us to identify a critical time $t^\ast(\beta)$ where $\ell_{,x}(t,x_2(t,\beta))$ would diverge, provided that $\ell_0'(\beta)\neq0$. Not to burden our presentation with too many case distinctions, we shall preferentially focus on the possible degeneracy of forward characteristics. When both forward and backward characteristics become degenerate,  the actual critical time will be the least between $t^\ast(\alpha)$ and $t^\ast(\beta)$, for all admissible $\alpha$ and $\beta$ for which these times exist.
\end{remark}
In the following section, we shall derive estimates for $t^\ast(\alpha)$ and record without explicit proof the corresponding ones for $t^\ast(\beta)$.

\section{Critical Time Estimates }\label{sec:critical_time}
Here, we build upon the method illustrated in the preceding section to analyze the formation of singularities in the smooth solutions of problem \eqref{eq:wave_system} for a broad class of initial data $\twist_0$; we provide an estimate for the critical time $t^\ast$ for these singularities to occur, which  depends only on $\twist_0$.
More particularly, these conditions will be shown to depend only on $r_0$: they encompass a large class of initial data. 
\begin{theorem}
	\label{th:suff_cond}
	Consider the global Cauchy problem \eqref{eq:diagonal_system} with initial condition $r_0(x)=-\ell_0(x)\in \mathcal{C}^1$ as given in \eqref{eq:IC_riemann}. Assume that $r_0$ is bounded and has a finite limit as $x\to+\infty$,
	\begin{equation}
		\label{eq:r_0_+infty}
		\lim_{x\to\infty}r_0(x)=:r_0(+\infty)\in\mathbb{R}.
	\end{equation}
	If there exists at least one $\alpha\in\mathbb{R}$ such that $r_0$ satisfies the condition
	\begin{equation}
		\label{eq:sufficient_condition}
		\sgn\left(r_0'(\alpha)\left(r_0(\alpha)+r_0(+\infty)\right)\right)=-1,
	\end{equation}
	then the solution $r(t,x)$ to \eqref{eq:diagonal_system} will develop a singularity along the characteristic curve $x=x_1(t,\alpha)$ in a finite time $t^\ast(\alpha)$. Similarly, if $r_0$ has a finite limit as $x\to-\infty$,
	\begin{equation}
		\label{eq:r_0_-infty}
		\lim_{x\to-\infty}r_0(x)=:r_0(-\infty)\in\mathbb{R},
	\end{equation}
	and if there exists at least one $\beta\in\mathbb{R}$ such that $r_0(x)$ satisfies the condition
	\begin{equation}
		\label{eq:sufficient_condition_minus_infinity}
		\sgn\left(r_0'(\beta)\left(r_0(\beta)+r_0(-\infty)\right)\right)=+1,
	\end{equation}
	then the solution $\ell(t,x)$ to \eqref{eq:diagonal_system} will develop a singularity along the characteristic curve $x=x_2(t,\beta)$ in a finite time $t^\ast(\beta)$.
\end{theorem}
\begin{proof}
	We focus on the occurrence of singularities along \emph{forward} characteristics, which obey \eqref{eq:x_1}. By use of Proposition~\ref{prop:beta_t_alpha}, we first estimate $c_1(t,\alpha)$ in \eqref{eq:infinitesimal_compression_ratio_x1}. Key to this end is to consider the point $(\tau,x_1(\tau,\alpha))$   as the endpoint at time $\tau$ of the \emph{backward} characteristic starting from $\beta(\tau,\alpha)$.  
	Hence, by \eqref{eq:l_constant} and \eqref{eq:alpha_beta_implicit_definitions}, we can express \eqref{eq:infinitesimal_compression_ratio_x1} as
	\begin{equation}
		\label{eq:solution_ode_x_1_alpha_critical}
		c_1(t,\alpha)=\sqrt{\frac{k(r_0(\alpha)-\ell_0(\beta(t,\alpha)))}{k(2r_0(\alpha))}}g(t,\alpha),
	\end{equation}
	where with the aid of  \eqref{eq:IC_riemann} the function $g$ is defined as 
	\begin{equation}\label{eq:g_def}
		g(t,\alpha):= 1+r_0'(\alpha)\sqrt{k(2r_0(\alpha))}\int_0^{t} f\left(r_0(\alpha)+r_0(\beta(\tau,\alpha))\right) \dd \tau.
	\end{equation}
	Thus, for $c_1$ to vanish at $t=t^\ast(\alpha)$, it must be
	\begin{equation}
		\label{eq:g_vanishing}
		g(t^\ast(\alpha),\alpha)=0. 
	\end{equation}
	For $t=0$, $g(0,\alpha)=1$, whereas for $t\to\infty$, the integral in \eqref{eq:g_def} can be easily estimated  as $\beta(t,\alpha)$  diverges to $+\infty$ at least logarithmically  in consequence of \eqref{eq:beta_confined}. Thus,
whenever 
	\begin{equation}
		\label{eq:condition_instability}
		r_0'(\alpha)f(r_0(\alpha)+r_0(+\infty))<0,
	\end{equation}
\nigh{$g(t,\alpha)$ diverges to $-\infty$ as $t\to+\infty$, and so}
	there always exists a sufficiently large time $t^\ast(\alpha)$ such that $g(t^\ast(\alpha),\alpha)$ vanishes. Since $\sgn(f(\eta)\eta)=+1$ for all $\eta\neq0$,  \eqref{eq:condition_instability} is equivalent to \eqref{eq:sufficient_condition}.  
\end{proof}
\nigh{
\begin{remark}
	\label{rmk:generalization}
In this paper, we are especially interested in the homogeneous initial condition \eqref{eq:wave_system_c} for $\twist_{,t}$. Had we replaced \eqref{eq:wave_system_c} with
\begin{equation}
	\label{eq:wave_system_c_star}
	\twist_{,t}(0,t)=w_1(x),
\end{equation}
\eqref{eq:IC_riemann} would have become
\begin{equation}
\label{eq:IC_riemann_star}
r(0,x)=r_0(x)=w_1(x)-L(w_0'(x)),\quad \ell(0,x)=\ell_0(x)=w_1(x)+L(w_0'(x)),
\end{equation}
so that in general $r_0(x)\neq-\ell_0(x)$. It is not difficult to show that in such a general setting the same qualitative conclusions reached in Theorem~\ref{th:suff_cond} would follow from the assumptions,
\begin{subequations}
\begin{align}
	\label{eq:generalized_assumptions}
	\exists\ \alpha\in\mathbb{R}\quad\text{such that}\quad \sgn(r_0'(\alpha)(r_0(\alpha)-\ell_0(+\infty)))=-1\quad\text{with}\quad\ell_0(+\infty):=\lim_{x\to+\infty}\ell_0(x),\\
	\exists\ \beta\in\mathbb{R}\quad\text{such that}\quad \sgn(\ell_0'(\beta)(\ell_0(\beta)-r_0(-\infty)))=+1\quad\text{with}\quad r_0(-\infty):=\lim_{x\to-\infty}r_0(x).
\end{align}
\end{subequations}
which generalize \eqref{eq:sufficient_condition} and \eqref{eq:sufficient_condition_minus_infinity}.
\end{remark}
}
\begin{remark}
	\label{rmk:symmetry}
It follows from \eqref{eq:sufficient_condition} and \eqref{eq:sufficient_condition_minus_infinity} that if $r_0(x)$ is even or odd (and correspondingly $r_0'(x)$ is  odd or even), then for a singularity occurring along a forward characteristic originating at $\alpha$, another singularity  also occurs  along the backward characteristic originating from $\beta= -\alpha$, and $t^\ast(\alpha)=t^\ast(\beta)$.
\end{remark}
An estimate of the critical time $t^\ast(\alpha)$  at which the singularity occurs can be provided under more restrictive assumptions on the behavior of $r_0(x)$ for $x>\alpha$, with $\alpha$ satisfying \eqref{eq:sufficient_condition}. We establish these estimates in the following Corollaries.
\begin{corollary}
\label{cor:estimate_T_crit}
If there exists $\alpha\in\mathbb{R}$ such that, in addition to \eqref{eq:sufficient_condition}, $r_0$ also satisfies
\begin{equation}
\label{eq:t_crit_estimate}
\sgn\left(r_0'(x)r_0(\alpha)\right)=-1 \quad \text{for every}\quad x\geq\alpha,
\end{equation}
then the critical time $t^\ast(\alpha)$ can be estimated as
\begin{equation}
\label{eq:T_crit_hp_1}
t^\ast(\alpha)\le\frac{1}{\sqrt{k(2r_0(\alpha))}|r_0'(\alpha)|\gamma^+_1(\alpha)} \quad\text{with}\quad \gamma^+_1(\alpha):=\min\left\{\left|f(2r_0(\alpha))\right|,\left|f(r_0(\alpha)+r_0(+\infty))\right|\right\}.
\end{equation}
\end{corollary}
\begin{proof}
Under the hypothesis on $r_0(x)$ stated in Theorem \ref{th:suff_cond}, for  $\alpha\in\mathbb{R}$ satisfying \eqref{eq:sufficient_condition} the forward characteristic $x=x_1(t,\alpha)$ becomes infinitely compressive, i.e. \eqref{eq:compressive_x1} holds, after the time $t^\ast(\alpha)$ such that $g(t^\ast(\alpha),\alpha)=0$, where $g$ is defined as in \eqref{eq:g_def}.
Since by \eqref{eq:beta_confined} $\beta(t,\alpha)\geq\alpha$ for every $t\geq0$, \eqref{eq:t_crit_estimate} and \eqref{eq:d_t_beta} imply that $r_0(\alpha)+r_0(\beta(\tau,\alpha))$ is monotonic and keeps the same sign for every $\tau\geq0$, and so also does $f(r_0(\alpha)+r_0(\beta(\tau,\alpha)))$: by \eqref{eq:sufficient_condition}, the asymptotic limit $f(r_0(\alpha)+r_0(+\infty))$ approached for $\tau\to\infty$ has the same sign as $f(r_0(\alpha)+r_0(\beta(0,\alpha)))$. Thus,
\begin{align}
g(t,\alpha)&= 1-\sqrt{k(2r_0(\alpha))}|r_0'(\alpha)|\int_0^t|f(r_0(\alpha)+r_0(\beta(\tau,\alpha)))|\dd\tau\nonumber\\
&\leq1-\sqrt{k(2r_0(\alpha))}|r_0'(\alpha)|\min\left\{\left|f(2r_0(\alpha))\right|,\left|f(r_0(\alpha)+r_0(+\infty))\right|\right\}t.\label{eq:g_estimate}
\end{align}
Since $g(0,\alpha)=1$, it follows from \eqref{eq:g_estimate} that $g(t,\alpha)<0$ when the inequality in  \eqref{eq:T_crit_hp_1} is violated. This complete the proof of the Corollary.
\end{proof}
\begin{remark}
	\label{rmk:backward_first_estimate}
	A parallel argument applied to backward characteristics proves that if there exists $\beta\in\mathbb{R}$ such that, in addition to \eqref{eq:sufficient_condition_minus_infinity}, $r_0$  also satisfies
	\begin{equation}
		\label{eq:t_crit_estimate_minus_infinity}
		\sgn\left(r_0'(x)r_0(\beta)\right)=+1 \quad \text{for every}\quad x\leq\beta,
	\end{equation}
	then the critical time $t^\ast(\beta)$ can be estimated as
	\begin{equation}
		\label{eq:T_crit_hp_1_minus_infinity}
		t^\ast(\beta)\leq\frac{1}{\sqrt{k(2r_0(\beta))}|r_0'(\beta)|\gamma^-_1(\beta)} \quad\text{with}\quad \gamma^-_1(\beta):=\min\left\{\left|f(2r_0(\beta))\right|,\left|f(r_0(\beta)+r_0(-\infty))\right|\right\}.
	\end{equation}
\end{remark}
\begin{corollary}
\label{cor:estimate_T_crit_caseb}
If there exist $\alpha\in\mathbb{R}$ and $\varepsilon>0$, possibly depending on $\alpha$,  such that, in addition to \eqref{eq:sufficient_condition}, $r_0$ also satisfies
\begin{equation}
\label{eq:norm_argument}
\sgn\left(r_0'(\alpha)r_0(\alpha)\right)=-1, \quad \text{and} \quad |r_0(\alpha)+r_0(x)|>\varepsilon>0 \quad \text{for every}\quad x\geq\alpha,
\end{equation}
then
\begin{equation}
\label{eq:T_crit_hp_2}
t^\ast(\alpha)\leq\frac{1}{\sqrt{k[2r_0(\alpha)]}|r_0'(\alpha)|\gamma^+_2(\alpha)} \quad\text{with}\quad \gamma^+_2(\alpha):=\min\left\{\left|f\left(\eta\right)\right|:  \eta\in[\varepsilon,\sup_{x\geq\alpha}|r_0(\alpha)+r_0(x)|]\right\}.
\end{equation}
\end{corollary}
\begin{proof}
Unlike the case considered in Corollary~\ref{cor:estimate_T_crit}, here $r_0'(x)$ can change its sign for $x\geq\alpha$. However,  $|r_0(\alpha)+r_0(\beta(\tau,\alpha))|$ cannot vanish, as it must range between $\varepsilon$ and $\sup_{x\geq\alpha}|r_0(\alpha)+r_0(x)|$. Accordingly, reasoning as in the proof of Corollary~~\ref{cor:estimate_T_crit}, we find that
\begin{equation}
\label{eq:g_inf_uniform}
g(t,\alpha)\leq1-|r_0'(\alpha)|\sqrt{k(2r_0(\alpha))}\min\left\{\left|f\left(\eta\right)\right|: \eta\in[\varepsilon,\sup_{x\geq\alpha}|r_0(\alpha)+r_0(x)|]\right\}t,
\end{equation}
which leads us to \eqref{eq:T_crit_hp_2}. 
\end{proof}
\begin{remark}
	\label{rmk:backward_second_estimate}
	A similar argument can be applied to the backward characteristics of family \eqref{eq:x_2} to prove that if there exist $\beta\in\mathbb{R}$ and $\varepsilon>0$, possibly depending on $\beta$, such that, in addition to \eqref{eq:sufficient_condition_minus_infinity}, $r_0$ also satisfies
	\begin{equation}
		\label{eq:norm_argument_minus_infinity}
		\sgn\left(r_0'(\beta)r_0(\beta)\right)=+1 \quad \text{and}\quad |r_0(\beta)+r_0(x)|>\varepsilon>0 \quad \text{for every}\quad x\leq\beta,
	\end{equation}
	then
	\begin{equation}
		\label{eq:T_crit_hp_2_minus_infinity}
		t^\ast(\beta)\leq\frac{1}{\sqrt{k(2r_0(\beta))}|r_0'(\beta)|\gamma^-_2(\beta)}\quad\text{with} \quad \gamma^-_2(\beta):=\min\left\{\left|f\left(\eta\right)\right|: \, \eta\in[\varepsilon,\sup_{x\leq\beta}|r_0(\beta)+r_0(x)|]\right\}.
	\end{equation}
\end{remark}
By minimizing $t^\ast(\alpha)$ and $t^\ast(\beta)$ among all $\alpha$ and $\beta$ for which a singularity occurs, we can derive an estimate for the singular time $t^\ast$ at which a regular solution of system \eqref{eq:diagonal_system} breaks down. In the following Proposition, we collect in a single formal inequality for $t^\ast$ the partial estimates in Corollaries~\ref{cor:estimate_T_crit} and \ref{cor:estimate_T_crit_caseb} and in Remarks~\ref{rmk:backward_first_estimate} and \ref{rmk:backward_second_estimate} above.
\begin{proposition}
	\label{prop:estimate_summary}
	Under the hypotheses of Theorem~\ref{th:suff_cond}, the critical time $t^\ast$ for the existence of a solution of class $\mathcal{C}^1$ to system \eqref{eq:diagonal_system} can be estimated as 
	\begin{equation}
		\label{eq:critical_time_estimate}
		t^\ast\leq t_\mathrm{c}:=\inf_{\alpha\in\mathbb{R}}\inf_{\gamma(\alpha)}\frac{1}{\sqrt{k(2r_0(\alpha))}|r_0'(\alpha)|\gamma(\alpha)},
	\end{equation}
where
\begin{equation}\label{eq:Gamma_definition}
\gamma(\alpha):=
\begin{cases}
	\min\{|f(2r_0(\alpha))|,|f(r_0(\alpha)+r_0(+\infty))|\}&if\
	 \sgn(r_0'(x)r_0(\alpha))=-1\ \forall\ x\geq\alpha,\\
	\min\{|f(2r_0(\alpha))|,|f(r_0(\alpha)+r_0(-\infty))|\}&if\
	 \sgn(r_0'(x)r_0(\alpha))=+1\ \forall\ x\leq\alpha,\\
	 \min\{|f(\eta)|:\eta\in[\varepsilon,\sup_{x\leq\alpha}|r_0(\alpha)+r_0(x)|]\}&if\
	 \sgn(r_0'(\alpha)r_0(\alpha))=+1\ \text{\emph{and}}\ |r_0(\alpha)+r_0(x)|>\varepsilon>0\ \forall\ x\geq\alpha,\\
	 \min\{|f(\eta)|:\eta\in[\varepsilon,\sup_{x\leq\alpha}|r_0(\alpha)+r_0(x)|]\}&if\
	 \sgn(r_0'(\alpha)r_0(\alpha))=-1\ \text{\emph{and}}\ |r_0(\alpha)+r_0(x)|>\varepsilon>0\ \forall\ x\leq\alpha,\\
	 0& \text{\emph{otherwise}}.
\end{cases}
\end{equation}
\end{proposition}
\begin{remark}
	\label{rmk:Gamma=0}
	When $\gamma(\alpha)=0$, we conventionally set  equal to $+\infty$ the argument of the double $\inf$ on the right-hand side of \eqref{eq:critical_time_estimate}.
\end{remark}
\begin{remark}
	\label{rmk:never_void}
	By \eqref{eq:r_0_+infty} and \eqref{eq:r_0_-infty}, $\gamma(\alpha)>0$, at least for $\alpha\to+\infty$ or $\alpha\to-\infty$. Thus, the \emph{upper} estimate $t_\mathrm{c}$ for $t^\ast$ in \eqref{eq:critical_time_estimate} is never  $+\infty$ and a finite critical time always exist at which a regular solution of \eqref{eq:diagonal_system} breaks down.
\end{remark}
In the following section, we shall see a number of applications of our method where the estimate \eqref{eq:critical_time_estimate},  despite its complicated appearance, is proven effective and delivers critical times very close to those calculated numerically.

\section{Applications}\label{sec:applications}
As illustrative examples, we consider initial profiles $\twist_0$ for the twist angle that exhibit a strong concentration of distortion around $x=0$, which fades away at infinity without ever vanishing.
Thus, by \eqref{eq:Q_definition}, the more distorted core propagates faster than the distant tails, possibly overtaking them:  intuitively, this should prompt the creation of a singularity in a finite time. We shall see here how such an intuitive prediction is indeed confirmed by the estimate \eqref{eq:critical_time_estimate}.

Specifically, we shall consider two types of initial profiles $w_0$, namely, a \emph{kink} and a \emph{bump}. In either cases, we shall both estimate the critical time $t^\ast$ and identify the characteristics along which a singularity first occurs.

Numerical solutions of the global Cauchy problem \eqref{eq:wave_system} will also be provided: they are shown to be in good agreement with the theoretical predictions.

\subsection{Kink}\label{sec:arctan}
\begin{figure}[]
	\centering
	\begin{subfigure}[c]{0.35\linewidth}
		\centering
		\includegraphics[width=\linewidth]{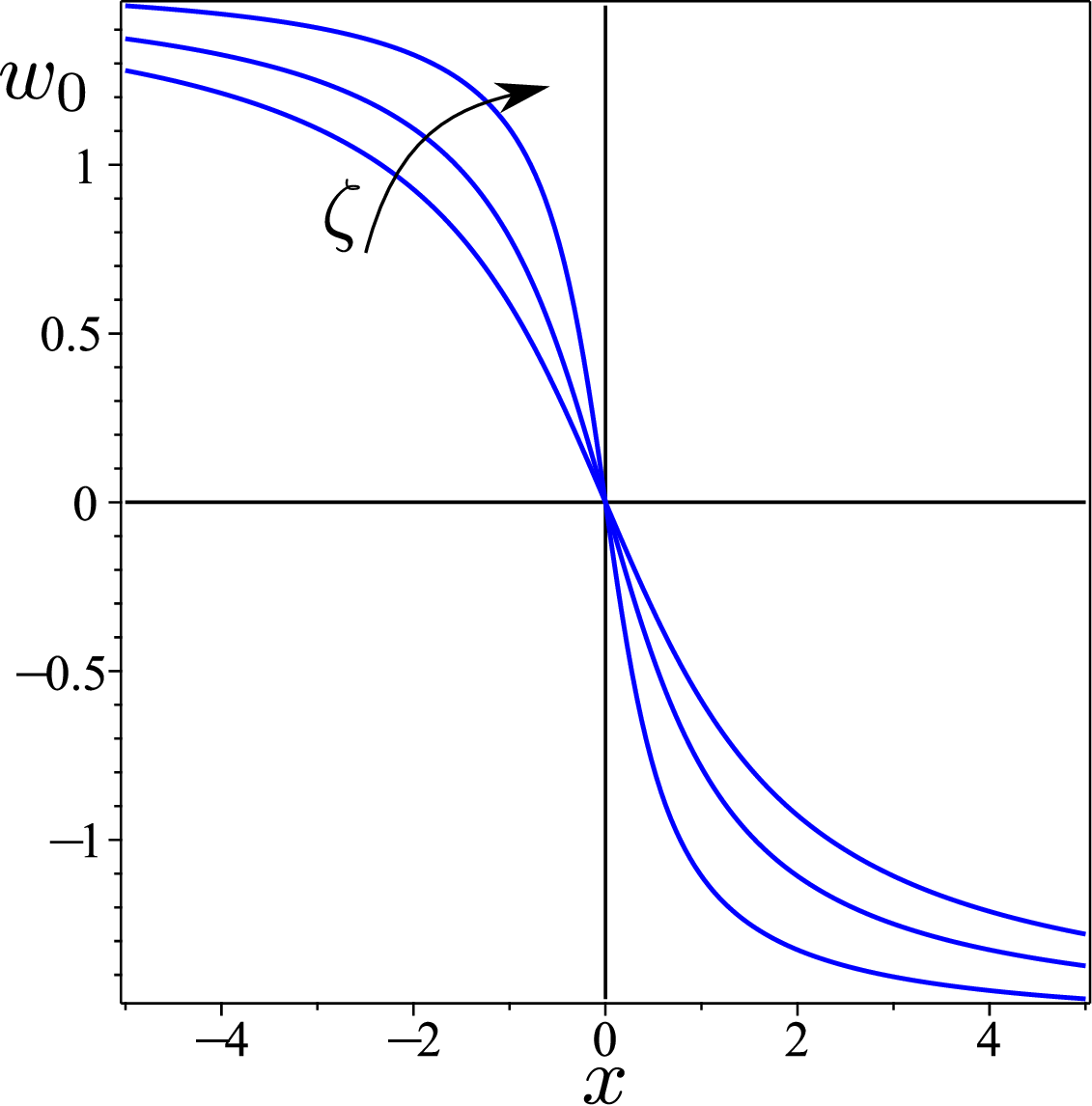}
		\caption{$\kappa=\pi/2$, $\zeta=1/2, \, 1, \, 3/2$.} 
		\label{fig:w_0arctanzeta}
	\end{subfigure}
	\begin{subfigure}[c]{0.35\linewidth}
		\centering
		\includegraphics[width=\linewidth]{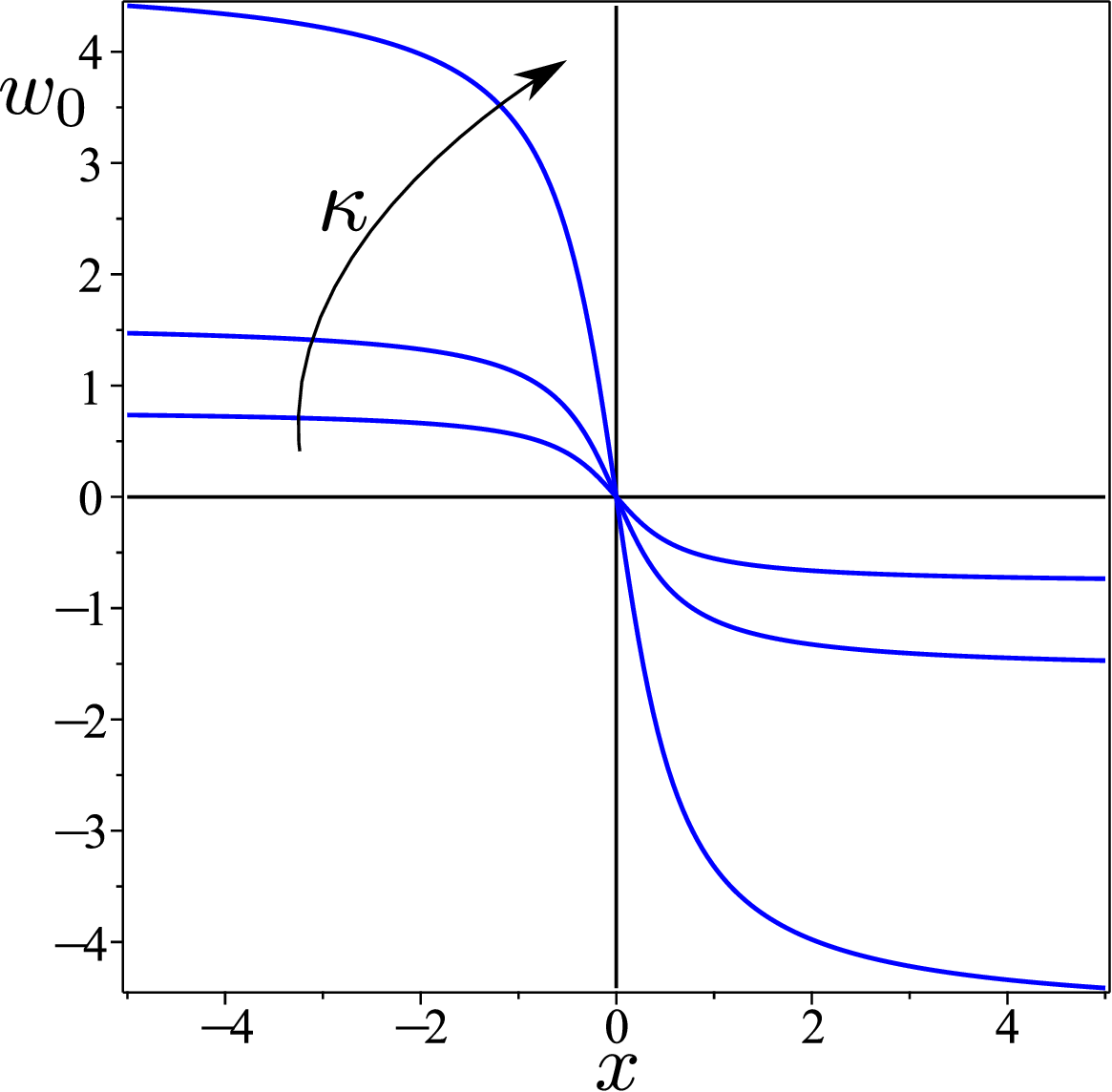}
		\caption{$\zeta=1/2$, $\kappa=\pi/4, \, \pi/2, \, 3\pi/2$.} 
		\label{fig:w_0arctankappa}
	\end{subfigure}
\caption{Initial profile of the twist angle $w_0$ represented by  \eqref{eq:w_0_arctan} for several values of the parameters $\kappa$ and $\zeta$, which describe the amplitude and width of the kink, respectively.}
	\label{fig:w_0arctan}
\end{figure}

We consider the following  initial twist  profile
\begin{equation}
\label{eq:w_0_arctan}
\twist_0(\kappa,\zeta;x):=-\frac{2\kappa}{\pi}\arctan\frac{x}{\zeta},
\end{equation}
where $\kappa$ and $\zeta$ are positive parameters. This is a \emph{kink} representing a smooth transition of the twist angle between two asymptotic values depending on $\kappa$, through an effective width around the origin  depending on $\zeta$.
Figure~\ref{fig:w_0arctan} illustrates the graphs of the initial profile $\twist_0$ in \eqref{eq:w_0_arctan} for different values of $\zeta$ and $\kappa$: either increasing  $\zeta$ or decreasing  $\kappa$ makes the initial profile \emph{less} distorted (whereas either decreasing $\zeta$ or increasing $\kappa$ makes it \emph{more} distorted).

\begin{figure}[]
	\centering
	\begin{subfigure}[c]{0.35\linewidth}
		\centering
		\includegraphics[width=\linewidth]{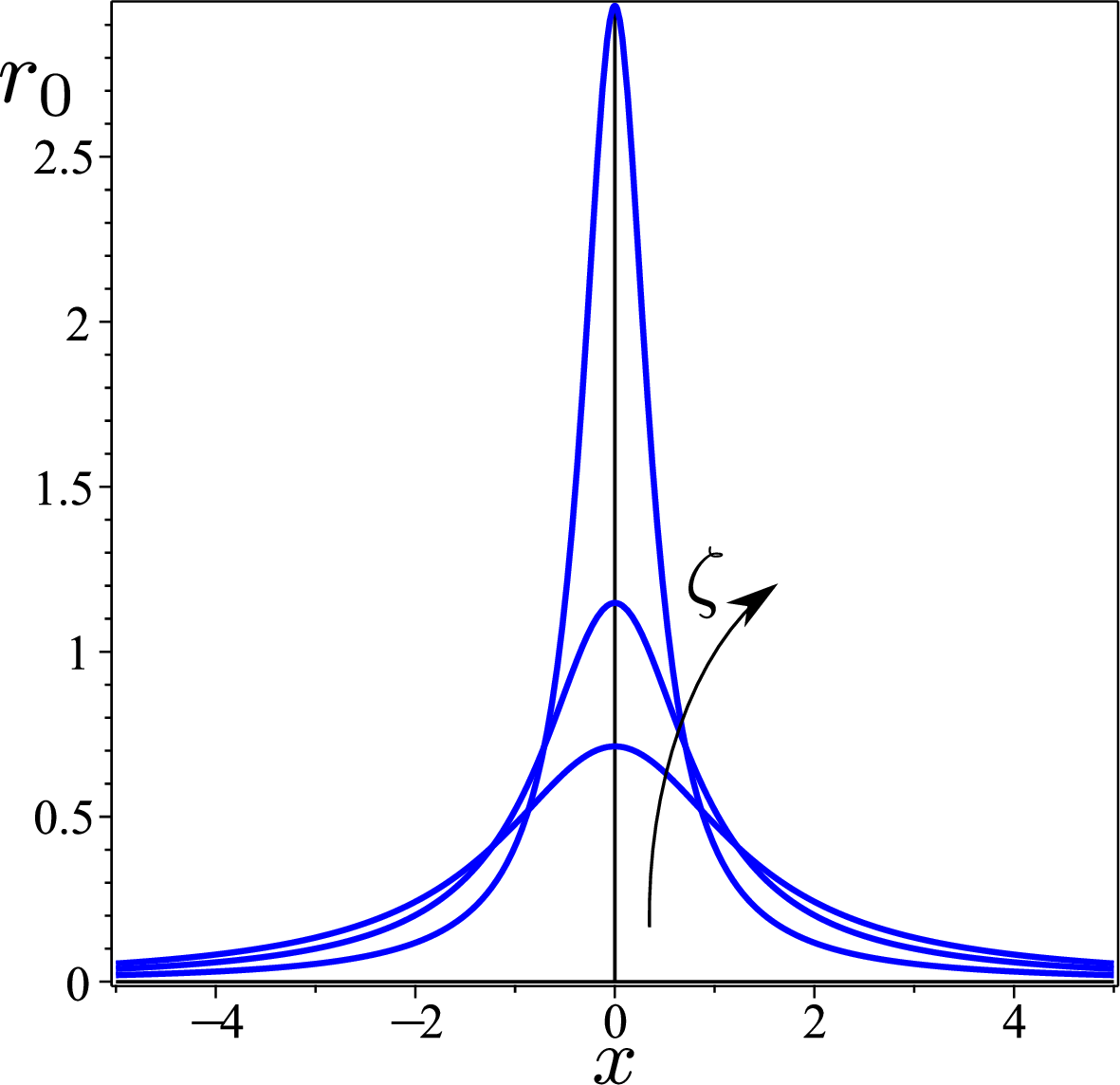}
		\caption{$\kappa=\pi/2$, $\zeta=1/2, \, 1, \, 3/2$.} 
		\label{fig:w_0arctanzeta}
	\end{subfigure}
	\begin{subfigure}[c]{0.35\linewidth}
		\centering
		\includegraphics[width=\linewidth]{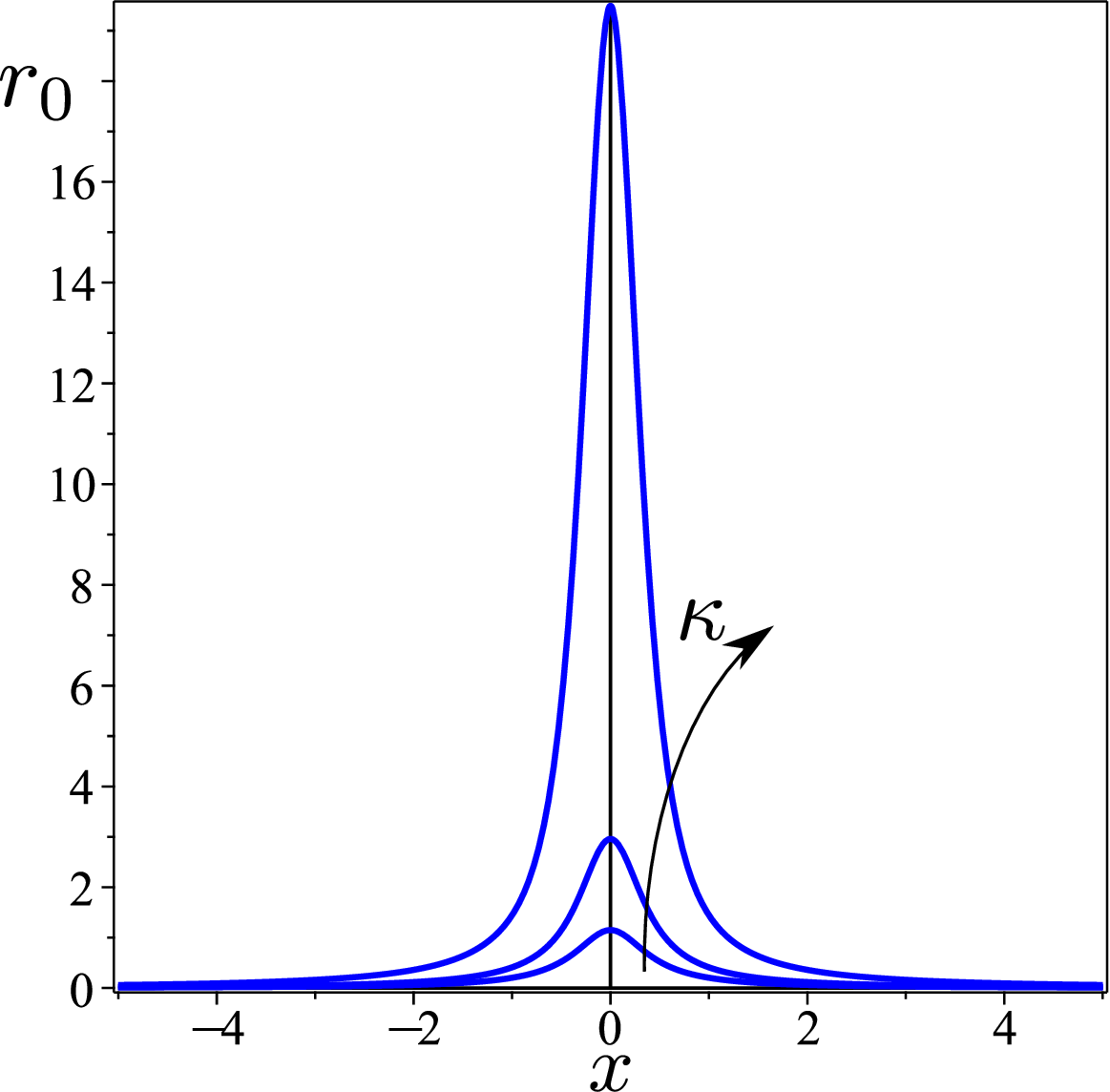}
		\caption{$\zeta=1/2$, $\kappa=\pi/4, \, \pi/2, \, 3\pi/2$.} 
		\label{fig:w_0arctankappa}
	\end{subfigure}
\caption{Initial profile of the Riemann invariant $r_0$ in \eqref{eq:r_0arctan} corresponding to  the initial twist $w_0$ illustrated in Fig.~\ref{fig:w_0arctan}.}
	\label{fig:r_0arctan}
\end{figure}

By \eqref{eq:IC_riemann} and \eqref{eq:L_u_1}, $r_0$ is given by
\begin{equation}
\label{eq:r_0arctan}
r_0(\kappa,\zeta;x)=\frac{1}{2}\left(\xi\sqrt{1+\xi^2}+\hbox{arcsinh}\,\xi\right), \quad \, \hbox{where} \quad \xi:=\frac{2\kappa\zeta}{x^2+\zeta^2}.
\end{equation}
The  graph of $r_0$ is illustrated in Fig.~\ref{fig:r_0arctan} for different values of $\kappa$ and $\zeta$. Since $r_0$ is an even function,  by Theorem~\ref{th:suff_cond},  we  need only consider forward characteristics: if a singularity arises along a forward characteristic originating at $\alpha$, a singularity will also occur at the same critical time along the (symmetric) backward characteristic originating at $\beta=-\alpha$.

\begin{figure}[]
	\centering
	\begin{subfigure}[c]{0.35\linewidth}
		\centering
		\includegraphics[width=\linewidth]{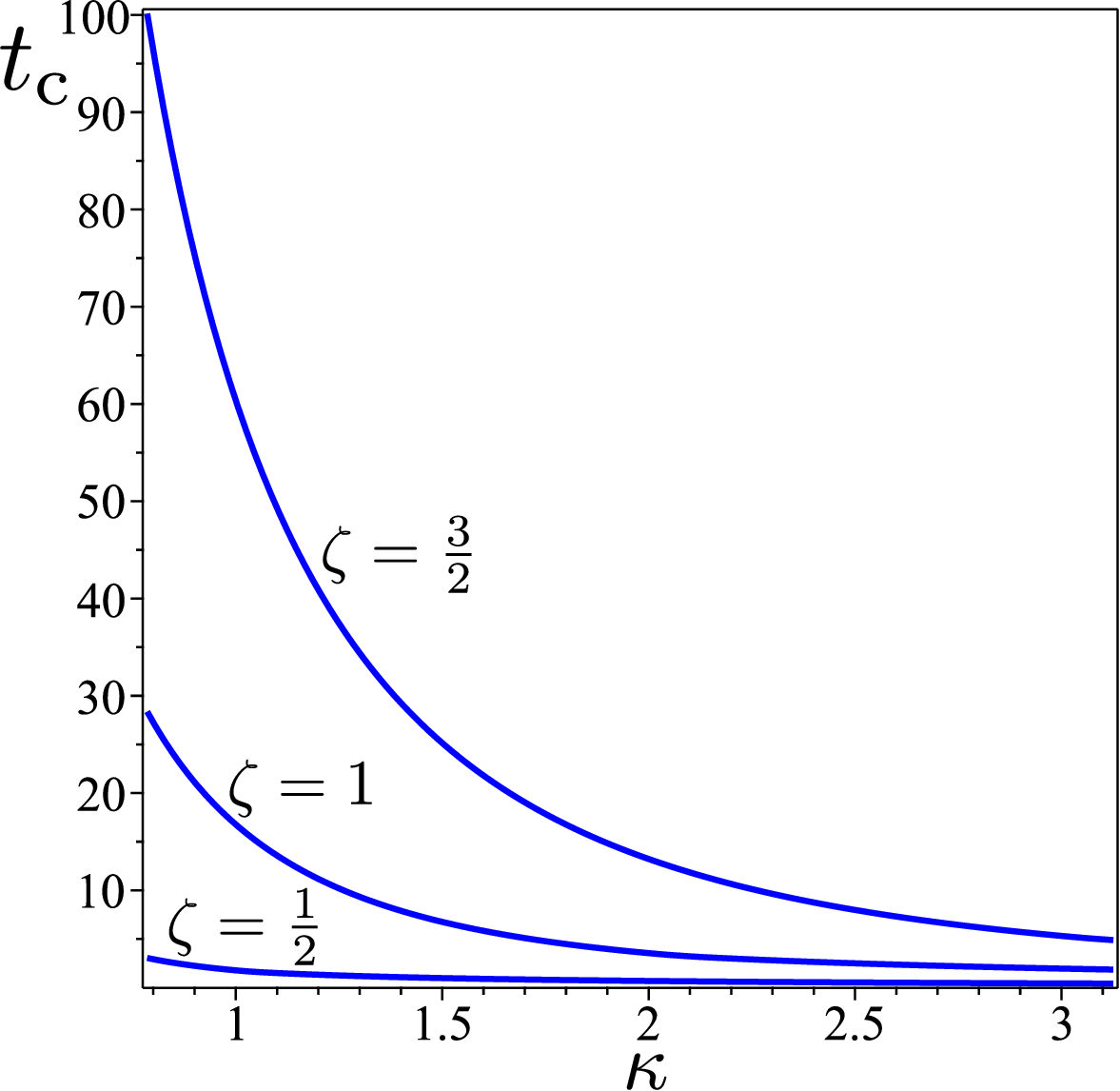}
		\caption{$\kappa\in[\pi/4,3\pi/2]$, $\zeta=1/2, \, 1, \, 3/2$.} 
		\label{fig:T_crit_arctanzeta}
	\end{subfigure}
	\begin{subfigure}[c]{0.35\linewidth}
		\centering
		\includegraphics[width=\linewidth]{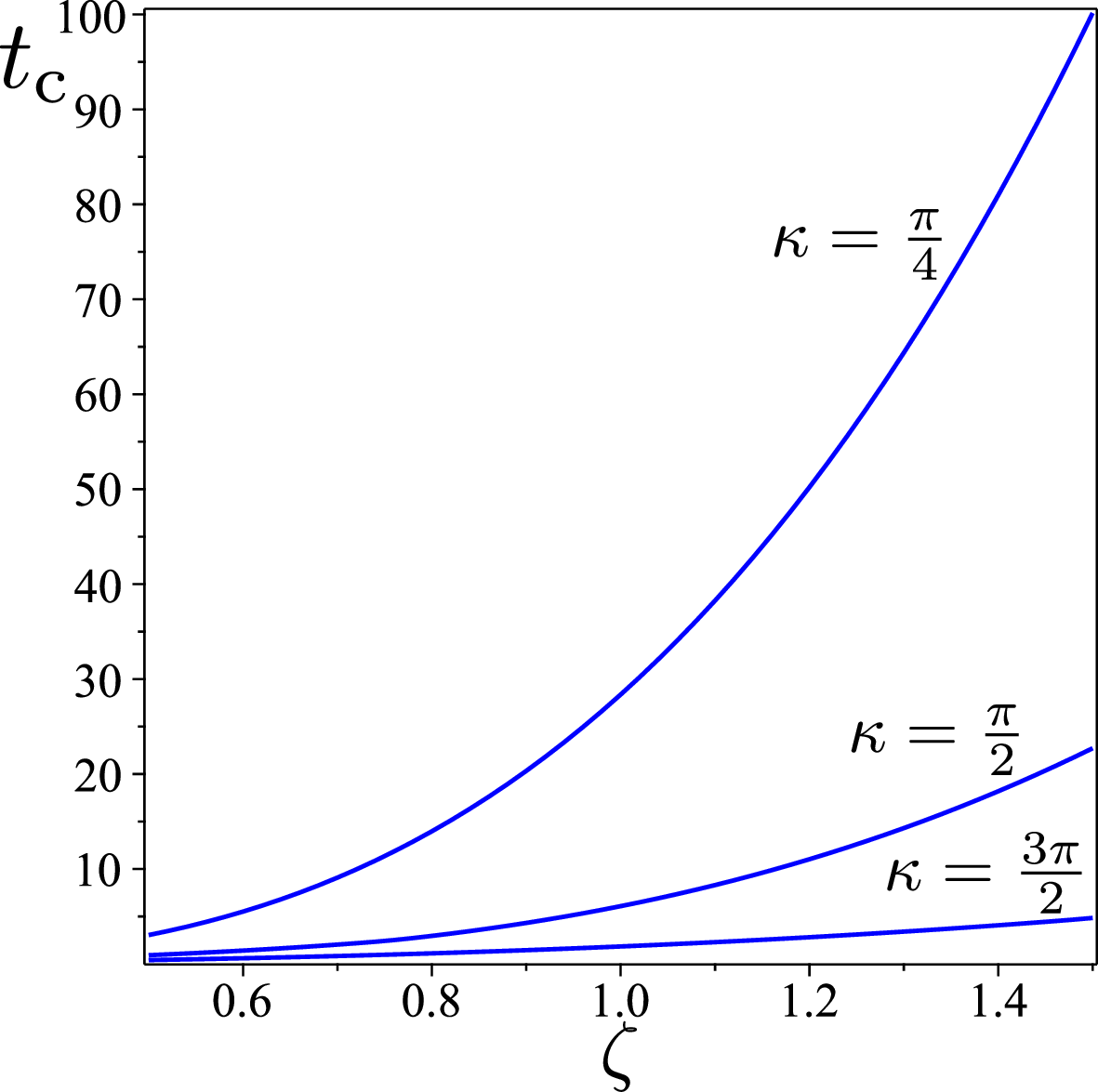}
		\caption{$\zeta\in[1/2,3/2]$, $\kappa=\pi/4, \, \pi/2, \, 3\pi/2$.} 
		\label{fig:T_crit_arctankappa}
	\end{subfigure}
\caption{The time $t_\mathrm{c}$ is the upper estimate in \eqref{eq:T_crit_arctan} for the critical time $t^\ast$ of a regular solution of the global Cauchy problem \eqref{eq:wave_system} with initial profile \eqref{eq:w_0_arctan}. In particular, for $\kappa=\pi/2$ and $\zeta=1/2$,  $t_\mathrm{c}\doteq0.9$.}
	\label{fig:T_crit_arctan}
\end{figure}

Here, $r_0(+\infty)=0$ and condition \eqref{eq:sufficient_condition} is satisfied for every $\alpha>0$. Consequently, a singularity occurs in a finite time for each of these values. Thus, the upper estimate $t_\mathrm{c}$ for the critical time $t^\ast$ in \eqref{eq:critical_time_estimate} reduces to
\begin{equation}
\label{eq:T_crit_arctan}
t_\mathrm{c}=\inf_{\alpha>0}\frac{1}{\sqrt{k[2r_0(\alpha)]}|r_0'(\alpha)|\gamma(\alpha)} \quad\text{with}\quad \gamma(\alpha)=\min\left\{f(2r_0(\alpha)),f(r_0(\alpha))\right\}.
\end{equation}
Figure~\ref{fig:T_crit_arctan} shows how $t_\mathrm{c}$ in \eqref{eq:T_crit_arctan} depends on both $\kappa$ and $\zeta$: it decreases with $\kappa$ and increases with $\zeta$. This behavior is in accord with intuition: as $\kappa$ decreases or $\zeta$ increases, the initial profile becomes more spread out and less prominent, suggesting that the kink's core propagates more slowly, thus delaying the shock formation.
 
\begin{figure}[]
	\centering
	\begin{subfigure}[c]{0.35\linewidth}
		\centering
		\includegraphics[width=\linewidth]{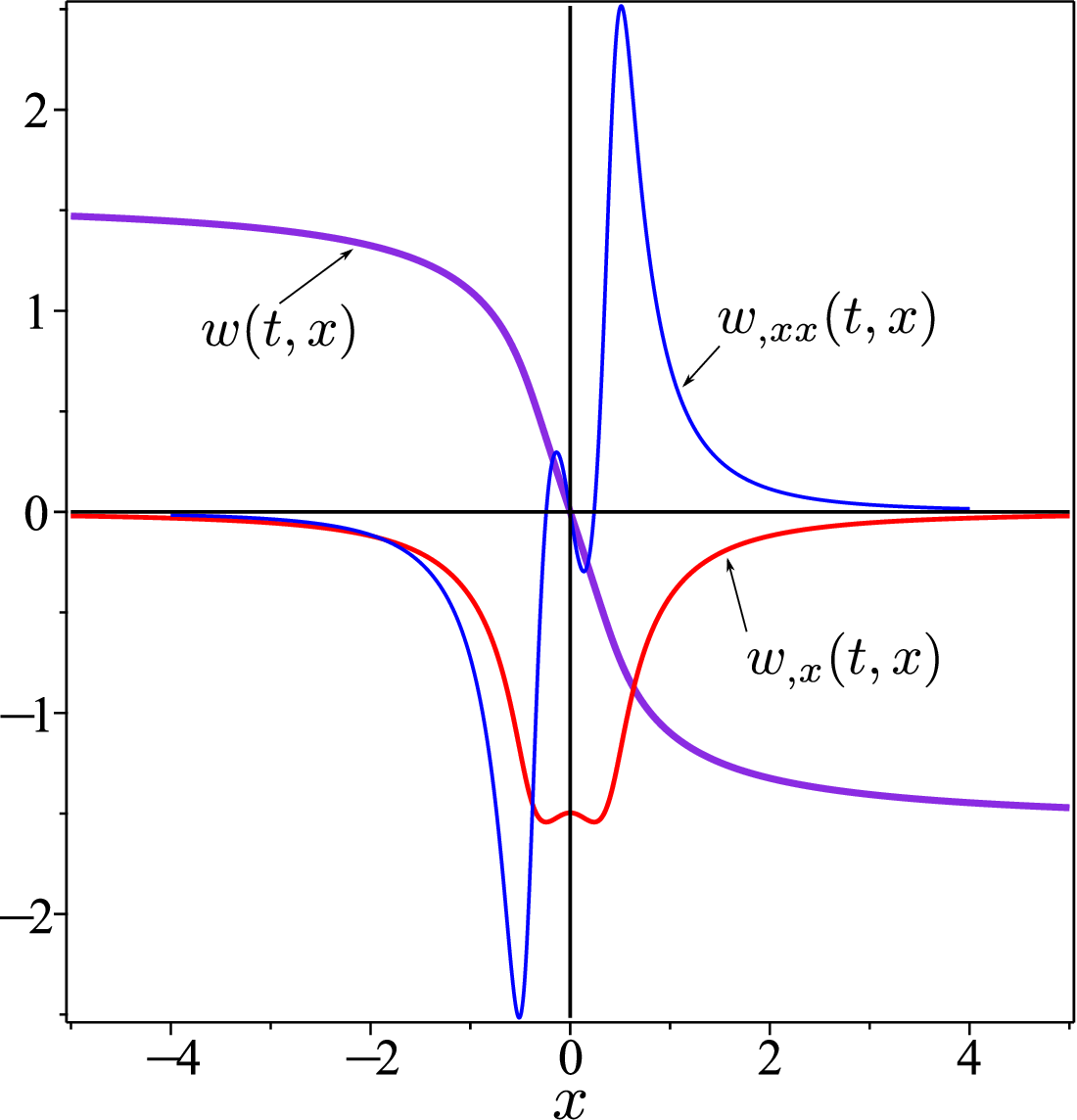}
		\caption{$t=0.15$} 
		\label{fig:t_015}
	\end{subfigure}
	\quad
	\begin{subfigure}[c]{0.35\linewidth}
		\centering
		\includegraphics[width=\linewidth]{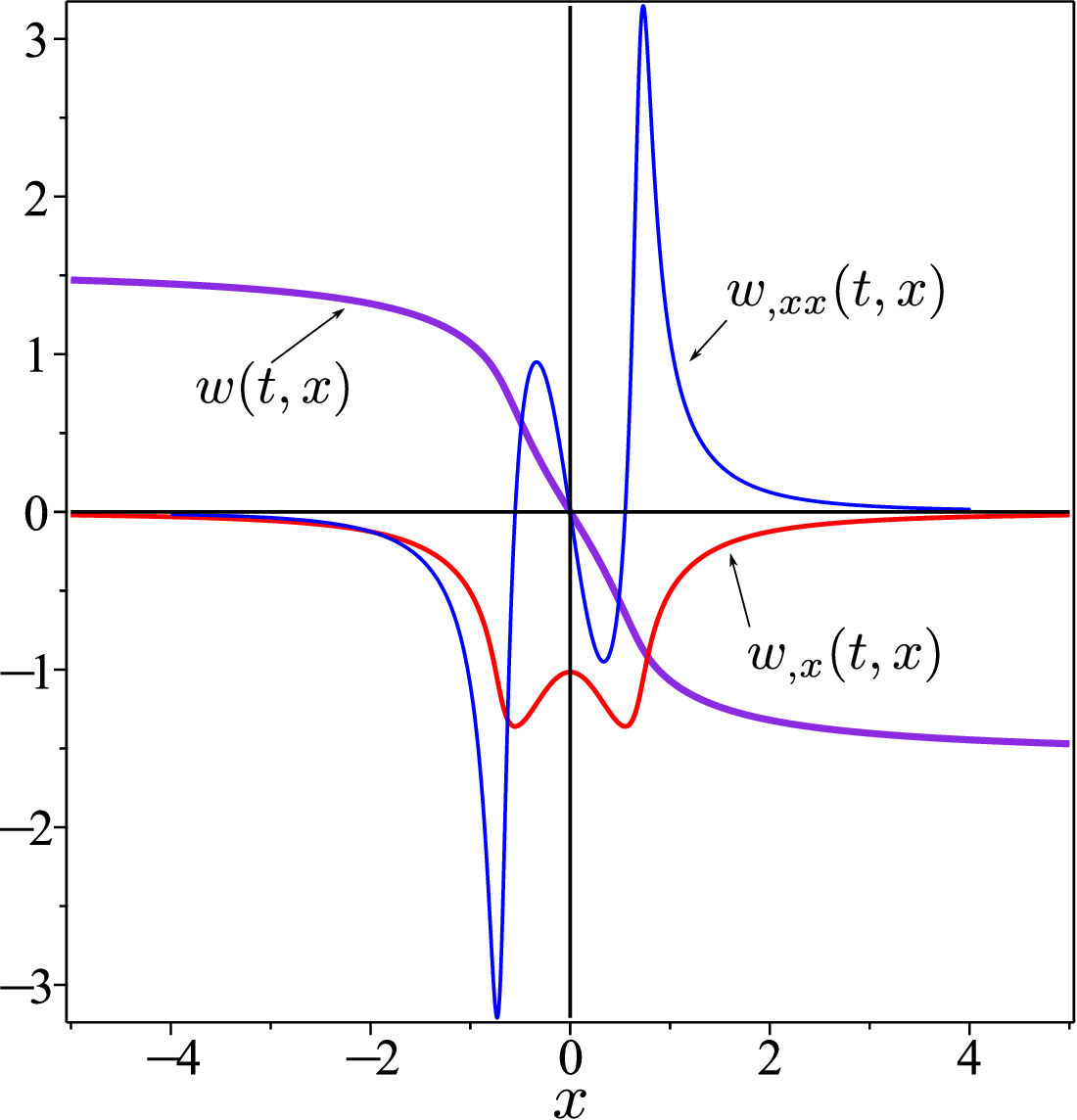}
		\caption{$t=0.30$} 
		\label{fig:t_03}
	\end{subfigure}
	\begin{subfigure}[c]{0.35\linewidth}
		\centering
		\includegraphics[width=\linewidth]{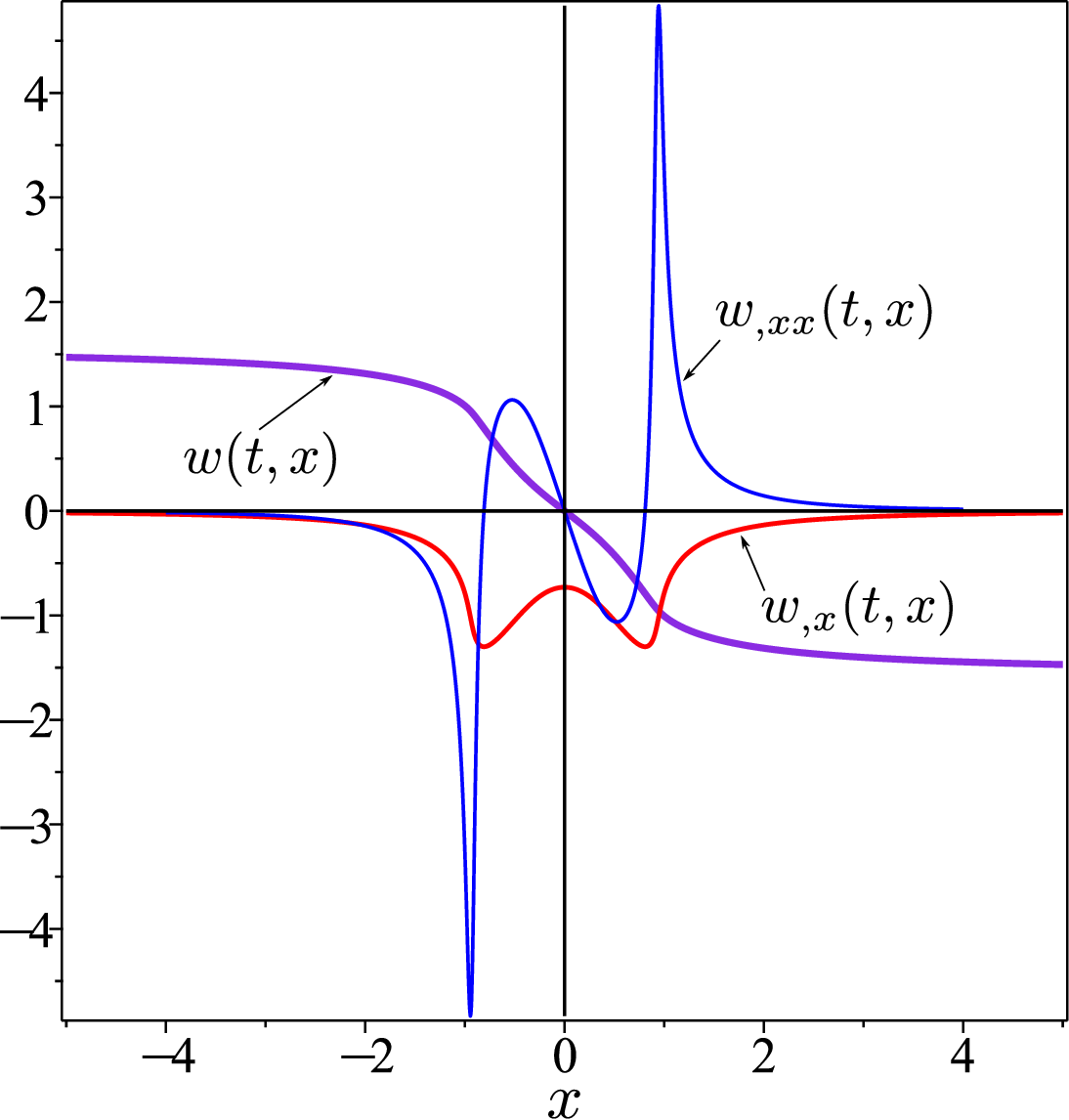}
		\caption{$t=0.45$} 
		\label{fig:t_045}
	\end{subfigure}
	\quad
	\begin{subfigure}[c]{0.35\linewidth}
		\centering
		\includegraphics[width=\linewidth]{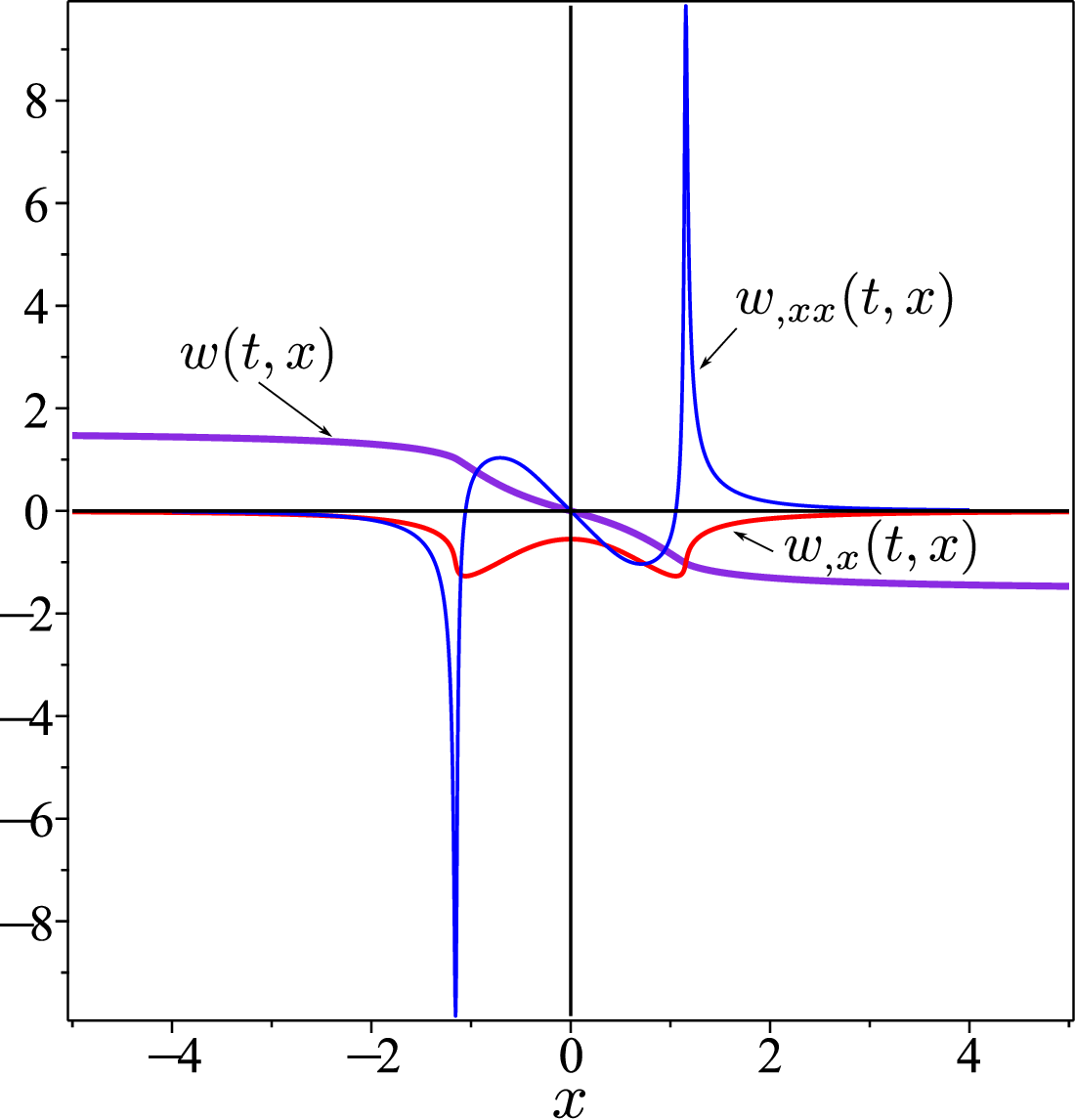}
		\caption{$t=0.60$} 
		\label{fig:t_06}
	\end{subfigure}
\caption{Graphs of the numerical solution $\twist(t,x)$ and its spatial derivatives $\twist_{,x}(t,x)$ and $\twist_{,xx}(t,x)$ for the initial profile $w_0$ in \eqref{eq:w_0_arctan} with $\kappa = \pi/2$, $\zeta=1/2$. The time sequence suggests that $w_{,xx}$ tends to develop two antisymmetric spikes, whereas $w_{,x}$ tends to develop symmetric jumps.}
	\label{fig:num_sol_arctan}
\end{figure}

\begin{figure}[]
	\centering
	\begin{subfigure}[c]{0.4\linewidth}
		\centering
		\includegraphics[width=\linewidth]{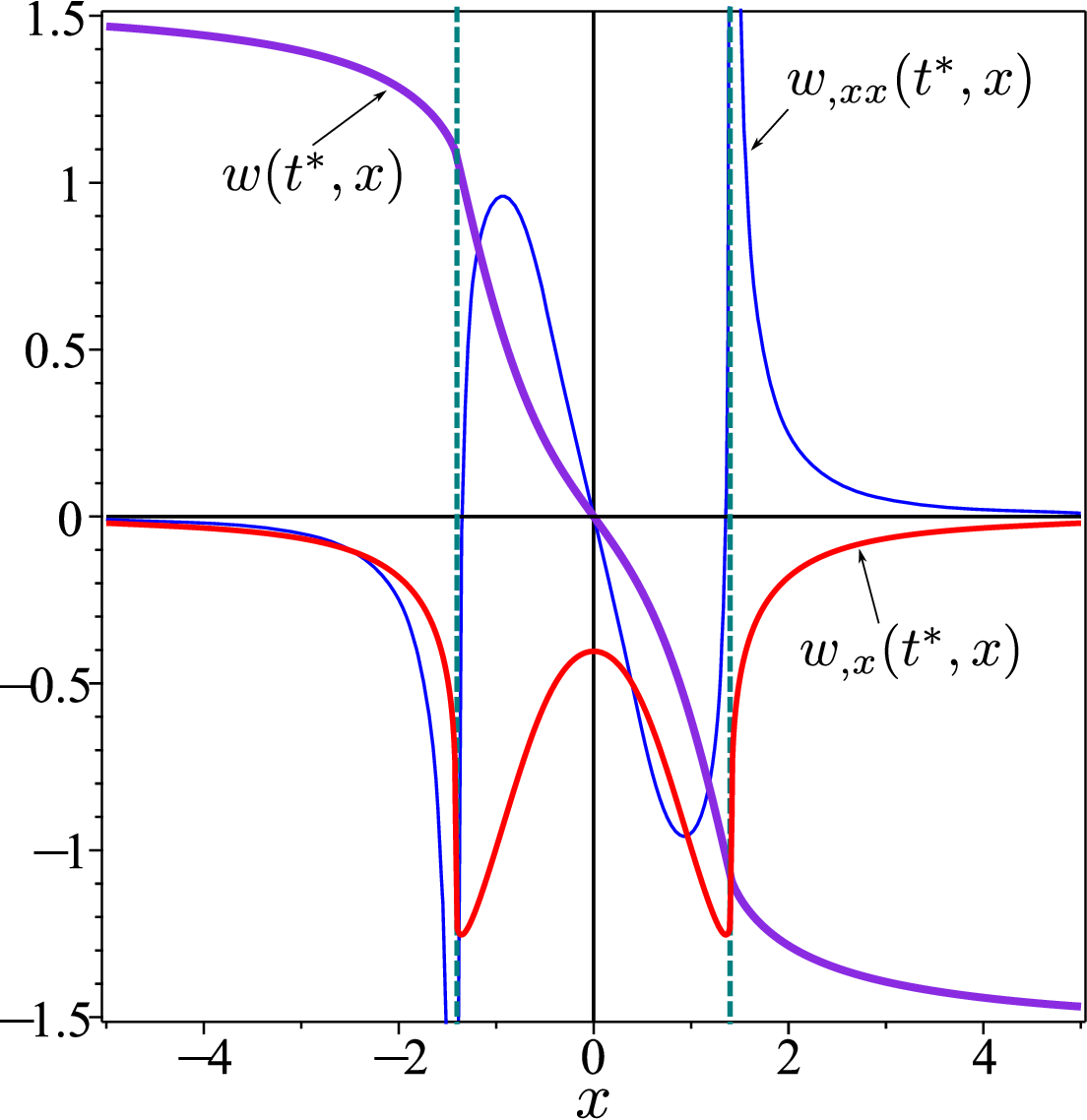}
		\caption{Graphs of the numerical solution $\twist(t^\ast,x)$ and its spatial derivatives $\twist_{,x}(t^\ast,x)$ and $\twist_{,xx}(t^\ast,x)$ at the critical time $t^\ast\doteq0.8$.} 
		\label{fig:num:sol_t_crit_arctan}
	\end{subfigure}
	\qquad
	\begin{subfigure}[c]{0.36\linewidth}
		\centering
		\includegraphics[width=\linewidth]{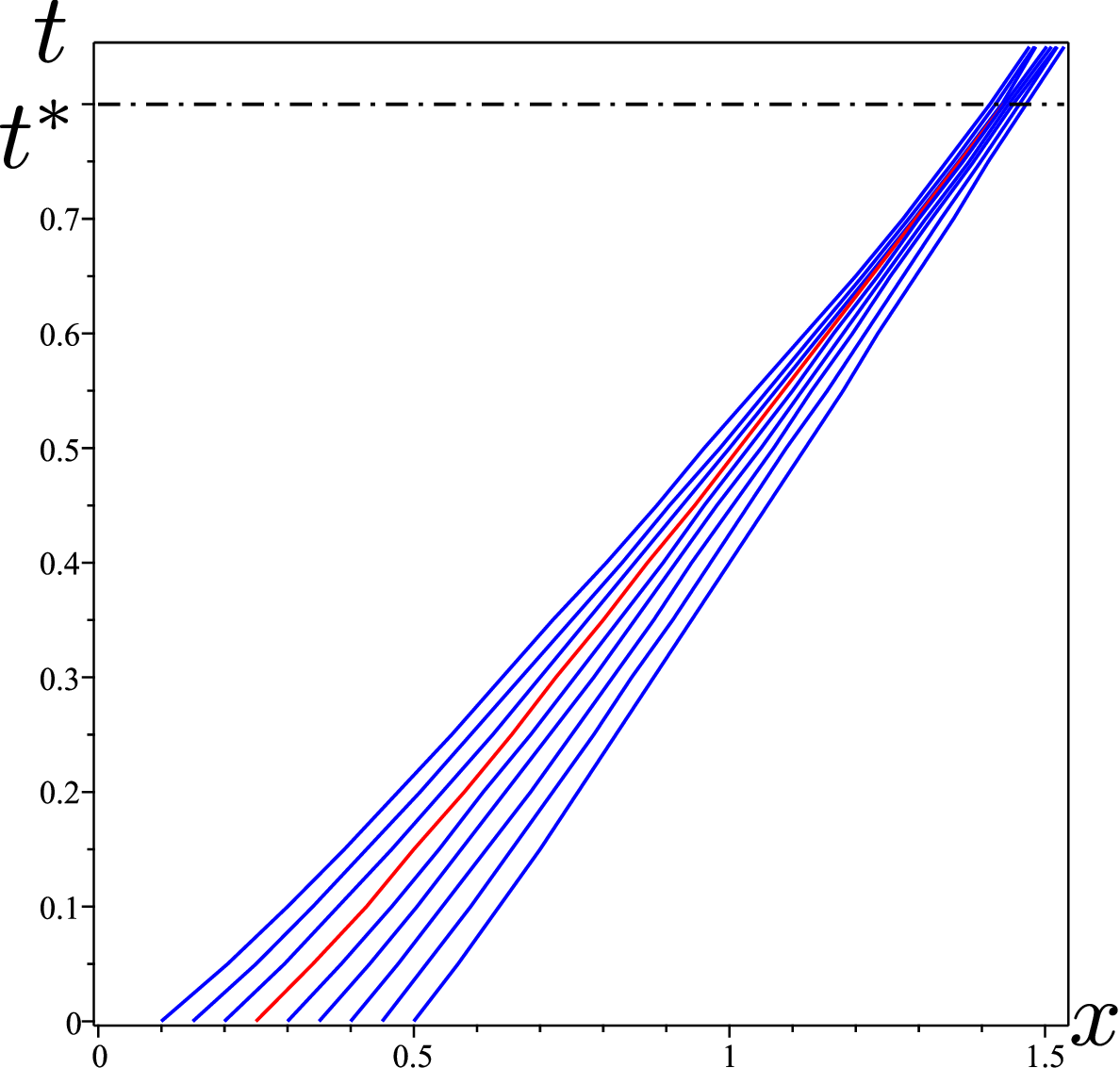}
		\caption{Forward characteristic curves  originating at different values of $\alpha>0$, each developing  a singularity in a finite time. The one starting at $\alpha\doteq0.27$ (marked in red) becomes infinitely compressive (that is, $c_1=0$) before all others.} 
		\label{fig:num:sol_t_crit_arctan_characteristics}
	\end{subfigure}
\caption{Singularity emerging  at the critical time $t^\ast\doteq0.8$ along  the solution of \eqref{eq:wave_system} with initial profile \eqref{eq:w_0_arctan} for $\kappa = \pi/2$, and $\zeta=1/2$.}
	\label{fig:T_crit_snapshot_arctan}
\end{figure}

\begin{figure}[] 
	\centering
	\includegraphics[width=0.4\linewidth]{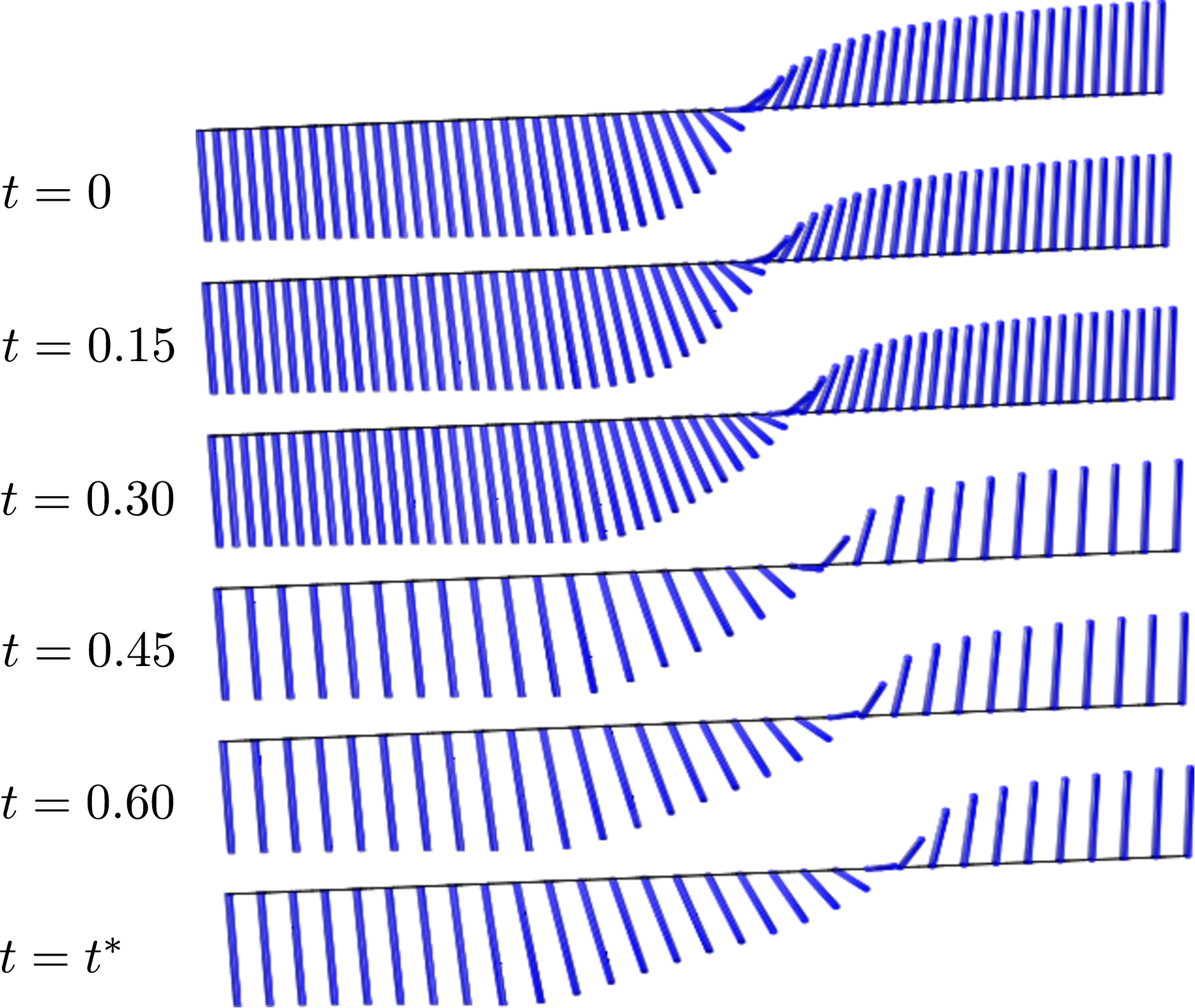}
	\caption{\nigh{Sketches of the directors corresponding to the solutions shown in Figs.~\ref{fig:num_sol_arctan} and \ref{fig:num:sol_t_crit_arctan}.}}
	\label{fig:sketch_arctan}
\end{figure}

Numerical solutions of the Cauchy problem \eqref{eq:wave_system} \nigh{were obtained with Maple 2021 (Maplesoft, a division of Waterloo Maple Inc., Waterloo, Ontario); they}  corroborate our theoretical predictions. We present here the case where in \eqref{eq:w_0_arctan} $\kappa=\pi/2$ and $\zeta=1/2$. The initial profile generates two symmetric waves propagating in opposite directions. We computed the conserved quantities $M$ and $E$ introduced  in Propositions~\ref{prop:mass_conservation} and \ref{prop:conservation_law}, respectively, and used them to monitor the accuracy of our numerical solutions. Our calculations indicate the existence of a critical time, estimated as $t^\ast\doteq0.8$, at which the solution exhibits a singularity. Figure~\ref{fig:num_sol_arctan} illustrates a typical  numerical solution $\twist(t,x)$ and its spatial derivatives $\twist_{,x}(t,x)$ and $\twist_{,xx}(t,x)$ for  a sequence of times $t$ in the interval $[0,t^\ast)$. A  snapshot at $t=t^\ast$ is shown in Fig.~\ref{fig:num:sol_t_crit_arctan}.
The observed behavior is in good agreement with our theory: a shock is formed in a finite time, at which $\twist_{,x}$ becomes discontinuous, and  second derivatives diverge. \nigh{Sketches of the directors corresponding to the solutions shown in Figs.~\ref{fig:num_sol_arctan} and \ref{fig:num:sol_t_crit_arctan} are provided in Fig.~\ref{fig:sketch_arctan}.}

The critical time identified numerically agrees with our  theoretical upper estimate $t_\mathrm{c}\doteq0.9$. The infimum in \eqref{eq:critical_time_estimate} is correspondingly attained for $\alpha\doteq0.26$. Figure~\ref{fig:num:sol_t_crit_arctan_characteristics} shows a set of forward characteristics determined numerically: they all start from points around $\alpha=0.26$ and become infinitesimally compressive (that is, with $c_1=0$) in a finite time: the one that becomes so before the others (at $t=t^\ast$) starts at $\alpha\doteq0.27$, again in good agreement with theory. 

\subsection{Bump}\label{sec:gaussian}
\begin{figure}[]
	\centering
	\begin{subfigure}[c]{0.35\linewidth}
		\centering
		\includegraphics[width=\linewidth]{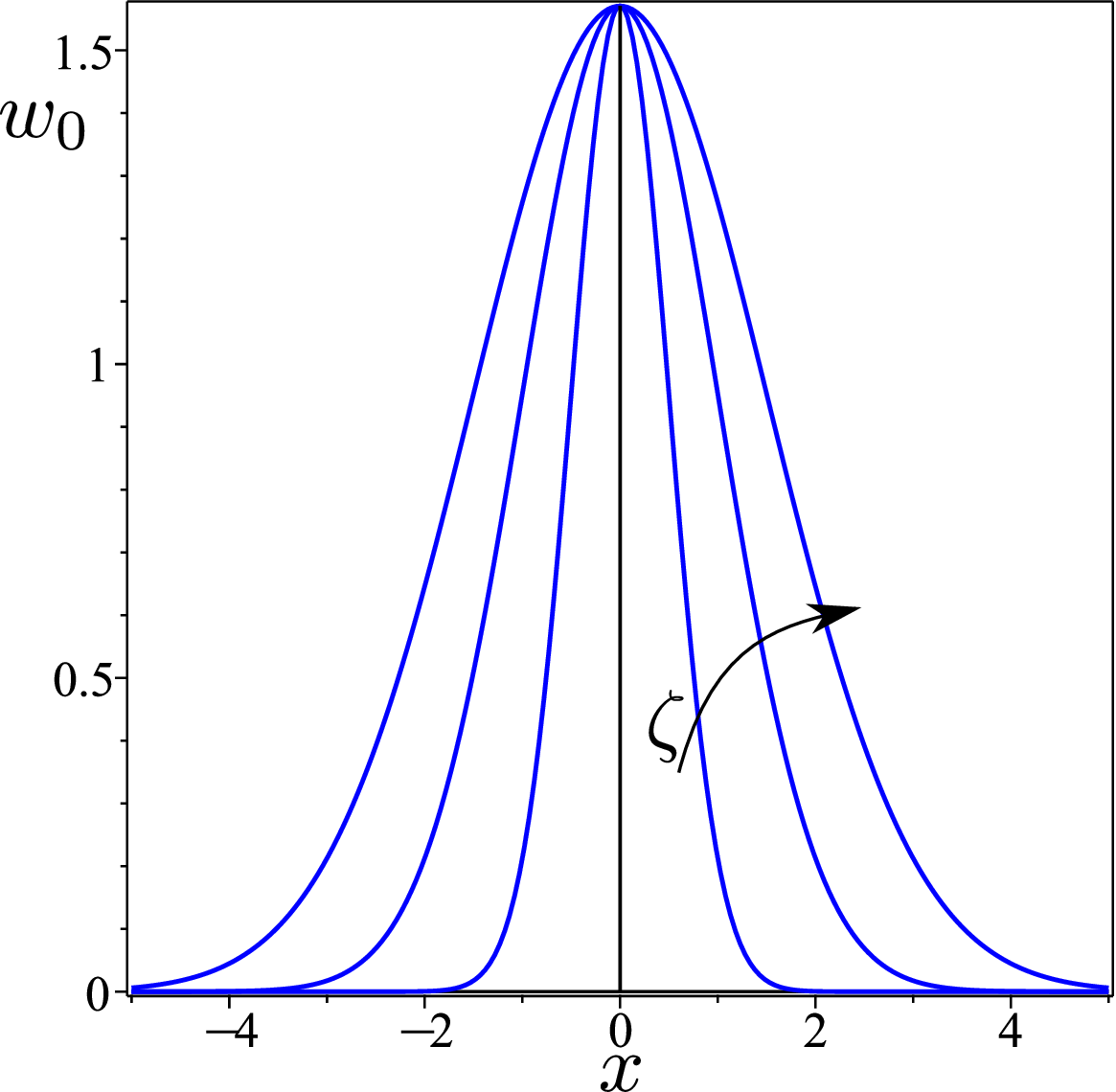}
		\caption{$\kappa=\pi/2$, $\zeta=1/2, \, 1, \, 3/2$.} 
		\label{fig:w_0gaussianzeta}
	\end{subfigure}
	\quad
	\begin{subfigure}[c]{0.35\linewidth}
		\centering
		\includegraphics[width=\linewidth]{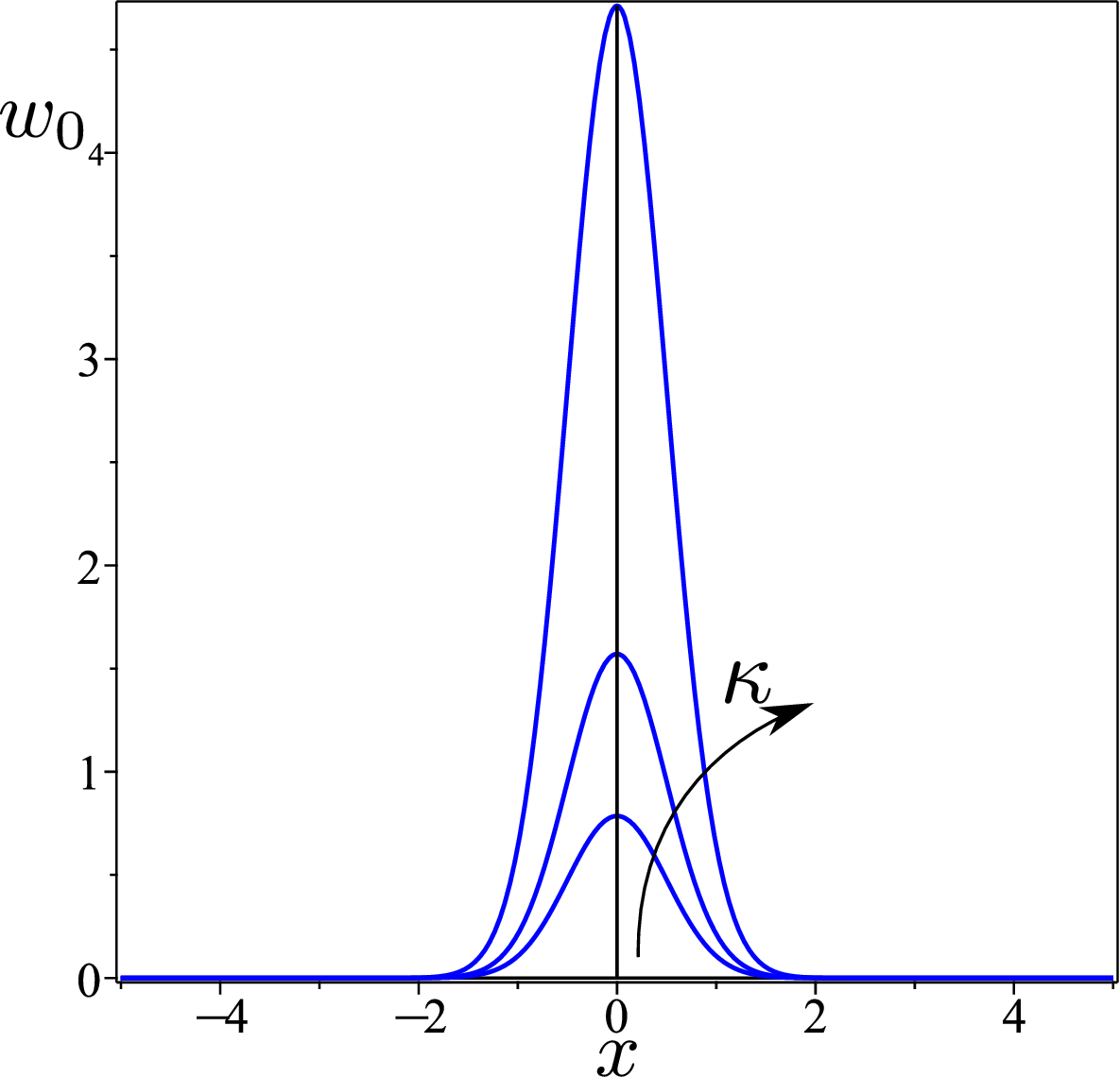}
		\caption{$\zeta=1/2$, $\kappa=\pi/4, \, \pi/2, \, 3\pi/2$.} 
		\label{fig:w_0gaussiankappa}
	\end{subfigure}
\caption{Initial profile of the twist angle $w_0$ represented by \eqref{eq:w_0_gaussian} for several values of the parameters $\kappa$ and $\zeta$, which describe the amplitude and width of the bump, respectively.}
	\label{fig:w_0gaussian}
\end{figure}

We next consider a  Gaussian profile for the initial twist angle $w_0$, which represents a localized \emph{bump} in a otherwise  nearly uniform director field,
\begin{equation}
\label{eq:w_0_gaussian}
\twist_0(\kappa,\zeta;x):=\kappa \esp^{-\frac{x^2}{2\zeta^2}}.
\end{equation}
Here, the amplitude $\kappa$ and the variance $\zeta$ are positive parameters that control the height of the bump and its width, respectively. Figure~\ref{fig:w_0gaussian} illustrates how upon increasing $\zeta$  (or decreasing $\kappa$), the initial distortion becomes less pronounced (and the ensuing smooth solution of \eqref{eq:wave_system} presumably longer lived).

\begin{figure}[]
	\centering
	\begin{subfigure}[c]{0.35\linewidth}
		\centering
		\includegraphics[width=\linewidth]{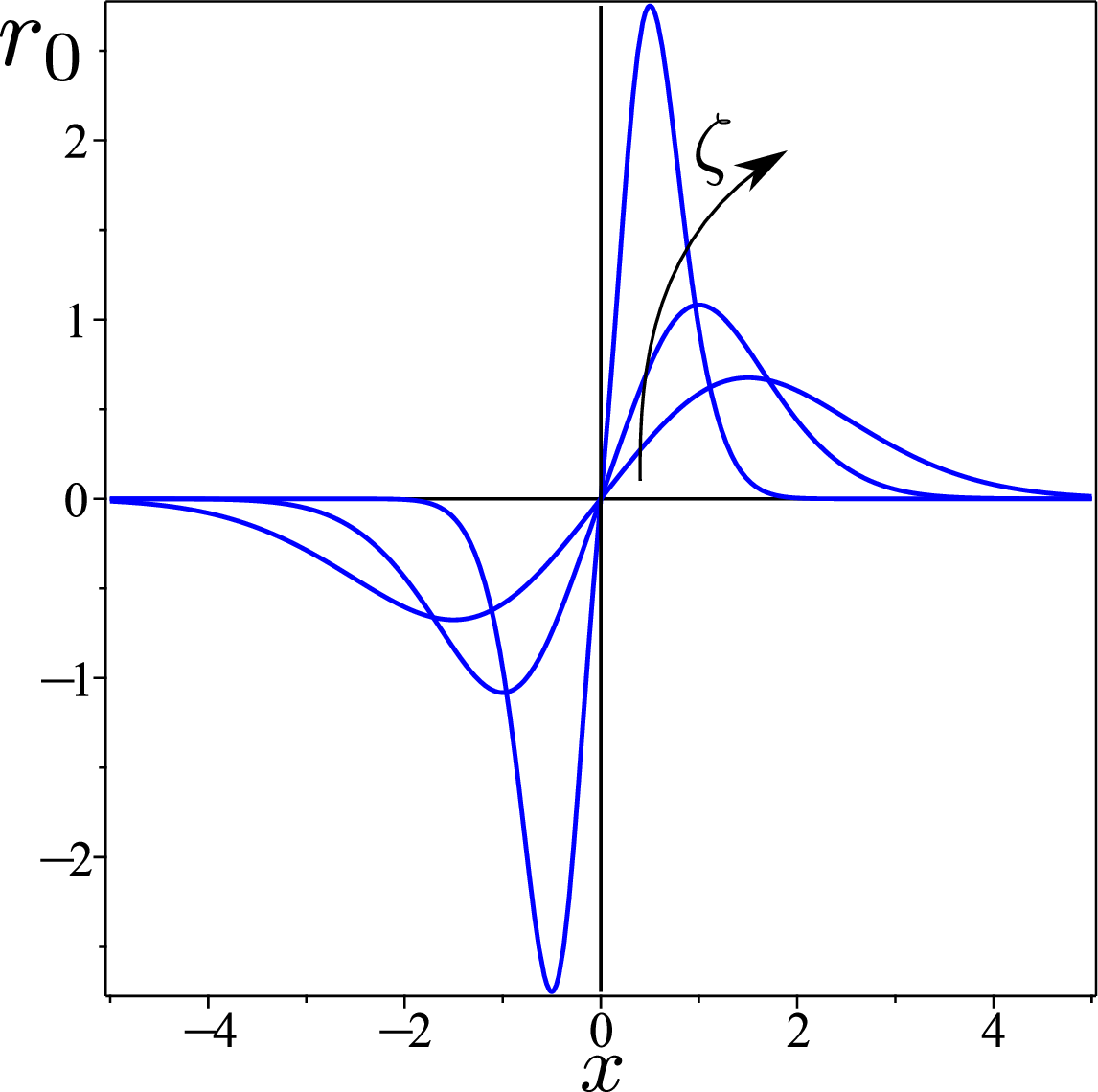}
		\caption{$\kappa=\pi/2$, $\zeta=1/2, \, 1, \, 3/2$.} 
		\label{fig:w_0gaussianzeta}
	\end{subfigure}
	\quad
	\begin{subfigure}[c]{0.35\linewidth}
		\centering
		\includegraphics[width=\linewidth]{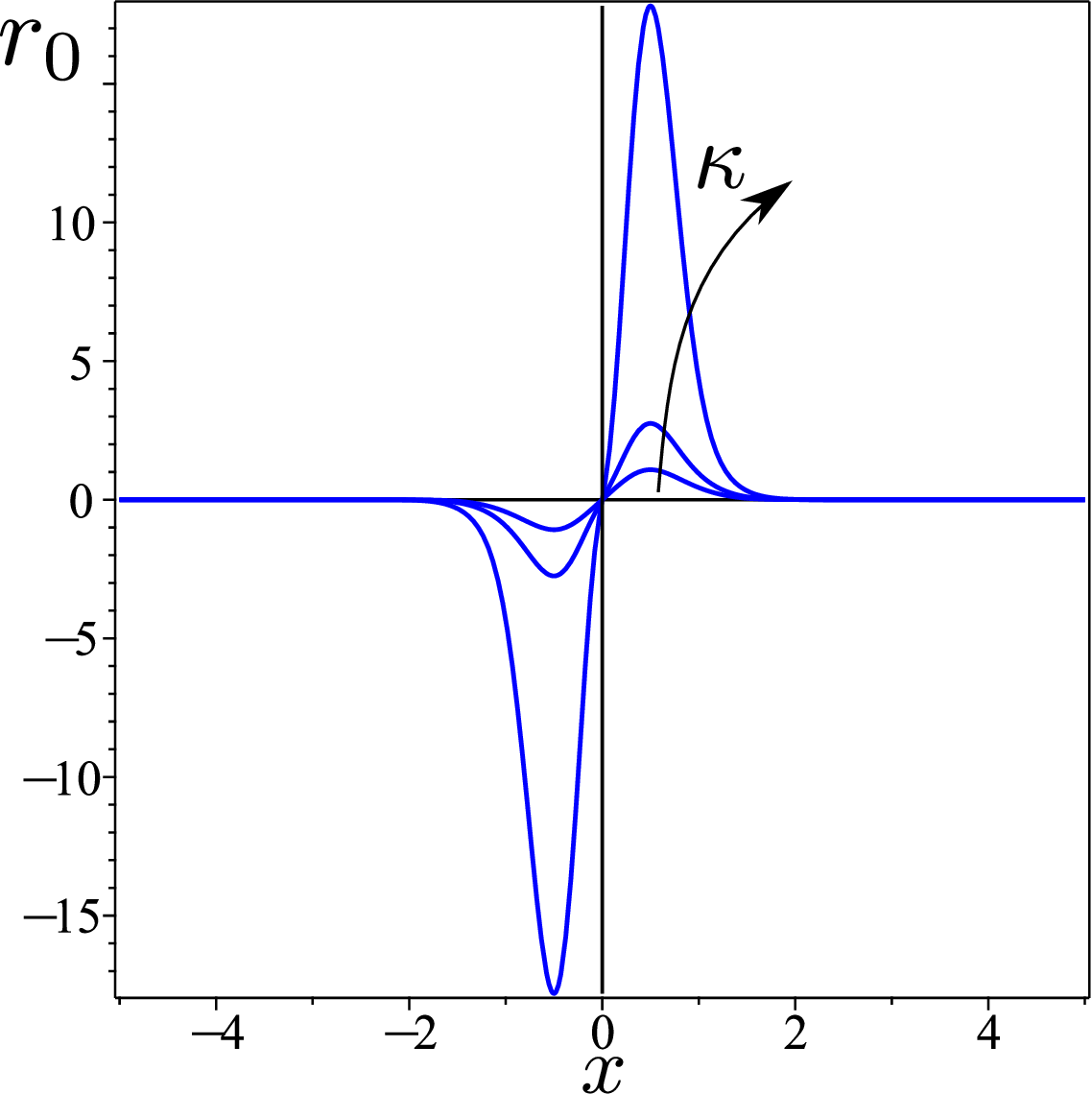}
		\caption{$\zeta=1/2$, $\kappa=\pi/4, \, \pi/2, \, 3\pi/2$.} 
		\label{fig:w_0gaussiankappa}
	\end{subfigure}
	\caption{Initial profile of the Riemann invariant $r_0$ in \eqref{eq:r_0gaussian} corresponding to the initial twist $w_0$ illustrated in Fig.~\ref{fig:w_0gaussian}.}
	\label{fig:r_0gaussian}
\end{figure}

By \eqref{eq:IC_riemann} and \eqref{eq:L_u_1}, $r_0$ is given by
\begin{equation}
\label{eq:r_0gaussian}
r_0(\kappa,\zeta;x)=\frac{1}{2}\left(\xi\sqrt{1+\xi^2}+\hbox{arcsinh}\,\xi\right), \quad \text{where} \quad \xi:=-\frac{x}{\zeta^2}\twist_0(x,\kappa,\zeta)=-\frac{\kappa}{\zeta^2}x \esp^{-\frac{x^2}{2\zeta^2}}
\end{equation}
and its graph is illustrated in Fig.~\ref{fig:r_0gaussian} for different values of $\kappa$ and $\zeta$. As $r_0$ is an odd function, by Theorem~\ref{th:suff_cond}, it suffices to look for the occurrence of singularities along the forward characteristics.
Since $r_0(+\infty)=0$, the values of $\alpha\in\mathbb{R}$ for which $r_0$ satisfies condition \eqref{eq:sufficient_condition} can be found for both  $\alpha\in(-\zeta,0)$ and $\alpha>\zeta$. Singularities develop in a finite time along every characteristic starting from these values of $\alpha$, but only for $\alpha>\zeta$ is $t_\mathrm{c}$ in \eqref{eq:critical_time_estimate} finite.

\begin{figure}[]
	\centering
	\begin{subfigure}[c]{0.35\linewidth}
		\centering
		\includegraphics[width=\linewidth]{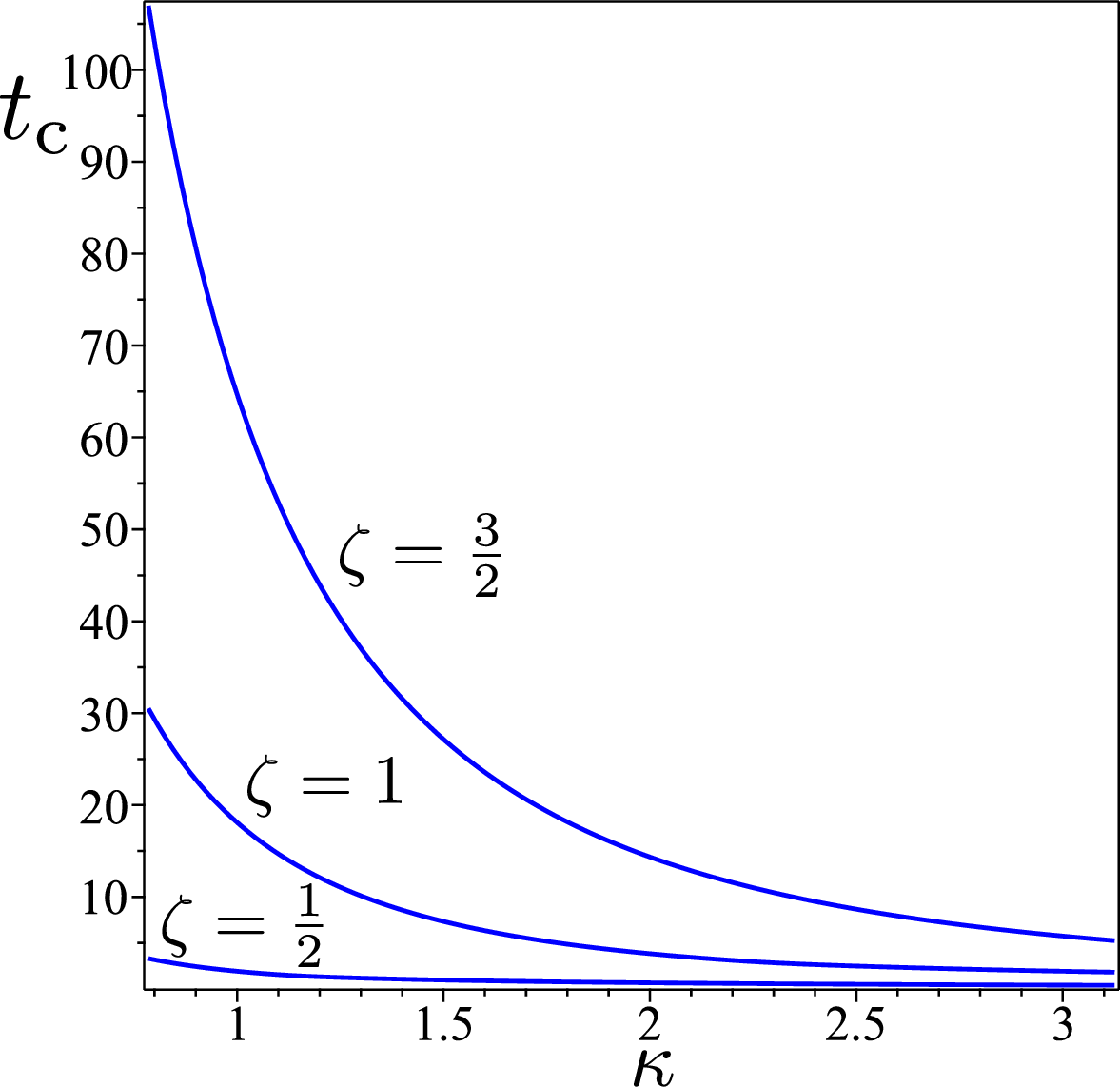}
		\caption{$\kappa\in[\pi/4,3\pi/2]$, $\zeta=1/2, \, 1, \, 3/2$.} 
		\label{fig:T_crit_gaussianzeta}
	\end{subfigure}
	\quad
	\begin{subfigure}[c]{0.35\linewidth}
		\centering
		\includegraphics[width=\linewidth]{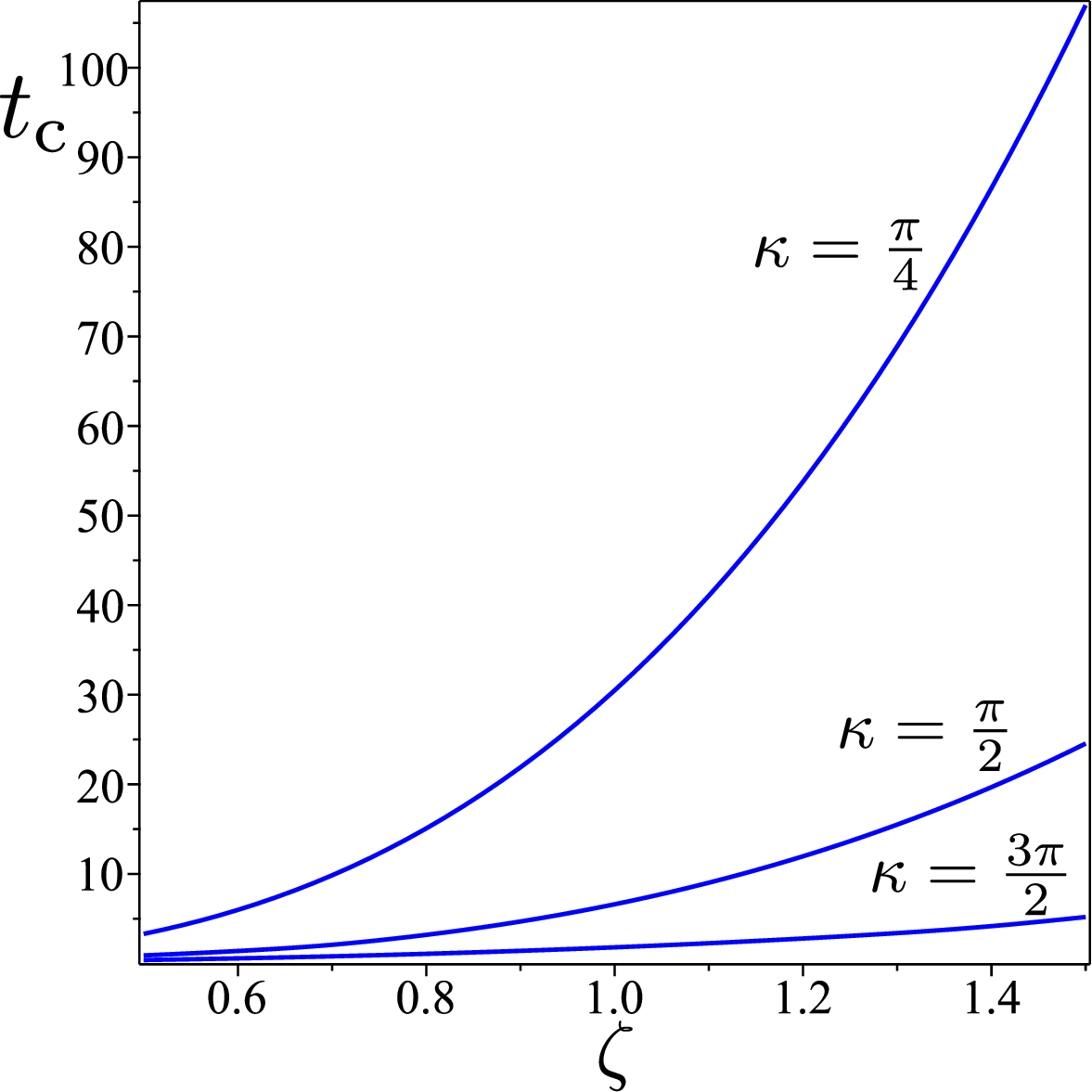}
		\caption{$\zeta\in[1/2,3/2]$, $\kappa=\pi/4, \, \pi/2, \, 3\pi/2$.} 
		\label{fig:T_crit_gaussiankappa}
	\end{subfigure}
\caption{The time $t_\mathrm{c}$ delivered by \eqref{eq:critical_time_estimate} is the upper estimate of the critical time $t^\ast$ of a regular solution of the global Cauchy problem \eqref{eq:wave_system}, with initial profile \eqref{eq:w_0_gaussian}. In particular, for $\kappa=\pi/2$ and $\zeta=1/2$, $t_\mathrm{c}\doteq0.9$.}
	\label{fig:T_crit_gaussian}
\end{figure}

Fig.~\ref{fig:T_crit_gaussian} illustrates how $t_\mathrm{c}$ depends on both $\kappa$ and $\zeta$; as expected, $t_\mathrm{c}$ decreases upon increasing $\kappa$ or decreasing $\zeta$, both actions corresponding to an enhancement of distortion in the initial twist profile.

\begin{figure}[]
	\centering
	\begin{subfigure}[c]{0.35\linewidth}
		\centering
		\includegraphics[width=\linewidth]{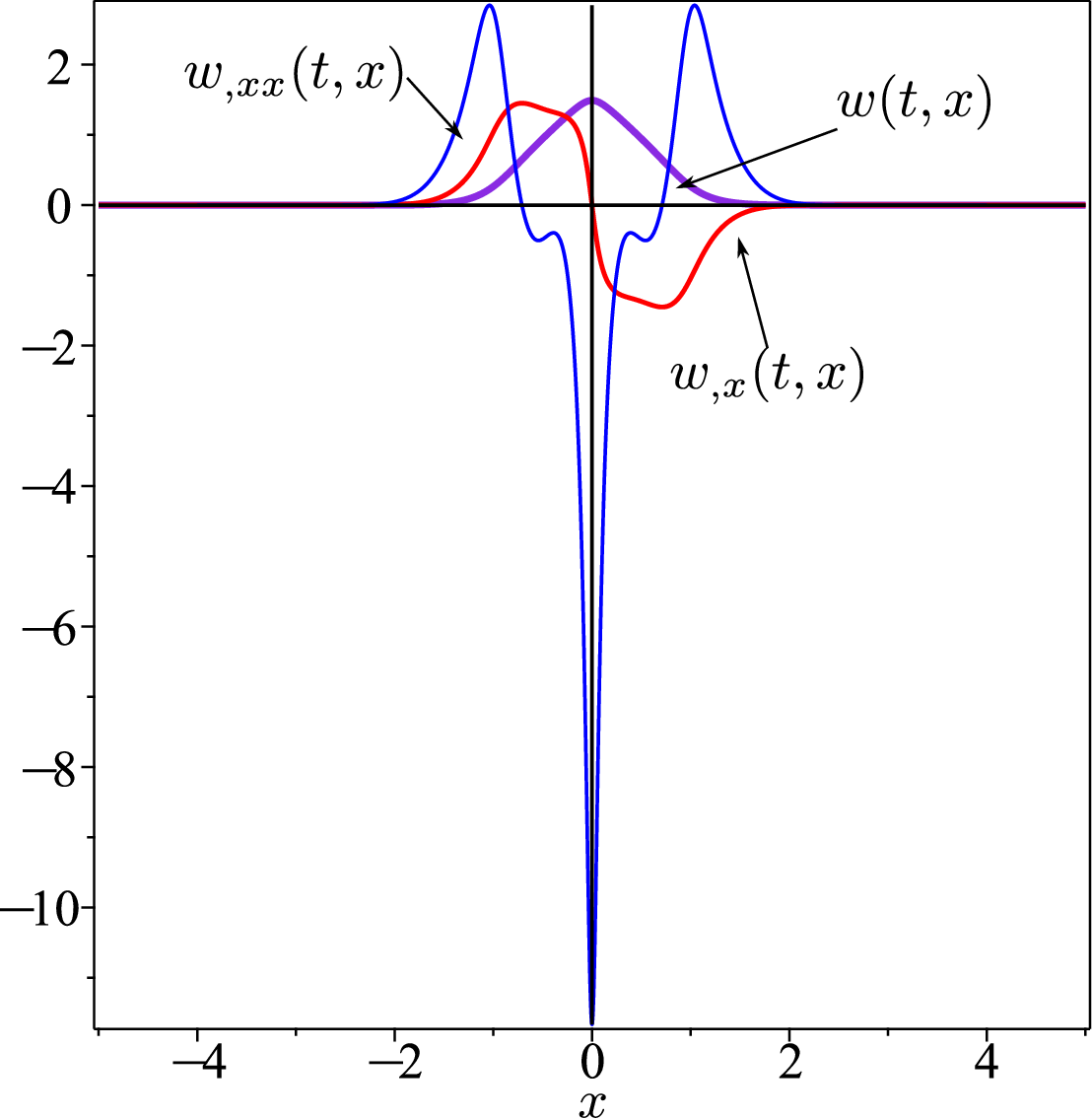}
		\caption{$t=0.15$} 
		\label{fig:t_015}
	\end{subfigure}
	\quad
	\begin{subfigure}[c]{0.35\linewidth}
		\centering
		\includegraphics[width=\linewidth]{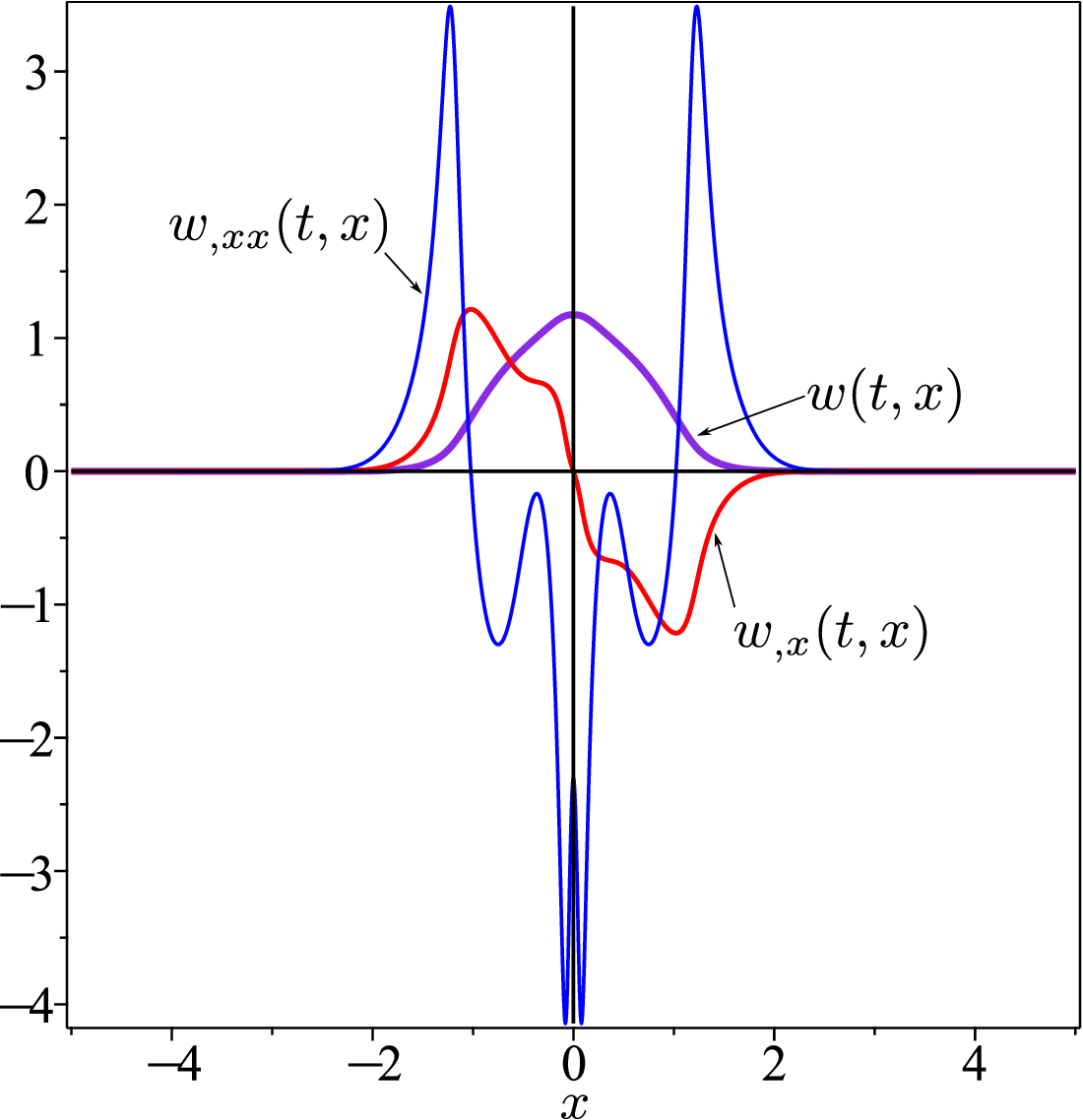}
		\caption{$t=0.30$} 
		\label{fig:t_03}
	\end{subfigure}
	\begin{subfigure}[c]{0.35\linewidth}
		\centering
		\includegraphics[width=\linewidth]{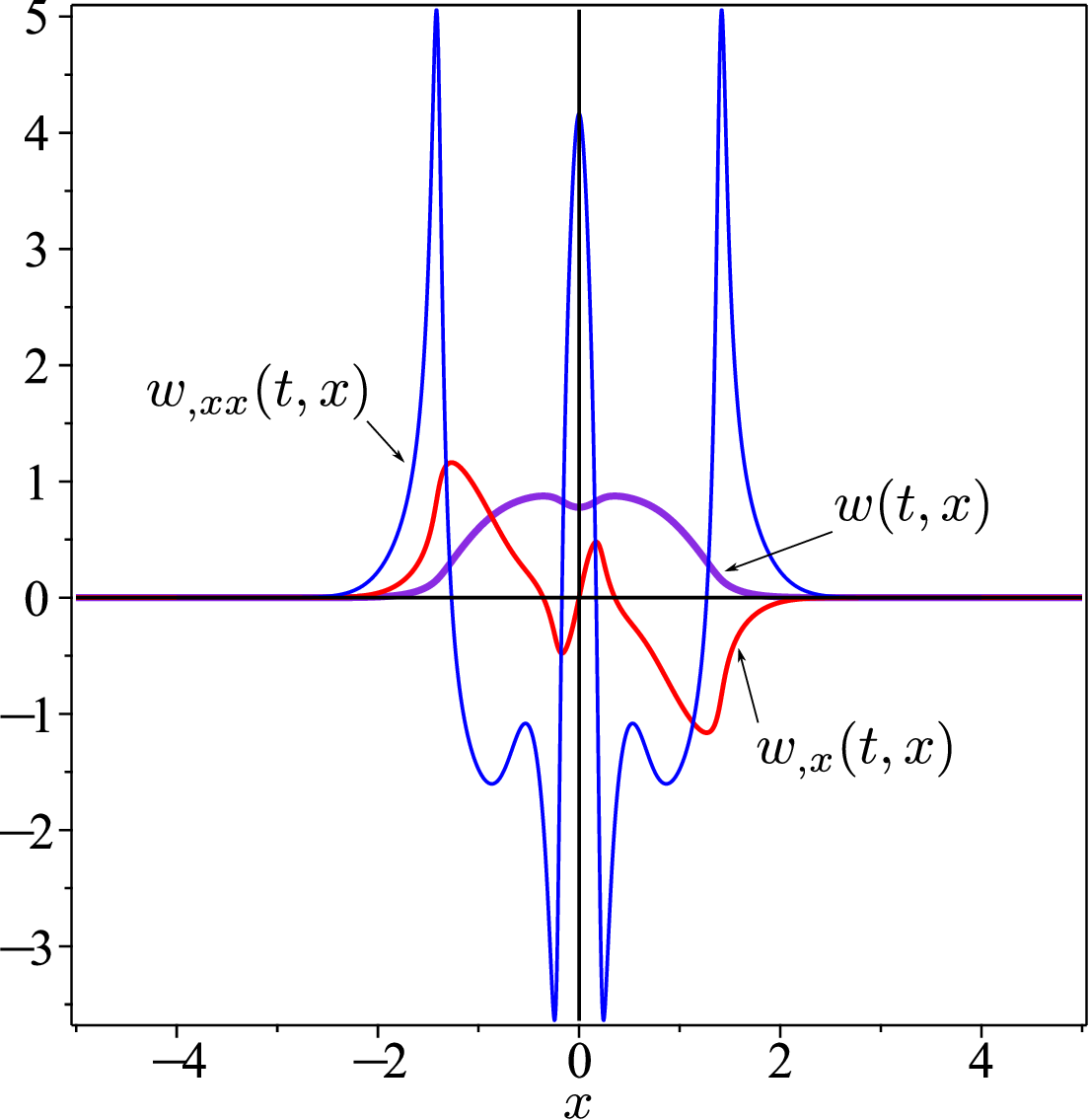}
		\caption{$t=0.45$} 
		\label{fig:t_045}
	\end{subfigure}
	\quad
	\begin{subfigure}[c]{0.35\linewidth}
		\centering
		\includegraphics[width=\linewidth]{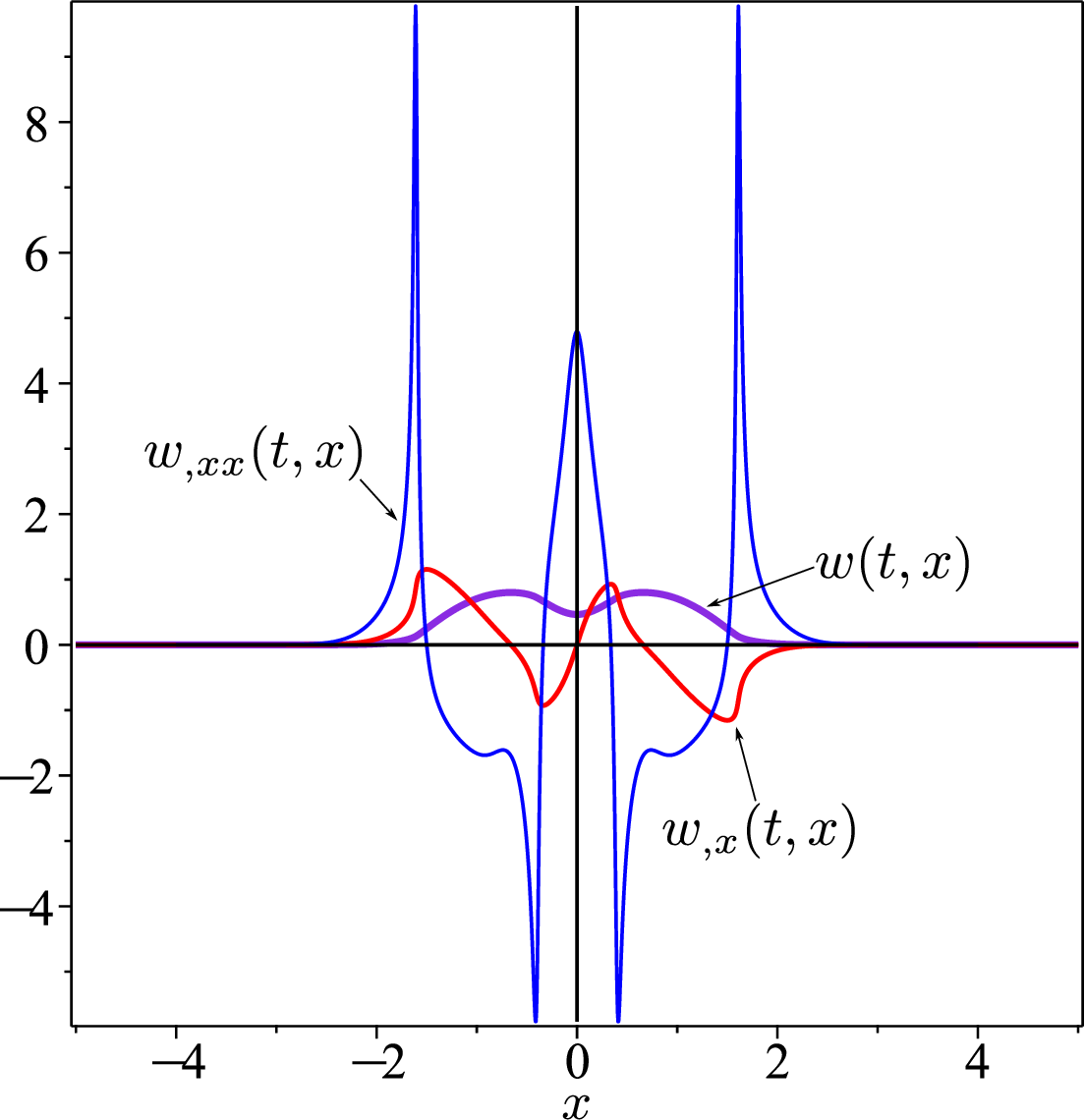}
		\caption{$t=0.60$} 
		\label{fig:t_06}
	\end{subfigure}
	\caption{Graphs of the numerical solution $\twist(t,x)$ and its spatial derivatives $w_{,x}(t,x)$ and $w_{,xx}(t,x)$ for the initial profile in  \eqref{eq:w_0_gaussian} with $\kappa = \pi/2$, $\zeta=1/2$. The time sequence suggests that $w_{,xx}$ tends to develop two symmetric spikes, whereas $w_{,x}$ tends to develop antisymmetric jumps.}
	\label{fig:num_sol_gaussian}
\end{figure}

\begin{figure}[]
	\centering
	\begin{subfigure}[c]{0.41\linewidth}
		\centering
		\includegraphics[width=\linewidth]{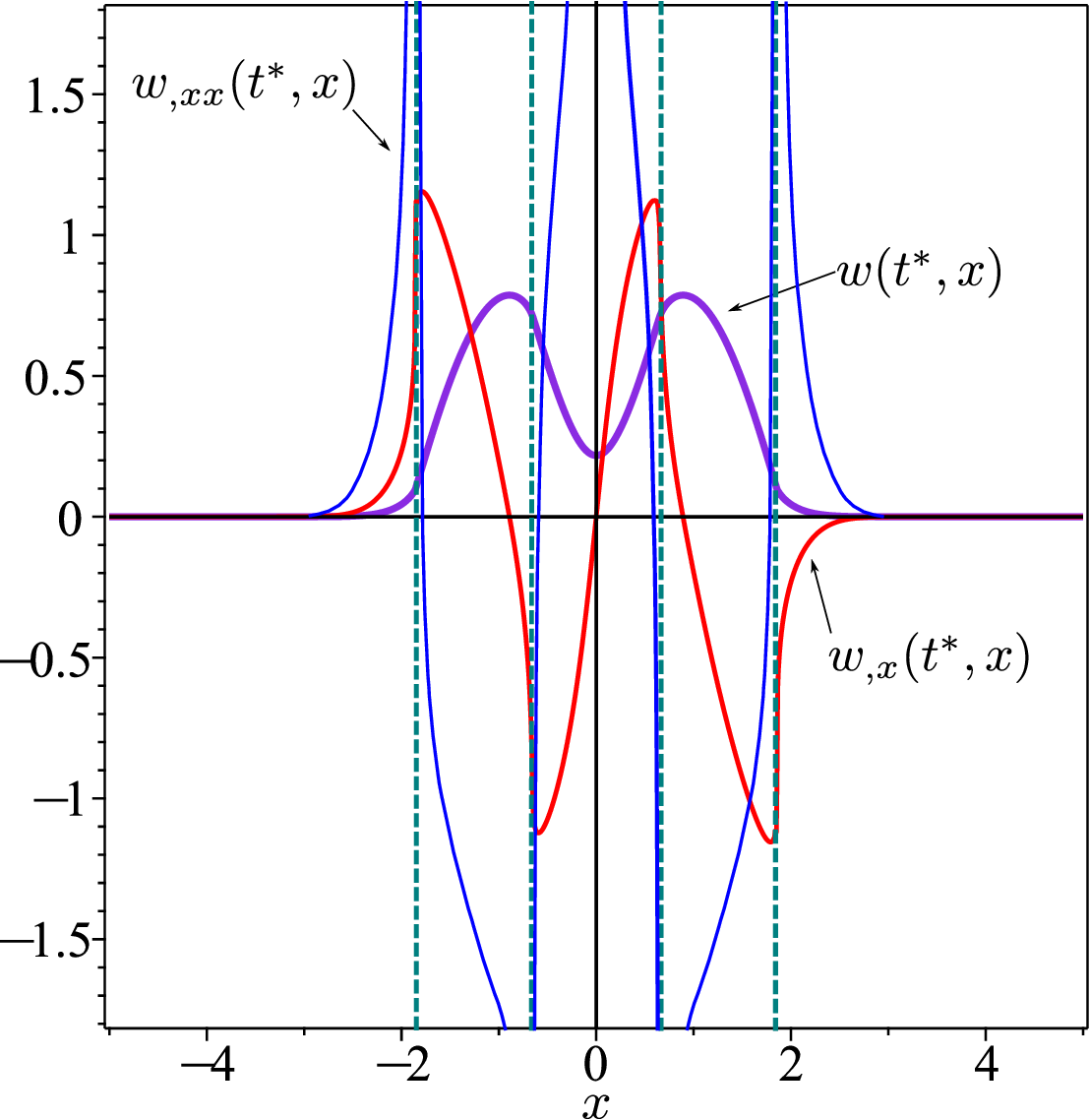}
		\caption{Graphs of the numerical solution $\twist(t^\ast,x)$ and its spatial  derivatives $\twist_{,x}(t^\ast,x)$ and $\twist_{,xx}(t^\ast,x)$ at the critical time $t^\ast\doteq0.8$.} 
		\label{fig:num:sol_t_crit_gaussian}
	\end{subfigure}
	\qquad
	\begin{subfigure}[c]{0.3\linewidth}
		\centering
		\includegraphics[width=\linewidth]{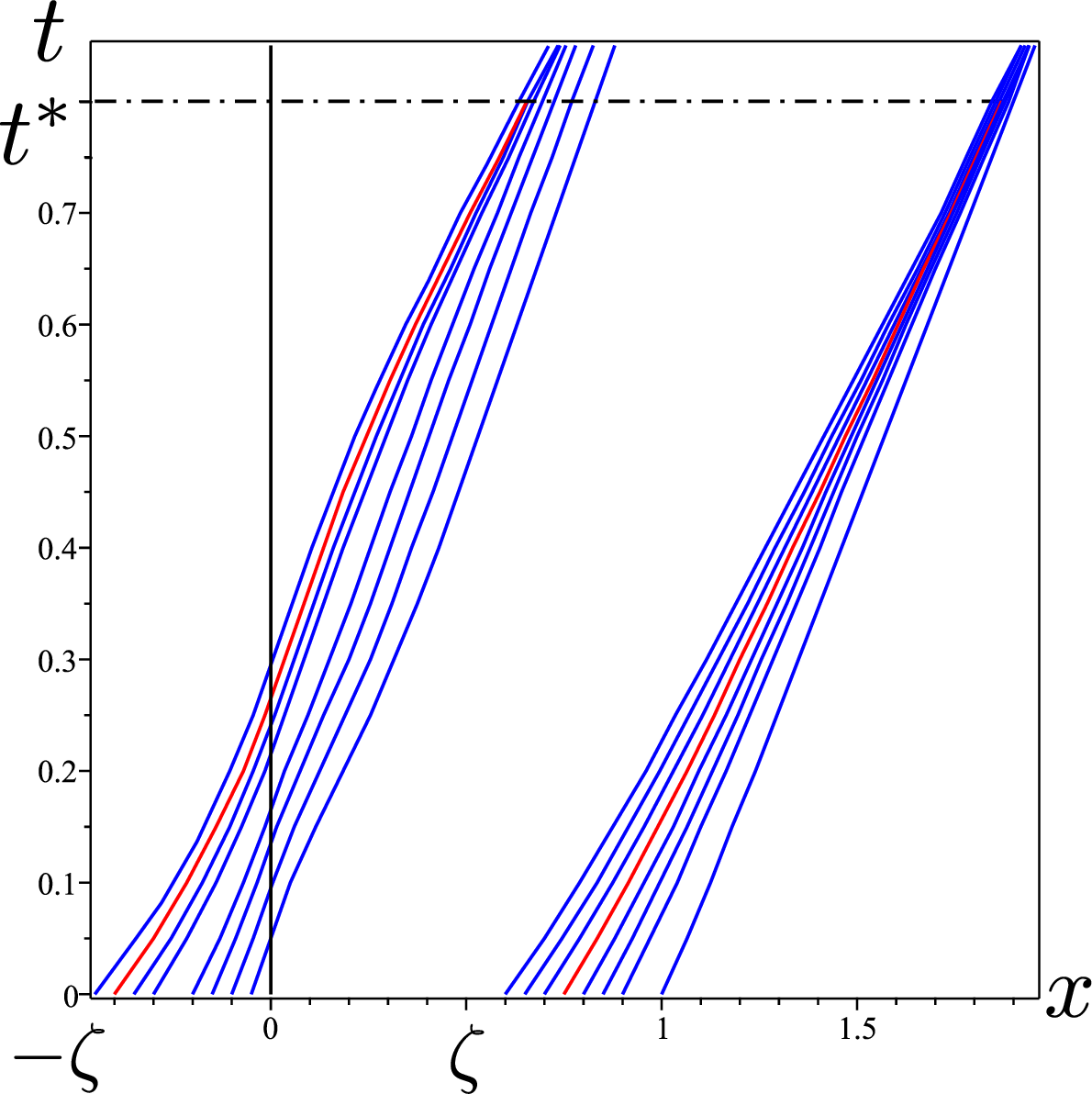}
		\caption{Forward characteristic curves originating at different values of $\alpha\in(-\zeta,0)$ and $\alpha>\zeta$, each developing a  singularity  in a finite time. The ones starting at $\alpha\doteq-0.4$ and $\alpha\doteq0.75$ become infinitesimally compressive (that is, $c_1=0$) at approximately the same time, $t\approx t^\ast$, before all others.} 
		\label{fig:num:sol_t_crit_gaussian_characteristics}
	\end{subfigure}
	\caption{Singularity emerging at the critical time $t^\ast\doteq0.8$ along the solution of \eqref{eq:wave_system} for the initial profile \eqref{eq:w_0_gaussian} with $\kappa = \pi/2$, $\zeta=1/2$.}
	\label{fig:T_crit_snapshot_gaussian}
\end{figure}

\begin{figure}[h] 
	\centering
	\includegraphics[width=0.4\linewidth]{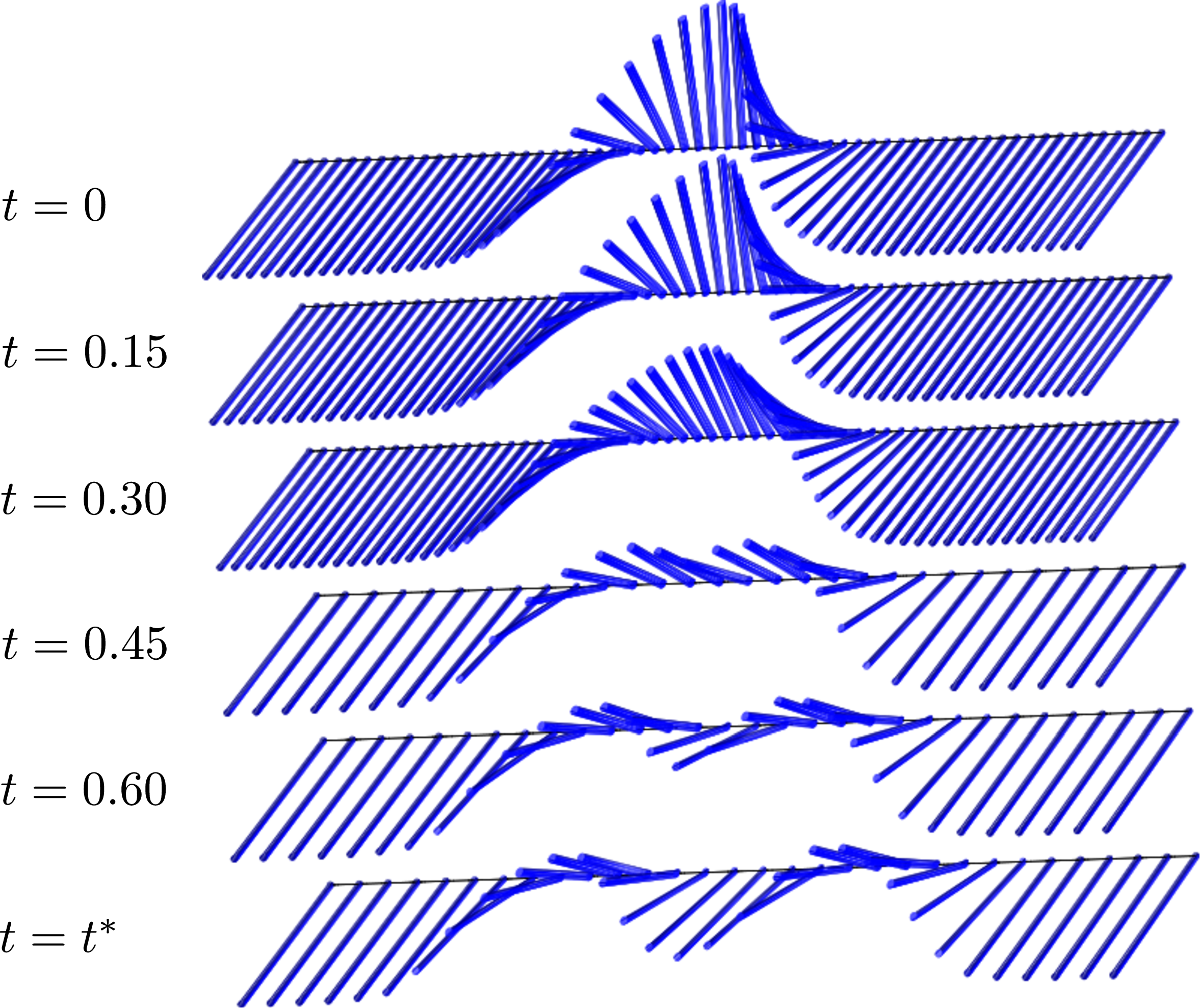}
	\caption{\nigh{Sketches of the directors corresponding to the solutions shown in Figs.~\ref{fig:num_sol_gaussian} and \ref{fig:num:sol_t_crit_gaussian}.}}
	\label{fig:sketch_gaussian}
\end{figure}

Numerical solutions of the global Cauchy problem \eqref{eq:wave_system} with initial profile \eqref{eq:w_0_gaussian} confirm our theoretical predictions. Specifically, for $\kappa=\pi/2$ and $\zeta=1/2$, Fig.~\ref{fig:num_sol_gaussian} depicts $\twist(t,x)$ and its spatial derivatives $\twist_{,x}(t,x)$ and $\twist_{,xx}(t,x)$ for times in the interval  $[0,t^\ast)$, with $t^\ast\doteq0.8$. 
The graphs at time $t=t^\ast$ are illustrated in Fig. \ref{fig:num:sol_t_crit_gaussian}.
\nigh{Sketches of the directors corresponding to the solutions shown in Figs.~\ref{fig:num_sol_gaussian} and \ref{fig:num:sol_t_crit_gaussian} are provided in Fig.~\ref{fig:sketch_gaussian}.}
Our numerical results indicate that a singularity occurs  along forward characteristics  originating at $\alpha\doteq-0.4$  and $\alpha\doteq 0.75$, which become infinitesimally compressive (that is, with $c_1=0$) at approximately the same time $t\approx t^\ast$, within our numerical accuracy (see Fig.~\ref{fig:num:sol_t_crit_gaussian_characteristics}). These findings agree well with our theory, as \eqref{eq:critical_time_estimate} delivers  $t_\mathrm{c}\doteq0.9$, with the infimum actually attained at  $\alpha\doteq0.78$.

\section{Conclusions}\label{sec:conclusions}
It has long been known in liquid crystal science that for quadratic elastic energy densities (such as that delivered by the classical Frank-Oseen formula), all weak twist shock waves decay in a finite time, and so they are physically irrelevant. If, on the contrary, the elastic energy density grows more than quadratically in the twist measure, weaker twist shock waves (bearing discontinuities in derivatives of order higher than two) decay rapidly, whereas acceleration shock waves persist and can even evolve in ordinary shock waves (bearing discontinuities in the first derivatives). 

This may just appear as a mere mathematical curiosity, as no firm physical ground is available to justify an energy density more than quadratic in the measure of twist for ordinary liquid crystals. The picture changes, however, when one considers chromonic phases, for which a number of studies (both theoretical and experimental) have suggested that a quartic twist energy may be plausible.

For these materials, shock waves may thus result from the evolution of acceleration waves. One may also say that strong shock waves are the only ones eventually surviving. The quest of this paper went the opposite way. We asked whether a shock wave could arise in a finite time from a \emph{smooth} solution of the (non-linear) wave equation associated with a quartic twist energy. Answering generically for the positive this question would identify shock waves as \emph{attractors} in the propagation of twist waves in chromonics.

We considered the global Cauchy problem with zero initial velocity, for which we identified (in Theorem~\ref{th:suff_cond}) mild assumptions on the initial distortion profile that guarantee that the ensuing smooth solution breaks down in a finite time $t^\ast$, giving way to the formation of a shock. Whenever these hypotheses are satisfied, an upper estimate for the critical time $t^\ast$ can be given (see Proposition~\ref{prop:estimate_summary}), which numerical calculations performed in a number of exemplary cases confirmed to be rather accurate.

In particular, we proved that under the assumptions of Theorem~\ref{th:suff_cond} \emph{all} initial profiles break down in a finite time. This conclusion is somehow reminiscent of the result of the classical theory for hyperbolic equations stating that (under appropriate assumptions) initial profiles with compact support always break down in a finite time. However, this is just a superficial similarity, as the assumptions of the classical theory do \emph{not} apply to our case, and our analysis was built on much more recent results.

The generic breaking of smooth twist waves in chromonic liquid crystals  that this paper \nigh{predicted  is} a possible experimental signature of the validity of the elastic quartic twist theory adopted here.

We have confined attention to the conservative wave equation, which can be justified on physical grounds only when \nigh{$\lambda\ll1$}. The dissipative case is more realistic, but far more difficult: one relevant issue would be whether dissipation could prevent the formation of a shock or not. Most of the mathematical methodology applied in this paper would not be directly applicable to the dissipative case. In the near future we plan to address this issue by developing appropriate mathematical tools.

\begin{acknowledgments}
Both authors are members of the Italian \emph{Gruppo Nazionale per la Fisica Matematica} (GNFM), which is part of INdAM, the Italian National Institute for Advanced Mathematics. S.P. gratefully acknowledges partial financial support provided for this work by GNFM.
\end{acknowledgments}

\appendix

\section{Governing Equations}\label{sec:twist_waves_LCs}
Our aim here is to derive equations \eqref{eq:pressure}, \eqref{eq:lagrange_multiplier} and \eqref{eq:balance_torques_wave} from \eqref{eq:momentum_balance} and \eqref{eq:balance_torques} when $W=\WQT$ as in \eqref{eq:quartic_free_energy_density} and $\n$ is represented as in \eqref{eq:director_tw}.

First, it easily follows from \eqref{eq:director_tw} that 
\begin{equation}
	\label{eq:nabla_n_twist}
	\nabla\n=\twist_{,x}\nper\otimes\e_x,
\end{equation}
where $\n_\perp:=\e_x\times\n$. The equations in \eqref{eq:distortion_measures} follow immediately from \eqref{eq:nabla_n_twist}.

Second, we remark that derivatives such as $\frac{\partial W}{\partial\nabla\n}$, $\frac{\partial R}{\partial\mathbf{D}}$, and $\frac{\partial R}{\partial\mathring{\n}}$ are to be interpreted in the intrinsic sense (as explained, for example, in \cite[p.\,133]{virga:variational}). Thus, the first is a tensor whose transpose annihilates $\n$, the second is a traceless, symmetric tensor, and the third is a vector orthogonal to $\n$. In particular, since
\begin{equation}
	\label{eq:twist_derivative}
	\frac{\partial T}{\partial\nabla\n}=\Wn,
\end{equation}
where $\Wn$ is the skew-symmetric tensor associated with $\n$,\footnote{$\Wn$ acts on a generic vector $\vv$ as  $\Wn\vv = \n\times\vv.$} by \eqref{eq:distortion_measures} and \eqref{eq:quartic_free_energy_density}, we obtain from \eqref{eq:momentum_balance_tw} that the Cauchy stress tensor $\mathbf{T}$ can be written as 
\begin{equation}
	\label{eq:stress_tensor_n_n_perp}
	\mathbf{T}=-p\bm{\mathrm{I}}-K_{22}\twist_{,x}^2\left(1+a^2\twist_{,x}^2\right)\e_x\otimes\e_x+\twist_{,t}(\mu_2\nper\otimes\n+\mu_3\n\otimes\nper),
\end{equation}
which follows from 
\begin{equation}
	\label{eq:W_QT_nabla_n}
	\frac{\partial\WQT}{\partial\nabla\n}=-K_{22}\twist_{,x}^2(1+a^2\twist_{,x}^2)\Wn,
\end{equation}
once use has also been made of the identities
\begin{equation}
	\label{eq:identities_appendix}
	\frac12\frac{\partial}{\partial\nabla\n}\tr(\nabla\n)^2=\Pn(\nabla\n)^{\mathsf{T}}\quad\text{and}\quad\partial_{t}\n=\twist_{,t}\nper,
\end{equation}
where $\Pn:=\mathbf{I}-\n\otimes\n$ is the projector onto the plane orthogonal to $\n$.
Furthermore, since \eqref{eq:director_tw} implies that
\begin{equation}
	\label{eq:div_n_appendix}
	\diver\n=\diver\nper=0\quad\text{and}\quad(\nabla\n)\nper=(\nabla\nper)\n=\bm{0},
\end{equation}
equation \eqref{eq:pressure} follows at once from \eqref{eq:balance_momentum_reduced} upon computing $\diver\mathbf{T}$ from \eqref{eq:stress_tensor_n_n_perp}.

Similarly, since in general
\begin{equation}
	\label{eq:partial_T_n}
	\frac{\partial T}{\partial\n}=\Pn\curl\n,
\end{equation}
use of \eqref{eq:nabla_n_twist} leads us to 
\begin{equation}
	\label{eq:partial_W_QT_n}
	\frac{\partial\WQT}{\partial\n}=\bm{0},
\end{equation}
whenever $\n$ is as in \eqref{eq:director_tw}. Moreover, since
\begin{equation}
	\label{eq:div_W_appendix}
	\diver\Wn=-\curl\n=\twist_{,x}\n,
\end{equation}
where the second equation requires \eqref{eq:nabla_n_twist}, we can write the balance equation \eqref{eq:balance_torques} in the explicit form
\begin{equation}
	\label{eq:balance_torques_n_n_perp}
	\sigma(\twist_{,tt}\nper-\twist_{,t}^2\n)+\gamma_1\twist_{,t}\nper+K_{22}\left[\twist_{,x}^2\left(1+a^2\twist_{,x}^2\right)\n-\twist_{,xx}\left(1+3a^2\twist_{,x}^2\right)\nper\right]=\mu\n.
\end{equation}
Equations \eqref{eq:lagrange_multiplier} and \eqref{eq:balance_torques_wave} in the main text then follow by projecting \eqref{eq:balance_torques_n_n_perp} along $\n$ and $\nper$, respectively.

\section{Proof of Auxiliary Results}\label{sec:manfrin}
In this Appendix, we present for completeness the proofs of two Propositions stated in the main text.

\begin{proof}[Proof of Proposition~\ref{prop:infinitesimal_compression_ratios}]
We differentiate both sides of the first equation in \eqref{eq:x_1} with respect to $\alpha$, and we find that
\begin{equation}
\label{eq:d_alpha_tx_1}
\frac{\partial}{\partial\alpha}\left(\frac{\dd x_1}{\dd t}(t,\alpha)\right)=\frac{\partial}{\partial\alpha}k\left(r_0(\alpha)-\ell(t,x_1(t,\alpha))\right)=k'\left(r_0(\alpha)-\ell(t,x_1(t,\alpha))\right)\left(r_0'(\alpha)-\ell_{,x}(t,x_1(t,\alpha))\frac{\partial x_1}{\partial\alpha}(t,\alpha)\right).
\end{equation}
We now derive an alternative expression for $\ell_{,x}(t,x_1(t,\alpha))$: by \eqref{eq:l_syst} and \eqref{eq:x_1} we obtain that 
\begin{equation}
\label{eq:l_x_t}
\frac{\dd }{\dd t}\ell(t,x_1(t,\alpha))=\ell_{,t}(t,x_1(t,\alpha))+\ell_{,x}(t,x_1(t,\alpha))\frac{\dd x_1}{\dd t}(t,\alpha)=2k\left(r_0(\alpha)-\ell(t,x_1(t,\alpha))\right)\ell_{,x}(t,x_1(t,\alpha)),
\end{equation}
from which it follows that
\begin{equation}
\label{eq:l_x_explicit}
\ell_{,x}(t,x_1(t,\alpha))=\frac{1}{2k\left(r_0(\alpha)-\ell(t,x_1(t,\alpha))\right)}\frac{\dd }{\dd t}\ell(t,x_1(t,\alpha)).
\end{equation}
Letting
\begin{equation}
\label{eq:H_def}
H(\eta):=\int_0^\eta\frac{k'(y)}{2k(y)}\dd y=\frac{1}{2}\ln k(\eta),
\end{equation}
where also \eqref{eq:k_l_r} has been used, with the aid of \eqref{eq:l_x_explicit} we arrive at
\begin{equation}
\label{eq:H_r_l}
-k'\left(r_0(\alpha)-\ell(t,x_1(t,\alpha))\right)\ell_{,x}(t,x_1(t,\alpha))=\frac{\dd H}{\dd t}(r_0(\alpha)-\ell(t,x_1(t,\alpha)))
\end{equation}
because
\begin{equation}
\label{eq:d_t_H}
\frac{\dd}{\dd t}H\left(r_0(\alpha)-\ell(t,x_1(t,\alpha))\right)=-\frac{k'\left(r_0(\alpha)-\ell(t,x_1(t,\alpha))\right)}{2k\left(r_0(\alpha)-\ell(t,x_1(t,\alpha))\right)}\frac{\dd }{\dd t}\ell(t,x_1(t,\alpha)).
\end{equation}
Then, \eqref{eq:d_alpha_tx_1} reduces to
\begin{equation}
\label{eq:ode_x_1_alpha}
\frac{\dd}{\dd t}c_1(t,\alpha)=k'\left(r_0(\alpha)-\ell(t,x_1(t,\alpha))\right)r_0'(\alpha)+\frac{\dd}{\dd t}H\left(r_0(\alpha)-\ell(t,x_1(t,\alpha))\right)c_1(t,\alpha).
\end{equation}
This is an ODE in the general form
\begin{equation}
\label{eq:ode_gen}
\dot{c}_1-\dot{h}_1c_1=h,
\end{equation}
where
\begin{equation}
\label{eq:ode_ingredients}
 h_1(t,\alpha):=H\left(r_0(\alpha)-\ell(t,x_1(t,\alpha))\right) \quad\text{and}\quad h(t,\alpha):=k'\left(r_0(\alpha)-\ell(t,x_1(t,\alpha))\right)r_0'(\alpha).
\end{equation}
By letting $A(t,\alpha):=h_1(t,\alpha)-h_1(0,\alpha)$, we obtain the following solution of \eqref{eq:ode_gen},
\begin{equation}
\label{eq:solution_ode_gen}
c_1(t,\alpha)=\esp^{A(t,\alpha)}\left[c_1(0,\alpha)+\int_0^t\nigh{h(\tau,\alpha)}\esp^{-A(\tau,\alpha)}\dd \tau\right],
\end{equation}
and so
\begin{equation}
\label{eq:solution_ode_x_1_alpha}
c_1(t,\alpha)=\esp^{h_1(t,\alpha)-h_1(0,\alpha)}\left\{1+r_0'(\alpha)\int_0^tk'\left(r_0(\alpha)-\ell(\tau,x_1(\tau,\alpha))\right)\esp^{h_1(0,\alpha)-h_1(\tau,\alpha)}\dd \tau\right\},
\end{equation}
where, by \eqref{eq:ode_ingredients} and \eqref{eq:IC_riemann}, $h_1(0,\alpha)=H(r_0(\alpha)-\ell_0(\alpha))=H(2r_0(\alpha))$, while $c_1(0,\alpha)=1$.

We can show in the same way that
\begin{equation}
\label{eq:solution_ode_x_2_beta}
c_2(t,\beta)=\esp^{h_2(t,\beta)-h_2(0,\beta)}\left\{1+\ell_0'(\beta)\int_0^tk'\left(r_0(\tau,x_2(\tau,\beta))-\ell_0(\beta)\right)\esp^{h_2(0,\beta)-h_2(\tau,\beta)}\dd \tau\right\}.
\end{equation}
Here  $h_2(t,\beta):=H\left(r(t,x_2(t,\beta))-\ell_0(\beta)\right)$, and so $h_2(0,\beta)=H(2r_0(\beta))$, where  \eqref{eq:IC_riemann} has also been used.

Finally, we can write
\label{eq:exp_diff}
\begin{equation}\label{eq:exp_diff_h1_h2}
\esp^{h_1(t,\alpha)-h_1(0,\alpha)}=\sqrt{\frac{k(r_0(\alpha)-\ell(t,x_1(t,\alpha)))}{k(2r_0(\alpha))}}\quad\text{and}\quad
\esp^{h_2(t,\beta)-h_2(0,\beta)}=\sqrt{\frac{k(r(t,x_2(t,\beta))-\ell_0(\beta))}{k(2r_0(\beta))}}.
\end{equation}
thus obtaining equations \eqref{eq:infinitesimal_compression_ratios_formulae} in the main text.
\end{proof}

\begin{proof}[Proof of Proposition~\ref{prop:beta_t_alpha}]
The existence of the unique solution $\beta=\beta(t,\alpha)$ is a consequence of the fact that for any $\tau\in[0,t)$ the Cauchy problem
\begin{equation}
\label{eq:backward_characteristic}
\begin{cases}
\frac{\dd y_2}{\dd \tau}(\tau,t,\alpha)=-k(\tau,y_2(\tau,t,\alpha)), \\
y_2(t,t,\alpha)=x_1(t,\alpha),
\end{cases}
\end{equation}
has a unique solution $y_2(\tau,t,\alpha)$ for $\tau\in[0,t^\ast)$. We define $\beta(t,\alpha):=y_2(0,t,\alpha)$ and $y_2(\tau,t,\alpha)=x_2(\tau,\beta(t,\alpha))$, where $x_2$ is a solution of \eqref{eq:x_2}. Thus, $\beta(t,\alpha)$ represents the point from which the backward characteristic $x_2$ starts. By \eqref{eq:x_2},  we conclude that $\beta(t,\alpha)>\alpha$ for every $t\in(0,t^\ast)$, and $\beta(0,\alpha)=\alpha$.

By differentiating both sides of \eqref{eq:beta_of_alpha_implicit_definition} with respect to $t$, and using \eqref{eq:x_1} and \eqref{eq:x_2}, we obtain \eqref{eq:d_t_beta}.

Next, by integrating both sides of \eqref{eq:x_1} with respect to $t$ and using \eqref{eq:estimates}, we arrive at
\begin{subequations}
\label{eq:upperbound_x1_x2}
\begin{align}
x_1(t,\alpha)&=\alpha+\int_0^{t}k(r(\alpha)-\ell(\tau,x_1(\tau,\alpha)))\dd\tau\leq \alpha +\delta t,\label{eq:upperbound_x1}\\
x_2(t,\beta(t,\alpha))&=\beta(t,\alpha)-\int_0^{t}k(r(\tau,x_2(\tau,\beta(t,\alpha)))-\ell_0(\beta(t,\alpha)))\dd\tau\geq \beta(t,\alpha) -\delta t. \label{eq:upperbound_x2}
\end{align}
\end{subequations}
By \eqref{eq:beta_of_alpha_implicit_definition} and \eqref{eq:upperbound_x1}, \eqref{eq:upperbound_x2} gives 
\begin{equation}
\label{eq:upperbound_beta}
\beta(t,\alpha)\leq\alpha+2\delta t.
\end{equation} 

To establish a lower bound for $\beta(t,\alpha)$, we return to \eqref{eq:d_t_beta}. First, we establish an upper bound for $c_2$ by evaluating \eqref{eq:infinitesimal_compression_ratio_x2}  at $\beta=\beta(t,{\alpha})$. Since, by \eqref{eq:alpha_of_beta_implicit_definition} and \eqref{eq:r_constant},  $r(t,x_2(t,\beta(t,\alpha)))=r(t,x_1(t,\alpha))=r_0(\alpha)$, from \eqref{eq:infinitesimal_compression_ratio_x2} we obtain that
\begin{equation}
\label{eq:partial_x2_beta}
c_2(t,\beta(t,\alpha))=\sqrt{\frac{k(r_0(\alpha)-\ell_0(\beta(t,\alpha)))}{k(2r_0(\beta(t,\alpha)))}}\left\{1+\ell_0'(\beta(t,\alpha))\sqrt{k(2r_0(\beta(t,\alpha)))}
\int_0^t f\left(r(\tau,x_2(\tau,\beta(t,\alpha)))-\ell_0(\beta(t,\alpha))\right)\dd \tau\right\}.
\end{equation}

By \eqref{eq:estimates}, since $f(\eta)\leq f_0$, we have that
\begin{equation}
\label{eq:upperbound_partial_x2}
c_2(t,\beta(t,\alpha))\leq\sqrt{k(r_0(\alpha)-\ell_0(\beta(t,\alpha)))}\left(1+f_0||\ell_0'||_\infty t\right).
\end{equation}
Then, by integrating both sides of \eqref{eq:d_t_beta} with respect to $t$ and using \eqref{eq:upperbound_partial_x2} and \eqref{eq:estimates}, we arrive at
\begin{equation}
\label{eq:lowerbound_beta}
\begin{aligned}
\beta(t,\alpha)&=\beta(0,\alpha)+\int_0^t\frac{2k(r_0(\alpha)-\ell_0(\beta(\tau,\alpha)))}{c_2(\tau,\beta(\tau,\alpha))}\dd\tau
\geq\alpha+2\int_0^{t}\frac{\dd\tau}{1+f_0||\ell_0'||_\infty \tau}\nonumber\\
&=\alpha+\frac{2}{f_0||\ell_0'||_\infty}\ln\left(1+f_0||\ell_0'||_\infty t\right),
\end{aligned}
\end{equation}
where use has also been made of the identity $\beta(0,\alpha)=\alpha$ (see Remark~\ref{rmk:alpha_beta_nesting}). The properties of the function $\alpha(t,\beta)$ stated  in Proposition~\ref{prop:beta_t_alpha} can be established similarly by use of \eqref{eq:x_2} and \eqref{eq:infinitesimal_compression_ratio_x1}.
\end{proof}

\begin{thebibliography}{69}%
	\makeatletter
	\providecommand \@ifxundefined [1]{%
		\@ifx{#1\undefined}
	}%
	\providecommand \@ifnum [1]{%
		\ifnum #1\expandafter \@firstoftwo
		\else \expandafter \@secondoftwo
		\fi
	}%
	\providecommand \@ifx [1]{%
		\ifx #1\expandafter \@firstoftwo
		\else \expandafter \@secondoftwo
		\fi
	}%
	\providecommand \natexlab [1]{#1}%
	\providecommand \enquote  [1]{``#1''}%
	\providecommand \bibnamefont  [1]{#1}%
	\providecommand \bibfnamefont [1]{#1}%
	\providecommand \citenamefont [1]{#1}%
	\providecommand \href@noop [0]{\@secondoftwo}%
	\providecommand \href [0]{\begingroup \@sanitize@url \@href}%
	\providecommand \@href[1]{\@@startlink{#1}\@@href}%
	\providecommand \@@href[1]{\endgroup#1\@@endlink}%
	\providecommand \@sanitize@url [0]{\catcode `\\12\catcode `\$12\catcode
		`\&12\catcode `\#12\catcode `\^12\catcode `\_12\catcode `\%12\relax}%
	\providecommand \@@startlink[1]{}%
	\providecommand \@@endlink[0]{}%
	\providecommand \url  [0]{\begingroup\@sanitize@url \@url }%
	\providecommand \@url [1]{\endgroup\@href {#1}{\urlprefix }}%
	\providecommand \urlprefix  [0]{URL }%
	\providecommand \Eprint [0]{\href }%
	\providecommand \doibase [0]{https://doi.org/}%
	\providecommand \selectlanguage [0]{\@gobble}%
	\providecommand \bibinfo  [0]{\@secondoftwo}%
	\providecommand \bibfield  [0]{\@secondoftwo}%
	\providecommand \translation [1]{[#1]}%
	\providecommand \BibitemOpen [0]{}%
	\providecommand \bibitemStop [0]{}%
	\providecommand \bibitemNoStop [0]{.\EOS\space}%
	\providecommand \EOS [0]{\spacefactor3000\relax}%
	\providecommand \BibitemShut  [1]{\csname bibitem#1\endcsname}%
	\let\auto@bib@innerbib\@empty
	\bibitem [{\citenamefont {Shiyanovskii}\ \emph {et~al.}(2005)\citenamefont
		{Shiyanovskii}, \citenamefont {Schneider}, \citenamefont {Smalyukh},
		\citenamefont {Ishikawa}, \citenamefont {Niehaus}, \citenamefont {Doane},
		\citenamefont {Woolverton},\ and\ \citenamefont
		{Lavrentovich}}]{shiyanovskii:real-time}%
	\BibitemOpen
	\bibfield  {author} {\bibinfo {author} {\bibfnamefont {S.~V.}\ \bibnamefont
			{Shiyanovskii}}, \bibinfo {author} {\bibfnamefont {T.}~\bibnamefont
			{Schneider}}, \bibinfo {author} {\bibfnamefont {I.~I.}\ \bibnamefont
			{Smalyukh}}, \bibinfo {author} {\bibfnamefont {T.}~\bibnamefont {Ishikawa}},
		\bibinfo {author} {\bibfnamefont {G.~D.}\ \bibnamefont {Niehaus}}, \bibinfo
		{author} {\bibfnamefont {K.~J.}\ \bibnamefont {Doane}}, \bibinfo {author}
		{\bibfnamefont {C.~J.}\ \bibnamefont {Woolverton}},\ and\ \bibinfo {author}
		{\bibfnamefont {O.~D.}\ \bibnamefont {Lavrentovich}},\ }\bibfield  {title}
	{\bibinfo {title} {Real-time microbe detection based on director distortions
			around growing immune complexes in lyotropic chromonic liquid crystals},\
	}\href {https://doi.org/10.1103/PhysRevE.71.020702} {\bibfield  {journal}
		{\bibinfo  {journal} {Phys. Rev. E}\ }\textbf {\bibinfo {volume} {71}},\
		\bibinfo {pages} {020702} (\bibinfo {year} {2005})}\BibitemShut {NoStop}%
	\bibitem [{\citenamefont {Mushenheim}\ \emph
		{et~al.}(2014{\natexlab{a}})\citenamefont {Mushenheim}, \citenamefont
		{Trivedi}, \citenamefont {Tuson}, \citenamefont {Weibel},\ and\ \citenamefont
		{Abbott}}]{mushenheim:dynamic}%
	\BibitemOpen
	\bibfield  {author} {\bibinfo {author} {\bibfnamefont {P.~C.}\ \bibnamefont
			{Mushenheim}}, \bibinfo {author} {\bibfnamefont {R.~R.}\ \bibnamefont
			{Trivedi}}, \bibinfo {author} {\bibfnamefont {H.~H.}\ \bibnamefont {Tuson}},
		\bibinfo {author} {\bibfnamefont {D.~B.}\ \bibnamefont {Weibel}},\ and\
		\bibinfo {author} {\bibfnamefont {N.~L.}\ \bibnamefont {Abbott}},\ }\bibfield
	{title} {\bibinfo {title} {Dynamic self-assembly of motile bacteria in
			liquid crystals},\ }\href {https://doi.org/10.1039/C3SM52423J} {\bibfield
		{journal} {\bibinfo  {journal} {Soft Matter}\ }\textbf {\bibinfo {volume}
			{10}},\ \bibinfo {pages} {88} (\bibinfo {year}
		{2014}{\natexlab{a}})}\BibitemShut {NoStop}%
	\bibitem [{\citenamefont {Mushenheim}\ \emph
		{et~al.}(2014{\natexlab{b}})\citenamefont {Mushenheim}, \citenamefont
		{Trivedi}, \citenamefont {Weibel},\ and\ \citenamefont
		{Abbott}}]{mushenheim:using}%
	\BibitemOpen
	\bibfield  {author} {\bibinfo {author} {\bibfnamefont {P.~C.}\ \bibnamefont
			{Mushenheim}}, \bibinfo {author} {\bibfnamefont {R.~R.}\ \bibnamefont
			{Trivedi}}, \bibinfo {author} {\bibfnamefont {D.}~\bibnamefont {Weibel}},\
		and\ \bibinfo {author} {\bibfnamefont {N.}~\bibnamefont {Abbott}},\
	}\bibfield  {title} {\bibinfo {title} {Using liquid crystals to reveal how
			mechanical anisotropy changes interfacial behaviors of motile bacteria},\
	}\href {https://doi.org/https://doi.org/10.1016/j.bpj.2014.04.047} {\bibfield
		{journal} {\bibinfo  {journal} {Biophys. J.}\ }\textbf {\bibinfo {volume}
			{107}},\ \bibinfo {pages} {255} (\bibinfo {year}
		{2014}{\natexlab{b}})}\BibitemShut {NoStop}%
	\bibitem [{\citenamefont {Zhou}\ \emph
		{et~al.}(2014{\natexlab{a}})\citenamefont {Zhou}, \citenamefont {Sokolov},
		\citenamefont {Lavrentovich},\ and\ \citenamefont {Aranson}}]{zhou:living}%
	\BibitemOpen
	\bibfield  {author} {\bibinfo {author} {\bibfnamefont {S.}~\bibnamefont
			{Zhou}}, \bibinfo {author} {\bibfnamefont {A.}~\bibnamefont {Sokolov}},
		\bibinfo {author} {\bibfnamefont {O.~D.}\ \bibnamefont {Lavrentovich}},\ and\
		\bibinfo {author} {\bibfnamefont {I.~S.}\ \bibnamefont {Aranson}},\
	}\bibfield  {title} {\bibinfo {title} {Living liquid crystals},\ }\href
	{https://doi.org/10.1073/pnas.1321926111} {\bibfield  {journal} {\bibinfo
			{journal} {Proc. Natl. Acad. Sci. USA}\ }\textbf {\bibinfo {volume} {111}},\
		\bibinfo {pages} {1265} (\bibinfo {year} {2014}{\natexlab{a}})}\BibitemShut
	{NoStop}%
	\bibitem [{\citenamefont {Lydon}(1998{\natexlab{a}})}]{lydon:chromonic_1998}%
	\BibitemOpen
	\bibfield  {author} {\bibinfo {author} {\bibfnamefont {J.}~\bibnamefont
			{Lydon}},\ }\bibfield  {title} {\bibinfo {title} {Chromonic liquid crystal
			phases},\ }\href
	{https://doi.org/https://doi.org/10.1016/S1359-0294(98)80019-8} {\bibfield
		{journal} {\bibinfo  {journal} {Curr. Opin. Colloid Interface Sci.}\ }\textbf
		{\bibinfo {volume} {3}},\ \bibinfo {pages} {458} (\bibinfo {year}
		{1998}{\natexlab{a}})}\BibitemShut {NoStop}%
	\bibitem [{\citenamefont {Lydon}(1998{\natexlab{b}})}]{lydon:handbook}%
	\BibitemOpen
	\bibfield  {author} {\bibinfo {author} {\bibfnamefont {J.}~\bibnamefont
			{Lydon}},\ }\bibfield  {title} {\bibinfo {title} {Chromonics},\ }in\ \href
	{https://doi.org/https://doi.org/10.1002/9783527619276.ch15c} {\emph
		{\bibinfo {booktitle} {Handbook of Liquid Crystals: {L}ow Molecular Weight
				Liquid Crystals {II}}}},\ \bibinfo {editor} {edited by\ \bibinfo {editor}
		{\bibfnamefont {D.}~\bibnamefont {Demus}}, \bibinfo {editor} {\bibfnamefont
			{J.}~\bibnamefont {Goodby}}, \bibinfo {editor} {\bibfnamefont {G.~W.}\
			\bibnamefont {Gray}}, \bibinfo {editor} {\bibfnamefont {H.-W.}\ \bibnamefont
			{Spiess}},\ and\ \bibinfo {editor} {\bibfnamefont {V.}~\bibnamefont {Vill}}}\
	(\bibinfo  {publisher} {John Wiley \& Sons},\ \bibinfo {address} {Weinheim,
		Germany},\ \bibinfo {year} {1998})\ Chap.\ \bibinfo {chapter} {XVIII}, pp.\
	\bibinfo {pages} {981--1007}\BibitemShut {NoStop}%
	\bibitem [{\citenamefont {Lydon}(2010)}]{lydon:chromonic_2010}%
	\BibitemOpen
	\bibfield  {author} {\bibinfo {author} {\bibfnamefont {J.}~\bibnamefont
			{Lydon}},\ }\bibfield  {title} {\bibinfo {title} {Chromonic review},\ }\href
	{https://doi.org/10.1039/B926374H} {\bibfield  {journal} {\bibinfo  {journal}
			{J. Mater. Chem.}\ }\textbf {\bibinfo {volume} {20}},\ \bibinfo {pages}
		{10071} (\bibinfo {year} {2010})}\BibitemShut {NoStop}%
	\bibitem [{\citenamefont {Lydon}(2011)}]{lydon:chromonic}%
	\BibitemOpen
	\bibfield  {author} {\bibinfo {author} {\bibfnamefont {J.}~\bibnamefont
			{Lydon}},\ }\bibfield  {title} {\bibinfo {title} {Chromonic liquid
			crystalline phases},\ }\href {https://doi.org/10.1080/02678292.2011.614720}
	{\bibfield  {journal} {\bibinfo  {journal} {Liq. Cryst.}\ }\textbf {\bibinfo
			{volume} {38}},\ \bibinfo {pages} {1663} (\bibinfo {year}
		{2011})}\BibitemShut {NoStop}%
	\bibitem [{\citenamefont {Dierking}\ and\ \citenamefont {Martins
			Figueiredo~Neto}(2020)}]{dierking:novel}%
	\BibitemOpen
	\bibfield  {author} {\bibinfo {author} {\bibfnamefont {I.}~\bibnamefont
			{Dierking}}\ and\ \bibinfo {author} {\bibfnamefont {A.}~\bibnamefont {Martins
				Figueiredo~Neto}},\ }\bibfield  {title} {\bibinfo {title} {Novel trends in
			lyotropic liquid crystals},\ }\href {https://doi.org/10.3390/cryst10070604}
	{\bibfield  {journal} {\bibinfo  {journal} {Crystals}\ }\textbf {\bibinfo
			{volume} {10}},\ \bibinfo {pages} {604} (\bibinfo {year} {2020})}\BibitemShut
	{NoStop}%
	\bibitem [{\citenamefont {Oseen}(1933)}]{oseen:theory}%
	\BibitemOpen
	\bibfield  {author} {\bibinfo {author} {\bibfnamefont {C.~W.}\ \bibnamefont
			{Oseen}},\ }\bibfield  {title} {\bibinfo {title} {The theory of liquid
			crystals},\ }\href {https://doi.org/10.1039/TF9332900883} {\bibfield
		{journal} {\bibinfo  {journal} {Trans. Faraday Soc.}\ }\textbf {\bibinfo
			{volume} {29}},\ \bibinfo {pages} {883} (\bibinfo {year} {1933})}\BibitemShut
	{NoStop}%
	\bibitem [{\citenamefont {Frank}(1958)}]{frank:theory}%
	\BibitemOpen
	\bibfield  {author} {\bibinfo {author} {\bibfnamefont {F.~C.}\ \bibnamefont
			{Frank}},\ }\bibfield  {title} {\bibinfo {title} {On the theory of liquid
			crystals},\ }\href {https://doi.org/10.1039/DF9582500019} {\bibfield
		{journal} {\bibinfo  {journal} {Discuss. Faraday Soc.}\ }\textbf {\bibinfo
			{volume} {25}},\ \bibinfo {pages} {19} (\bibinfo {year} {1958})}\BibitemShut
	{NoStop}%
	\bibitem [{\citenamefont {Paparini}\ and\ \citenamefont
		{Virga}(2022{\natexlab{a}})}]{paparini:paradoxes}%
	\BibitemOpen
	\bibfield  {author} {\bibinfo {author} {\bibfnamefont {S.}~\bibnamefont
			{Paparini}}\ and\ \bibinfo {author} {\bibfnamefont {E.~G.}\ \bibnamefont
			{Virga}},\ }\bibfield  {title} {\bibinfo {title} {Paradoxes for chromonic
			liquid crystal droplets},\ }\href
	{https://doi.org/10.1103/PhysRevE.106.044703} {\bibfield  {journal} {\bibinfo
			{journal} {Phys. Rev. E}\ }\textbf {\bibinfo {volume} {106}},\ \bibinfo
		{pages} {044703} (\bibinfo {year} {2022}{\natexlab{a}})}\BibitemShut
	{NoStop}%
	\bibitem [{\citenamefont {Paparini}\ and\ \citenamefont
		{Virga}(2024{\natexlab{a}})}]{paparini:elastic}%
	\BibitemOpen
	\bibfield  {author} {\bibinfo {author} {\bibfnamefont {S.}~\bibnamefont
			{Paparini}}\ and\ \bibinfo {author} {\bibfnamefont {E.~G.}\ \bibnamefont
			{Virga}},\ }\bibfield  {title} {\bibinfo {title} {An elastic quartic twist
			theory for chromonic liquid crystals},\ }\href
	{https://doi.org/https://doi.org/10.1007/s10659-022-09983-4} {\bibfield
		{journal} {\bibinfo  {journal} {J. Elast.}\ }\textbf {\bibinfo {volume}
			{155}},\ \bibinfo {pages} {469} (\bibinfo {year}
		{2024}{\natexlab{a}})}\BibitemShut {NoStop}%
	\bibitem [{\citenamefont {Paparini}\ and\ \citenamefont
		{Virga}(2023)}]{paparini:spiralling}%
	\BibitemOpen
	\bibfield  {author} {\bibinfo {author} {\bibfnamefont {S.}~\bibnamefont
			{Paparini}}\ and\ \bibinfo {author} {\bibfnamefont {E.~G.}\ \bibnamefont
			{Virga}},\ }\bibfield  {title} {\bibinfo {title} {Spiralling defect cores in
			chromonic hedgehogs},\ }\href {https://doi.org/10.1080/02678292.2023.2190626}
	{\bibfield  {journal} {\bibinfo  {journal} {Liq. Cryst.}\ }\textbf {\bibinfo
			{volume} {50}},\ \bibinfo {pages} {1498} (\bibinfo {year}
		{2023})}\BibitemShut {NoStop}%
	\bibitem [{\citenamefont {Ciuchi}\ \emph {et~al.}(2024)\citenamefont {Ciuchi},
		\citenamefont {{De~Santo}}, \citenamefont {Paparini}, \citenamefont {Spina},\
		and\ \citenamefont {Virga}}]{ciuchi:inversion}%
	\BibitemOpen
	\bibfield  {author} {\bibinfo {author} {\bibfnamefont {F.}~\bibnamefont
			{Ciuchi}}, \bibinfo {author} {\bibfnamefont {M.~P.}\ \bibnamefont
			{{De~Santo}}}, \bibinfo {author} {\bibfnamefont {S.}~\bibnamefont
			{Paparini}}, \bibinfo {author} {\bibfnamefont {L.}~\bibnamefont {Spina}},\
		and\ \bibinfo {author} {\bibfnamefont {E.~G.}\ \bibnamefont {Virga}},\
	}\bibfield  {title} {\bibinfo {title} {Inversion ring in chromonic twisted
			hedgehogs: theory and experiment},\ }\href
	{https://doi.org/10.1080/02678292.2024.2313023} {\bibfield  {journal}
		{\bibinfo  {journal} {Liq. Cryst.}\ ,\ \bibinfo {pages} {1}} (\bibinfo {year}
		{2024})},\ \bibinfo {note} {{P}ublished online: 13 Feb 2024}\BibitemShut
	{NoStop}%
	\bibitem [{\citenamefont {Paparini}\ and\ \citenamefont
		{Virga}(2024{\natexlab{b}})}]{paparini:what}%
	\BibitemOpen
	\bibfield  {author} {\bibinfo {author} {\bibfnamefont {S.}~\bibnamefont
			{Paparini}}\ and\ \bibinfo {author} {\bibfnamefont {E.~G.}\ \bibnamefont
			{Virga}},\ }\bibfield  {title} {\bibinfo {title} {What a twist cell
			experiment tells about a quartic twist theory for chromonics},\ }\href
	{https://doi.org/10.1080/02678292.2024.2324465} {\bibfield  {journal}
		{\bibinfo  {journal} {Liq. Cryst.}\ }\textbf {\bibinfo {volume} {51}},\
		\bibinfo {pages} {993} (\bibinfo {year} {2024}{\natexlab{b}})}\BibitemShut
	{NoStop}%
	\bibitem [{\citenamefont {Ericksen}(1991)}]{ericksen:liquid}%
	\BibitemOpen
	\bibfield  {author} {\bibinfo {author} {\bibfnamefont {J.~L.}\ \bibnamefont
			{Ericksen}},\ }\bibfield  {title} {\bibinfo {title} {Liquid crystals with
			variable degree of orientation},\ }\href {https://doi.org/10.1007/BF00380413}
	{\bibfield  {journal} {\bibinfo  {journal} {Arch. Rational Mech. Anal.}\
		}\textbf {\bibinfo {volume} {113}},\ \bibinfo {pages} {97} (\bibinfo {year}
		{1991})}\BibitemShut {NoStop}%
	\bibitem [{\citenamefont {Sonnet}\ and\ \citenamefont
		{Virga}(2012)}]{sonnet:dissipative}%
	\BibitemOpen
	\bibfield  {author} {\bibinfo {author} {\bibfnamefont {A.}~\bibnamefont
			{Sonnet}}\ and\ \bibinfo {author} {\bibfnamefont {E.~G.}\ \bibnamefont
			{Virga}},\ }\href@noop {} {\emph {\bibinfo {title} {Dissipative Ordered
				Fluids: Theories for Liquid Crystals}}}\ (\bibinfo  {publisher} {Springer},\
	\bibinfo {address} {New York},\ \bibinfo {year} {2012})\BibitemShut {NoStop}%
	\bibitem [{\citenamefont {Ericksen}(5960)}]{ericksen:anisotropic}%
	\BibitemOpen
	\bibfield  {author} {\bibinfo {author} {\bibfnamefont {J.~L.}\ \bibnamefont
			{Ericksen}},\ }\bibfield  {title} {\bibinfo {title} {Anisotropic fluids},\
	}\href {https://doi.org/https://doi.org/10.1007/BF00281389} {\bibfield
		{journal} {\bibinfo  {journal} {Arch. Rational Mech. Anal.}\ }\textbf
		{\bibinfo {volume} {4}},\ \bibinfo {pages} {231} (\bibinfo {year}
		{1959/60})}\BibitemShut {NoStop}%
	\bibitem [{\citenamefont {Ericksen}(1961)}]{ericksen:conservation}%
	\BibitemOpen
	\bibfield  {author} {\bibinfo {author} {\bibfnamefont {J.~L.}\ \bibnamefont
			{Ericksen}},\ }\bibfield  {title} {\bibinfo {title} {Conservation laws for
			liquid crystals},\ }\href {https://doi.org/10.1122/1.548883} {\bibfield
		{journal} {\bibinfo  {journal} {Trans. Soc. Rheol.}\ }\textbf {\bibinfo
			{volume} {5}},\ \bibinfo {pages} {23} (\bibinfo {year} {1961})}\BibitemShut
	{NoStop}%
	\bibitem [{\citenamefont {Leslie}(1966)}]{leslie:some_1966}%
	\BibitemOpen
	\bibfield  {author} {\bibinfo {author} {\bibfnamefont {F.~M.}\ \bibnamefont
			{Leslie}},\ }\bibfield  {title} {\bibinfo {title} {Some constitutive
			equations for anisotropic fluids},\ }\href
	{https://doi.org/https://doi.org/10.1093/qjmam/19.3.357} {\bibfield
		{journal} {\bibinfo  {journal} {Quart. J. Mech. Appl. Math.}\ }\textbf
		{\bibinfo {volume} {19}},\ \bibinfo {pages} {357} (\bibinfo {year}
		{1966})}\BibitemShut {NoStop}%
	\bibitem [{\citenamefont {Leslie}(1968{\natexlab{a}})}]{leslie:some}%
	\BibitemOpen
	\bibfield  {author} {\bibinfo {author} {\bibfnamefont {F.~M.}\ \bibnamefont
			{Leslie}},\ }\bibfield  {title} {\bibinfo {title} {Some constitutive
			equations for liquid crystals},\ }\href
	{https://doi.org/https://doi.org/10.1007/BF00251810} {\bibfield  {journal}
		{\bibinfo  {journal} {Arch. Rational Mech. Anal.}\ }\textbf {\bibinfo
			{volume} {28}},\ \bibinfo {pages} {265} (\bibinfo {year}
		{1968}{\natexlab{a}})}\BibitemShut {NoStop}%
	\bibitem [{\citenamefont {Leslie}(1968{\natexlab{b}})}]{leslie:thermal}%
	\BibitemOpen
	\bibfield  {author} {\bibinfo {author} {\bibfnamefont {F.~M.}\ \bibnamefont
			{Leslie}},\ }\bibfield  {title} {\bibinfo {title} {Thermal effects in
			cholesteric liquid crystals},\ }\href
	{https://doi.org/10.1098/rspa.1968.0195} {\bibfield  {journal} {\bibinfo
			{journal} {Proc. Roy. Soc. London A}\ }\textbf {\bibinfo {volume} {307}},\
		\bibinfo {pages} {359} (\bibinfo {year} {1968}{\natexlab{b}})}\BibitemShut
	{NoStop}%
	\bibitem [{\citenamefont {Leslie}(1969)}]{leslie:continuum}%
	\BibitemOpen
	\bibfield  {author} {\bibinfo {author} {\bibfnamefont {F.~M.}\ \bibnamefont
			{Leslie}},\ }\bibfield  {title} {\bibinfo {title} {Continuum theory of
			cholesteric liquid crystals},\ }\href
	{https://doi.org/10.1080/15421406908084887} {\bibfield  {journal} {\bibinfo
			{journal} {Mol. Cryst. Liq. Cryst.}\ }\textbf {\bibinfo {volume} {7}},\
		\bibinfo {pages} {407} (\bibinfo {year} {1969})}\BibitemShut {NoStop}%
	\bibitem [{\citenamefont {Ericksen}(1969)}]{ericksen:continuum}%
	\BibitemOpen
	\bibfield  {author} {\bibinfo {author} {\bibfnamefont {J.~L.}\ \bibnamefont
			{Ericksen}},\ }\bibfield  {title} {\bibinfo {title} {Contimuum theory of
			liquid crystals of nematic type},\ }\href
	{https://doi.org/10.1080/15421406908084869} {\bibfield  {journal} {\bibinfo
			{journal} {Mol. Cryst. Liq. Cryst.}\ }\textbf {\bibinfo {volume} {7}},\
		\bibinfo {pages} {153} (\bibinfo {year} {1969})}\BibitemShut {NoStop}%
	\bibitem [{\citenamefont {Ericksen}(1968{\natexlab{a}})}]{ericksen:twist}%
	\BibitemOpen
	\bibfield  {author} {\bibinfo {author} {\bibfnamefont {J.~L.}\ \bibnamefont
			{Ericksen}},\ }\bibfield  {title} {\bibinfo {title} {Twist waves in liquid
			crystals},\ }\href {https://doi.org/10.1093/qjmam/21.4.463} {\bibfield
		{journal} {\bibinfo  {journal} {Q. J. Mech. Appl. Math.}\ }\textbf {\bibinfo
			{volume} {21}},\ \bibinfo {pages} {463} (\bibinfo {year}
		{1968}{\natexlab{a}})}\BibitemShut {NoStop}%
	\bibitem [{\citenamefont {Bishop}\ and\ \citenamefont
		{Schneider}(1978)}]{bishop:solitonsl}%
	\BibitemOpen
	\bibinfo {editor} {\bibfnamefont {A.~R.}\ \bibnamefont {Bishop}}\ and\
	\bibinfo {editor} {\bibfnamefont {T.}~\bibnamefont {Schneider}},\ eds.,\
	\href@noop {} {\emph {\bibinfo {title} {Solitons and Condensed Matter
				Physics. {P}roceedings of the Symposium on Nonlinear (Soliton) Structure and
				Dynamics in Condensed Matter. {O}xford, England, June 27-29, 1978}}},\
	\bibinfo {series} {Springer {S}eries in {S}olid-{S}tate {S}ciences},
	Vol.~\bibinfo {volume} {8}\ (\bibinfo  {publisher} {Springer-Verlag},\
	\bibinfo {address} {Berlin},\ \bibinfo {year} {1978})\BibitemShut {NoStop}%
	\bibitem [{\citenamefont {Helfrich}(1968)}]{helfrich:alignment}%
	\BibitemOpen
	\bibfield  {author} {\bibinfo {author} {\bibfnamefont {W.}~\bibnamefont
			{Helfrich}},\ }\bibfield  {title} {\bibinfo {title} {Alignment-inversion
			walls in nematic liquid crystals in the presence of a magnetic field},\
	}\href {https://doi.org/10.1103/PhysRevLett.21.1518} {\bibfield  {journal}
		{\bibinfo  {journal} {Phys. Rev. Lett.}\ }\textbf {\bibinfo {volume} {21}},\
		\bibinfo {pages} {1518} (\bibinfo {year} {1968})}\BibitemShut {NoStop}%
	\bibitem [{\citenamefont {{De Gennes}}(1971)}]{degennes:mouvements}%
	\BibitemOpen
	\bibfield  {author} {\bibinfo {author} {\bibfnamefont {P.~G.}\ \bibnamefont
			{{De Gennes}}},\ }\bibfield  {title} {\bibinfo {title} {Mouvements de parois
			dans un n\'ematique sous champ tournant},\ }\href
	{https://doi.org/10.1051/jphys:019710032010078900} {\bibfield  {journal}
		{\bibinfo  {journal} {J. Phys. France}\ }\textbf {\bibinfo {volume} {32}},\
		\bibinfo {pages} {789} (\bibinfo {year} {1971})}\BibitemShut {NoStop}%
	\bibitem [{\citenamefont {Brochard}(1972)}]{brochard:mouvements}%
	\BibitemOpen
	\bibfield  {author} {\bibinfo {author} {\bibfnamefont {F.}~\bibnamefont
			{Brochard}},\ }\bibfield  {title} {\bibinfo {title} {Mouvements de parois
			dans une lame mince n\'ematique},\ }\href
	{https://doi.org/10.1051/jphys:01972003305-6060700} {\bibfield  {journal}
		{\bibinfo  {journal} {J. Phys. France}\ }\textbf {\bibinfo {volume} {33}},\
		\bibinfo {pages} {607} (\bibinfo {year} {1972})}\BibitemShut {NoStop}%
	\bibitem [{\citenamefont {Leger}(1972{\natexlab{a}})}]{leger:observation}%
	\BibitemOpen
	\bibfield  {author} {\bibinfo {author} {\bibfnamefont {L.}~\bibnamefont
			{Leger}},\ }\bibfield  {title} {\bibinfo {title} {Observation of wall motions
			in nematics},\ }\href
	{https://doi.org/https://doi.org/10.1016/0038-1098(72)90588-1} {\bibfield
		{journal} {\bibinfo  {journal} {Solid State Commun.}\ }\textbf {\bibinfo
			{volume} {10}},\ \bibinfo {pages} {697} (\bibinfo {year}
		{1972}{\natexlab{a}})}\BibitemShut {NoStop}%
	\bibitem [{\citenamefont {Leger}(1972{\natexlab{b}})}]{leger:static}%
	\BibitemOpen
	\bibfield  {author} {\bibinfo {author} {\bibfnamefont {L.}~\bibnamefont
			{Leger}},\ }\bibfield  {title} {\bibinfo {title} {Static and dynamic
			behaviour of walls in nematics above a {F}reedericks transition},\ }\href
	{https://doi.org/https://doi.org/10.1016/0038-1098(72)90508-X} {\bibfield
		{journal} {\bibinfo  {journal} {Solid State Commun.}\ }\textbf {\bibinfo
			{volume} {11}},\ \bibinfo {pages} {1499} (\bibinfo {year}
		{1972}{\natexlab{b}})}\BibitemShut {NoStop}%
	\bibitem [{\citenamefont {Guozhen}(1982)}]{guozhen:experiments}%
	\BibitemOpen
	\bibfield  {author} {\bibinfo {author} {\bibfnamefont {Z.}~\bibnamefont
			{Guozhen}},\ }\bibfield  {title} {\bibinfo {title} {Experiments on director
			waves in nematic liquid crystals},\ }\href
	{https://doi.org/10.1103/PhysRevLett.49.1332} {\bibfield  {journal} {\bibinfo
			{journal} {Phys. Rev. Lett.}\ }\textbf {\bibinfo {volume} {49}},\ \bibinfo
		{pages} {1332} (\bibinfo {year} {1982})}\BibitemShut {NoStop}%
	\bibitem [{\citenamefont {Lei}\ \emph {et~al.}(1982)\citenamefont {Lei},
		\citenamefont {Changqing}, \citenamefont {Juelian}, \citenamefont {Lam},\
		and\ \citenamefont {Yun}}]{lei:soliton}%
	\BibitemOpen
	\bibfield  {author} {\bibinfo {author} {\bibfnamefont {L.}~\bibnamefont
			{Lei}}, \bibinfo {author} {\bibfnamefont {S.}~\bibnamefont {Changqing}},
		\bibinfo {author} {\bibfnamefont {S.}~\bibnamefont {Juelian}}, \bibinfo
		{author} {\bibfnamefont {P.~M.}\ \bibnamefont {Lam}},\ and\ \bibinfo {author}
		{\bibfnamefont {H.}~\bibnamefont {Yun}},\ }\bibfield  {title} {\bibinfo
		{title} {Soliton propagation in liquid crystals},\ }\href
	{https://doi.org/10.1103/PhysRevLett.49.1335} {\bibfield  {journal} {\bibinfo
			{journal} {Phys. Rev. Lett.}\ }\textbf {\bibinfo {volume} {49}},\ \bibinfo
		{pages} {1335} (\bibinfo {year} {1982})}\BibitemShut {NoStop}%
	\bibitem [{\citenamefont {Magyari}(1984)}]{magyari:inertia}%
	\BibitemOpen
	\bibfield  {author} {\bibinfo {author} {\bibfnamefont {E.}~\bibnamefont
			{Magyari}},\ }\bibfield  {title} {\bibinfo {title} {The inertia mode of the
			mechanically generated solitons in nematic liquid crystals},\ }\href
	{https://doi.org/https://doi.org/10.1007/BF01470205} {\bibfield  {journal}
		{\bibinfo  {journal} {Z. Phys. B Condensed Matter}\ }\textbf {\bibinfo
			{volume} {56}},\ \bibinfo {pages} {1} (\bibinfo {year} {1984})}\BibitemShut
	{NoStop}%
	\bibitem [{\citenamefont {Lei}\ \emph {et~al.}(1985)\citenamefont {Lei},
		\citenamefont {Changqing},\ and\ \citenamefont {Gang}}]{lei:generation}%
	\BibitemOpen
	\bibfield  {author} {\bibinfo {author} {\bibfnamefont {L.}~\bibnamefont
			{Lei}}, \bibinfo {author} {\bibfnamefont {S.}~\bibnamefont {Changqing}},\
		and\ \bibinfo {author} {\bibfnamefont {X.}~\bibnamefont {Gang}},\ }\bibfield
	{title} {\bibinfo {title} {Generation and detection of propagating solitons
			in shearing liquid crystals},\ }\href
	{https://doi.org/https://doi.org/10.1007/BF01008357} {\bibfield  {journal}
		{\bibinfo  {journal} {J. Stat. Phys.}\ }\textbf {\bibinfo {volume} {39}},\
		\bibinfo {pages} {633} (\bibinfo {year} {1985})}\BibitemShut {NoStop}%
	\bibitem [{\citenamefont {Fergason}\ and\ \citenamefont
		{Brown}(1968)}]{fergason:liquid}%
	\BibitemOpen
	\bibfield  {author} {\bibinfo {author} {\bibfnamefont {J.~L.}\ \bibnamefont
			{Fergason}}\ and\ \bibinfo {author} {\bibfnamefont {G.~H.}\ \bibnamefont
			{Brown}},\ }\bibfield  {title} {\bibinfo {title} {Liquid crystals and living
			systems},\ }\href {https://doi.org/https://doi.org/10.1007/BF02915335}
	{\bibfield  {journal} {\bibinfo  {journal} {J. Am. Oil Chem. Soc.}\ }\textbf
		{\bibinfo {volume} {45}},\ \bibinfo {pages} {120} (\bibinfo {year}
		{1968})}\BibitemShut {NoStop}%
	\bibitem [{\citenamefont {Selinger}(2018)}]{selinger:interpretation}%
	\BibitemOpen
	\bibfield  {author} {\bibinfo {author} {\bibfnamefont {J.~V.}\ \bibnamefont
			{Selinger}},\ }\bibfield  {title} {\bibinfo {title} {Interpretation of
			saddle-splay and the {O}seen-{F}rank free energy in liquid crystals},\ }\href
	{https://doi.org/10.1080/21680396.2019.1581103} {\bibfield  {journal}
		{\bibinfo  {journal} {Liq. Cryst. Rev.}\ }\textbf {\bibinfo {volume} {6}},\
		\bibinfo {pages} {129} (\bibinfo {year} {2018})}\BibitemShut {NoStop}%
	\bibitem [{\citenamefont {Pedrini}\ and\ \citenamefont
		{Virga}(2020)}]{pedrini:liquid}%
	\BibitemOpen
	\bibfield  {author} {\bibinfo {author} {\bibfnamefont {A.}~\bibnamefont
			{Pedrini}}\ and\ \bibinfo {author} {\bibfnamefont {E.~G.}\ \bibnamefont
			{Virga}},\ }\bibfield  {title} {\bibinfo {title} {Liquid crystal distortions
			revealed by an octupolar tensor},\ }\href
	{https://doi.org/10.1103/PhysRevE.101.012703} {\bibfield  {journal} {\bibinfo
			{journal} {Phys. Rev. E}\ }\textbf {\bibinfo {volume} {101}},\ \bibinfo
		{pages} {012703} (\bibinfo {year} {2020})}\BibitemShut {NoStop}%
	\bibitem [{\citenamefont {Ericksen}(1966)}]{ericksen:inequalities}%
	\BibitemOpen
	\bibfield  {author} {\bibinfo {author} {\bibfnamefont {J.~L.}\ \bibnamefont
			{Ericksen}},\ }\bibfield  {title} {\bibinfo {title} {Inequalities in liquid
			crystal theory},\ }\href {https://doi.org/10.1063/1.1761821} {\bibfield
		{journal} {\bibinfo  {journal} {Phys. Fluids}\ }\textbf {\bibinfo {volume}
			{9}},\ \bibinfo {pages} {1205} (\bibinfo {year} {1966})}\BibitemShut
	{NoStop}%
	\bibitem [{\citenamefont {Selinger}(2022)}]{selinger:director}%
	\BibitemOpen
	\bibfield  {author} {\bibinfo {author} {\bibfnamefont {J.~V.}\ \bibnamefont
			{Selinger}},\ }\bibfield  {title} {\bibinfo {title} {Director deformations,
			geometric frustration, and modulated phases in liquid crystals},\ }\href
	{https://doi.org/10.1146/annurev-conmatphys-031620-105712} {\bibfield
		{journal} {\bibinfo  {journal} {Ann. Rev. Condens. Matter Phys.}\ }\textbf
		{\bibinfo {volume} {13}} (\bibinfo {year} {2022})},\ \bibinfo {note} {{First}
		posted online on October 12, 2021. {V}olume publication date, March
		2022.}\BibitemShut {Stop}%
	\bibitem [{\citenamefont {Long}\ and\ \citenamefont
		{Selinger}(2023)}]{long:explicit}%
	\BibitemOpen
	\bibfield  {author} {\bibinfo {author} {\bibfnamefont {C.}~\bibnamefont
			{Long}}\ and\ \bibinfo {author} {\bibfnamefont {J.~V.}\ \bibnamefont
			{Selinger}},\ }\bibfield  {title} {\bibinfo {title} {Explicit demonstration
			of geometric frustration in chiral liquid crystals},\ }\href
	{https://doi.org/10.1039/D2SM01420C} {\bibfield  {journal} {\bibinfo
			{journal} {Soft Matter}\ }\textbf {\bibinfo {volume} {19}},\ \bibinfo {pages}
		{519} (\bibinfo {year} {2023})}\BibitemShut {NoStop}%
	\bibitem [{\citenamefont {Virga}(2019)}]{virga:uniform}%
	\BibitemOpen
	\bibfield  {author} {\bibinfo {author} {\bibfnamefont {E.~G.}\ \bibnamefont
			{Virga}},\ }\bibfield  {title} {\bibinfo {title} {Uniform distortions and
			generalized elasticity of liquid crystals},\ }\href
	{https://doi.org/10.1103/PhysRevE.100.052701} {\bibfield  {journal} {\bibinfo
			{journal} {Phys. Rev. E}\ }\textbf {\bibinfo {volume} {100}},\ \bibinfo
		{pages} {052701} (\bibinfo {year} {2019})}\BibitemShut {NoStop}%
	\bibitem [{\citenamefont {Paparini}\ and\ \citenamefont
		{Virga}(2022{\natexlab{b}})}]{paparini:stability}%
	\BibitemOpen
	\bibfield  {author} {\bibinfo {author} {\bibfnamefont {S.}~\bibnamefont
			{Paparini}}\ and\ \bibinfo {author} {\bibfnamefont {E.~G.}\ \bibnamefont
			{Virga}},\ }\bibfield  {title} {\bibinfo {title} {Stability against the odds:
			the case of chromonic liquid crystals},\ }\href
	{https://doi.org/10.1007/s00332-022-09833-6} {\bibfield  {journal} {\bibinfo
			{journal} {J. Nonlinear Sci.}\ }\textbf {\bibinfo {volume} {32}},\ \bibinfo
		{pages} {74} (\bibinfo {year} {2022}{\natexlab{b}})}\BibitemShut {NoStop}%
	\bibitem [{\citenamefont {Collings}\ \emph {et~al.}(2015)\citenamefont
		{Collings}, \citenamefont {Goldstein}, \citenamefont {Hamilton},
		\citenamefont {Mercado}, \citenamefont {Nieser},\ and\ \citenamefont
		{Regan}}]{collings:nature}%
	\BibitemOpen
	\bibfield  {author} {\bibinfo {author} {\bibfnamefont {P.~J.}\ \bibnamefont
			{Collings}}, \bibinfo {author} {\bibfnamefont {J.~N.}\ \bibnamefont
			{Goldstein}}, \bibinfo {author} {\bibfnamefont {E.~J.}\ \bibnamefont
			{Hamilton}}, \bibinfo {author} {\bibfnamefont {B.~R.}\ \bibnamefont
			{Mercado}}, \bibinfo {author} {\bibfnamefont {K.~J.}\ \bibnamefont
			{Nieser}},\ and\ \bibinfo {author} {\bibfnamefont {M.~H.}\ \bibnamefont
			{Regan}},\ }\bibfield  {title} {\bibinfo {title} {The nature of the assembly
			process in chromonic liquid crystals},\ }\href
	{https://doi.org/10.1080/21680396.2015.1025305} {\bibfield  {journal}
		{\bibinfo  {journal} {Liq. Cryst. Rev.}\ }\textbf {\bibinfo {volume} {3}},\
		\bibinfo {pages} {1} (\bibinfo {year} {2015})}\BibitemShut {NoStop}%
	\bibitem [{\citenamefont {Yu}\ \emph {et~al.}(2021)\citenamefont {Yu},
		\citenamefont {Chen}, \citenamefont {Li}, \citenamefont {Li}, \citenamefont
		{Huang}, \citenamefont {Bake},\ and\ \citenamefont {Tian}}]{yu:rotational}%
	\BibitemOpen
	\bibfield  {author} {\bibinfo {author} {\bibfnamefont {J.-J.}\ \bibnamefont
			{Yu}}, \bibinfo {author} {\bibfnamefont {L.-F.}\ \bibnamefont {Chen}},
		\bibinfo {author} {\bibfnamefont {G.-Y.}\ \bibnamefont {Li}}, \bibinfo
		{author} {\bibfnamefont {Y.-R.}\ \bibnamefont {Li}}, \bibinfo {author}
		{\bibfnamefont {Y.}~\bibnamefont {Huang}}, \bibinfo {author} {\bibfnamefont
			{M.}~\bibnamefont {Bake}},\ and\ \bibinfo {author} {\bibfnamefont
			{Z.}~\bibnamefont {Tian}},\ }\bibfield  {title} {\bibinfo {title} {Rotational
			viscosity of nematic lyotropic chromonic liquid crystals},\ }\href
	{https://doi.org/https://doi.org/10.1016/j.molliq.2021.117756} {\bibfield
		{journal} {\bibinfo  {journal} {J. Mol. Liq.}\ }\textbf {\bibinfo {volume}
			{344}},\ \bibinfo {pages} {117756} (\bibinfo {year} {2021})}\BibitemShut
	{NoStop}%
	\bibitem [{\citenamefont {Zhou}\ \emph
		{et~al.}(2014{\natexlab{b}})\citenamefont {Zhou}, \citenamefont {Neupane},
		\citenamefont {Nastishin}, \citenamefont {Baldwin}, \citenamefont
		{Shiyanovskii}, \citenamefont {Lavrentovich},\ and\ \citenamefont
		{Sprunt}}]{zhou:elasticity_2014}%
	\BibitemOpen
	\bibfield  {author} {\bibinfo {author} {\bibfnamefont {S.}~\bibnamefont
			{Zhou}}, \bibinfo {author} {\bibfnamefont {K.}~\bibnamefont {Neupane}},
		\bibinfo {author} {\bibfnamefont {Y.~A.}\ \bibnamefont {Nastishin}}, \bibinfo
		{author} {\bibfnamefont {A.~R.}\ \bibnamefont {Baldwin}}, \bibinfo {author}
		{\bibfnamefont {S.~V.}\ \bibnamefont {Shiyanovskii}}, \bibinfo {author}
		{\bibfnamefont {O.~D.}\ \bibnamefont {Lavrentovich}},\ and\ \bibinfo {author}
		{\bibfnamefont {S.}~\bibnamefont {Sprunt}},\ }\bibfield  {title} {\bibinfo
		{title} {Elasticity, viscosity, and orientational fluctuations of a lyotropic
			chromonic nematic liquid crystal disodium cromoglycate},\ }\href
	{https://doi.org/10.1039/C4SM00772G} {\bibfield  {journal} {\bibinfo
			{journal} {Soft Matter}\ }\textbf {\bibinfo {volume} {10}},\ \bibinfo {pages}
		{6571} (\bibinfo {year} {2014}{\natexlab{b}})}\BibitemShut {NoStop}%
	\bibitem [{\citenamefont {Zhou}\ \emph
		{et~al.}(2014{\natexlab{c}})\citenamefont {Zhou}, \citenamefont {Cervenka},\
		and\ \citenamefont {Lavrentovich}}]{zhou:ionic}%
	\BibitemOpen
	\bibfield  {author} {\bibinfo {author} {\bibfnamefont {S.}~\bibnamefont
			{Zhou}}, \bibinfo {author} {\bibfnamefont {A.~J.}\ \bibnamefont {Cervenka}},\
		and\ \bibinfo {author} {\bibfnamefont {O.~D.}\ \bibnamefont {Lavrentovich}},\
	}\bibfield  {title} {\bibinfo {title} {Ionic-content dependence of
			viscoelasticity of the lyotropic chromonic liquid crystal sunset yellow},\
	}\href {https://doi.org/10.1103/PhysRevE.90.042505} {\bibfield  {journal}
		{\bibinfo  {journal} {Phys. Rev. E}\ }\textbf {\bibinfo {volume} {90}},\
		\bibinfo {pages} {042505} (\bibinfo {year} {2014}{\natexlab{c}})}\BibitemShut
	{NoStop}%
	\bibitem [{\citenamefont {Golo}\ \emph {et~al.}(1984)\citenamefont {Golo},
		\citenamefont {Kats},\ and\ \citenamefont {Leman}}]{golo:chaos}%
	\BibitemOpen
	\bibfield  {author} {\bibinfo {author} {\bibfnamefont {V.~L.}\ \bibnamefont
			{Golo}}, \bibinfo {author} {\bibfnamefont {E.~I.}\ \bibnamefont {Kats}},\
		and\ \bibinfo {author} {\bibfnamefont {A.~A.}\ \bibnamefont {Leman}},\
	}\bibfield  {title} {\bibinfo {title} {Chaos and long-lived modes in the
			dynamics of nematic liquid crystals},\ }\href
	{http://jetp.ras.ru/cgi-bin/e/index/e/59/1/p84?a=list} {\bibfield  {journal}
		{\bibinfo  {journal} {Sov. Phys. JETP}\ }\textbf {\bibinfo {volume} {59}},\
		\bibinfo {pages} {84} (\bibinfo {year} {1984})},\ \bibinfo {note} {{E}nglish
		translation of Zh. Eksp. Teor. Fiz. \textbf{86}, 147--156 (1984)}\BibitemShut
	{NoStop}%
	\bibitem [{\citenamefont {Golo}\ and\ \citenamefont {Kats}(1984)}]{golo:new}%
	\BibitemOpen
	\bibfield  {author} {\bibinfo {author} {\bibfnamefont {V.~L.}\ \bibnamefont
			{Golo}}\ and\ \bibinfo {author} {\bibfnamefont {E.~I.}\ \bibnamefont
			{Kats}},\ }\bibfield  {title} {\bibinfo {title} {New type of orbital waves in
			nematic liquid crystals},\ }\href
	{http://jetp.ras.ru/cgi-bin/e/index/r/87/5/p1700?a=list} {\bibfield
		{journal} {\bibinfo  {journal} {Sov. Phys. JETP}\ }\textbf {\bibinfo {volume}
			{60}},\ \bibinfo {pages} {977} (\bibinfo {year} {1984})},\ \bibinfo {note}
	{{E}nglish translation of Zh. Eksp. Teor. Fiz., \textbf{87}, 1700--1712
		(1984)}\BibitemShut {NoStop}%
	\bibitem [{\citenamefont {Majda}(1984)}]{majda:compressible}%
	\BibitemOpen
	\bibfield  {author} {\bibinfo {author} {\bibfnamefont {A.}~\bibnamefont
			{Majda}},\ }\href@noop {} {\emph {\bibinfo {title} {Compressible Fluid Flow
				and Systems of Conservation Laws in Several Space Variables}}},\ \bibinfo
	{series} {Applied Mathematical Sciences}, Vol.~\bibinfo {volume} {53}\
	(\bibinfo  {publisher} {Springer-Verlag},\ \bibinfo {address} {New York},\
	\bibinfo {year} {1984})\BibitemShut {NoStop}%
	\bibitem [{\citenamefont
		{Ericksen}(1968{\natexlab{b}})}]{ericksen:propagation}%
	\BibitemOpen
	\bibfield  {author} {\bibinfo {author} {\bibfnamefont {J.~L.}\ \bibnamefont
			{Ericksen}},\ }\bibfield  {title} {\bibinfo {title} {Propagation of weak
			waves in liquid crystals of nematic type},\ }\href
	{https://doi.org/10.1121/1.1911101} {\bibfield  {journal} {\bibinfo
			{journal} {J. Acoust. Soc. Am.}\ }\textbf {\bibinfo {volume} {44}},\ \bibinfo
		{pages} {444} (\bibinfo {year} {1968}{\natexlab{b}})}\BibitemShut {NoStop}%
	\bibitem [{\citenamefont {Shahinpoor}(1975)}]{shahinpoor:finite}%
	\BibitemOpen
	\bibfield  {author} {\bibinfo {author} {\bibfnamefont {M.}~\bibnamefont
			{Shahinpoor}},\ }\bibfield  {title} {\bibinfo {title} {Finite twist waves in
			liquid crystals},\ }\href {https://doi.org/10.1093/qjmam/28.2.223} {\bibfield
		{journal} {\bibinfo  {journal} {Q. J. Mech. Appl. Math.}\ }\textbf {\bibinfo
			{volume} {28}},\ \bibinfo {pages} {223} (\bibinfo {year} {1975})}\BibitemShut
	{NoStop}%
	\bibitem [{\citenamefont {Shahinpoor}(1976)}]{shahinpoor:effect}%
	\BibitemOpen
	\bibfield  {author} {\bibinfo {author} {\bibfnamefont {M.}~\bibnamefont
			{Shahinpoor}},\ }\bibfield  {title} {\bibinfo {title} {Effect of material
			nonlinearity on the acceleration twist waves in liquid crystals},\ }\href
	{https://doi.org/10.1080/15421407608084351} {\bibfield  {journal} {\bibinfo
			{journal} {Mol. Cryst. Liq. Cryst.}\ }\textbf {\bibinfo {volume} {37}},\
		\bibinfo {pages} {121} (\bibinfo {year} {1976})}\BibitemShut {NoStop}%
	\bibitem [{\citenamefont {Chen}\ and\ \citenamefont
		{Zheng}(2013)}]{chen:singularity}%
	\BibitemOpen
	\bibfield  {author} {\bibinfo {author} {\bibfnamefont {G.}~\bibnamefont
			{Chen}}\ and\ \bibinfo {author} {\bibfnamefont {Y.}~\bibnamefont {Zheng}},\
	}\bibfield  {title} {\bibinfo {title} {Singularity and existence for a wave
			system of nematic liquid crystals},\ }\href
	{https://doi.org/https://doi.org/10.1016/j.jmaa.2012.08.048} {\bibfield
		{journal} {\bibinfo  {journal} {J. Math. Anal. Appl.}\ }\textbf {\bibinfo
			{volume} {398}},\ \bibinfo {pages} {170} (\bibinfo {year}
		{2013})}\BibitemShut {NoStop}%
	\bibitem [{\citenamefont {Glassey}\ \emph {et~al.}(1996)\citenamefont
		{Glassey}, \citenamefont {Hunter},\ and\ \citenamefont
		{Zheng}}]{glassey:singularities}%
	\BibitemOpen
	\bibfield  {author} {\bibinfo {author} {\bibfnamefont {R.~T.}\ \bibnamefont
			{Glassey}}, \bibinfo {author} {\bibfnamefont {J.~K.}\ \bibnamefont
			{Hunter}},\ and\ \bibinfo {author} {\bibfnamefont {Y.}~\bibnamefont
			{Zheng}},\ }\bibfield  {title} {\bibinfo {title} {Singularities of a
			variational wave equation},\ }\href
	{https://doi.org/https://doi.org/10.1006/jdeq.1996.0111} {\bibfield
		{journal} {\bibinfo  {journal} {J. Diff. Eq.}\ }\textbf {\bibinfo {volume}
			{129}},\ \bibinfo {pages} {49} (\bibinfo {year} {1996})}\BibitemShut
	{NoStop}%
	\bibitem [{\citenamefont {Chen}\ \emph {et~al.}(2013)\citenamefont {Chen},
		\citenamefont {Zhang},\ and\ \citenamefont {Zheng}}]{chen:energy}%
	\BibitemOpen
	\bibfield  {author} {\bibinfo {author} {\bibfnamefont {G.}~\bibnamefont
			{Chen}}, \bibinfo {author} {\bibfnamefont {P.}~\bibnamefont {Zhang}},\ and\
		\bibinfo {author} {\bibfnamefont {Y.}~\bibnamefont {Zheng}},\ }\bibfield
	{title} {\bibinfo {title} {Energy conservative solutions to a nonlinear wave
			system of nematic liquid crystals},\ }\href
	{https://doi.org/10.3934/cpaa.2013.12.1445} {\bibfield  {journal} {\bibinfo
			{journal} {Comm. Pure Appl. Anal.}\ }\textbf {\bibinfo {volume} {12}},\
		\bibinfo {pages} {1445} (\bibinfo {year} {2013})}\BibitemShut {NoStop}%
	\bibitem [{\citenamefont {Zabusky}(1962)}]{zabusky:exact}%
	\BibitemOpen
	\bibfield  {author} {\bibinfo {author} {\bibfnamefont {N.~J.}\ \bibnamefont
			{Zabusky}},\ }\bibfield  {title} {\bibinfo {title} {Exact solution for the
			vibrations of a nonlinear continuous model string},\ }\href
	{https://doi.org/10.1063/1.1724290} {\bibfield  {journal} {\bibinfo
			{journal} {J. Math. Phys.}\ }\textbf {\bibinfo {volume} {3}},\ \bibinfo
		{pages} {1028} (\bibinfo {year} {1962})}\BibitemShut {NoStop}%
	\bibitem [{\citenamefont {Ludford}(1952)}]{ludford:extension}%
	\BibitemOpen
	\bibfield  {author} {\bibinfo {author} {\bibfnamefont {G.~S.~S.}\
			\bibnamefont {Ludford}},\ }\bibfield  {title} {\bibinfo {title} {On an
			extension of {R}iemann’s method of integration, with applications to
			one-dimensional gas dynamics},\ }\href
	{https://doi.org/10.1017/S0305004100027900} {\bibfield  {journal} {\bibinfo
			{journal} {Math .Proc. Cambridge Phil. Soc.}\ }\textbf {\bibinfo {volume}
			{48}},\ \bibinfo {pages} {499} (\bibinfo {year} {1952})}\BibitemShut
	{NoStop}%
	\bibitem [{\citenamefont {Fermi}\ \emph {et~al.}(1955)\citenamefont {Fermi},
		\citenamefont {Pasta},\ and\ \citenamefont {Ulam}}]{fermi:studies}%
	\BibitemOpen
	\bibfield  {author} {\bibinfo {author} {\bibfnamefont {E.}~\bibnamefont
			{Fermi}}, \bibinfo {author} {\bibfnamefont {J.~R.}\ \bibnamefont {Pasta}},\
		and\ \bibinfo {author} {\bibfnamefont {S.}~\bibnamefont {Ulam}},\ }\href@noop
	{} {\emph {\bibinfo {title} {Studies of Non-linear Problems I}}},\ \bibinfo
	{type} {Tech. Rep.}\ \bibinfo {number} {LA 1940}\ (\bibinfo  {institution}
	{Los Alamos Sci. Lab. Rept.},\ \bibinfo {year} {1955})\ \bibinfo {note}
	{{T}he problem studied in this report is described briefly in \emph{A
			Collection of Mathematical Problems} by S. Ulam (Interscience Publishers,
		Inc., New York, 1960), Chap.~7, Sect.~8. It has also been reprinted in
		\emph{Collected Papers of Enrico Fermi}, Vol.~2, pp.~490--501 (The University
		of Chicago Press, 1965) and is available from
		\url{https://www.physics.utah.edu/~detar/phys6720/handouts/fpu/FermiCollectedPapers1965.pdf}}\BibitemShut
	{NoStop}%
	\bibitem [{\citenamefont {Lax}(1964)}]{lax:development}%
	\BibitemOpen
	\bibfield  {author} {\bibinfo {author} {\bibfnamefont {P.~D.}\ \bibnamefont
			{Lax}},\ }\bibfield  {title} {\bibinfo {title} {{Development of singularities
				of solutions of nonlinear hyperbolic partial differential equations}},\
	}\href {https://doi.org/10.1063/1.1704154} {\bibfield  {journal} {\bibinfo
			{journal} {J. Math. Phys.}\ }\textbf {\bibinfo {volume} {5}},\ \bibinfo
		{pages} {611} (\bibinfo {year} {1964})}\BibitemShut {NoStop}%
	\bibitem [{\citenamefont {MacCamy}\ and\ \citenamefont
		{Mizel}(1967)}]{maccamy:existence}%
	\BibitemOpen
	\bibfield  {author} {\bibinfo {author} {\bibfnamefont {R.~C.}\ \bibnamefont
			{MacCamy}}\ and\ \bibinfo {author} {\bibfnamefont {V.~J.}\ \bibnamefont
			{Mizel}},\ }\bibfield  {title} {\bibinfo {title} {Existence and nonexistence
			in the large of solutions of quasilinear wave equations},\ }\href
	{https://doi.org/https://doi.org/10.1007/BF00250932} {\bibfield  {journal}
		{\bibinfo  {journal} {Arch. Rational Mech. Anal.}\ }\textbf {\bibinfo
			{volume} {25}},\ \bibinfo {pages} {299} (\bibinfo {year} {1967})}\BibitemShut
	{NoStop}%
	\bibitem [{\citenamefont {Manfrin}(2000)}]{manfrin:note}%
	\BibitemOpen
	\bibfield  {author} {\bibinfo {author} {\bibfnamefont {R.}~\bibnamefont
			{Manfrin}},\ }\bibfield  {title} {\bibinfo {title} {A note on the formation
			of singularities for quasi-linear hyperbolic systems},\ }\href
	{https://doi.org/10.1137/S0036141098341526} {\bibfield  {journal} {\bibinfo
			{journal} {SIAM J. Math. Anal.}\ }\textbf {\bibinfo {volume} {32}},\ \bibinfo
		{pages} {261} (\bibinfo {year} {2000})}\BibitemShut {NoStop}%
	\bibitem [{\citenamefont {Chang}(1977)}]{chang:existence}%
	\BibitemOpen
	\bibfield  {author} {\bibinfo {author} {\bibfnamefont {P.~H.}\ \bibnamefont
			{Chang}},\ }\bibfield  {title} {\bibinfo {title} {On the existence of shock
			curves of quasilinear wave equations},\ }\href
	{https://doi.org/10.1512/iumj.1977.26.26049} {\bibfield  {journal} {\bibinfo
			{journal} {Indiana Univ. Math. J.}\ }\textbf {\bibinfo {volume} {26}},\
		\bibinfo {pages} {605} (\bibinfo {year} {1977})}\BibitemShut {NoStop}%
	\bibitem [{\citenamefont {Klainerman}\ and\ \citenamefont
		{Majda}(1980)}]{klainerman:formation}%
	\BibitemOpen
	\bibfield  {author} {\bibinfo {author} {\bibfnamefont {S.}~\bibnamefont
			{Klainerman}}\ and\ \bibinfo {author} {\bibfnamefont {A.}~\bibnamefont
			{Majda}},\ }\bibfield  {title} {\bibinfo {title} {Formation of singularities
			for wave equations including the nonlinear vibrating string},\ }\href
	{https://doi.org/https://doi.org/10.1002/cpa.3160330304} {\bibfield
		{journal} {\bibinfo  {journal} {Comm. Pure Appl. Math.}\ }\textbf {\bibinfo
			{volume} {33}},\ \bibinfo {pages} {241} (\bibinfo {year} {1980})}\BibitemShut
	{NoStop}%
	\bibitem [{\citenamefont {Douglis}(1952)}]{douglis:existence}%
	\BibitemOpen
	\bibfield  {author} {\bibinfo {author} {\bibfnamefont {A.}~\bibnamefont
			{Douglis}},\ }\bibfield  {title} {\bibinfo {title} {Some existence theorems
			for hyperbolic systems of partial differential equations in two independent
			variables},\ }\href {https://doi.org/https://doi.org/10.1002/cpa.3160050202}
	{\bibfield  {journal} {\bibinfo  {journal} {Comm. Pure Appl. Math.}\ }\textbf
		{\bibinfo {volume} {5}},\ \bibinfo {pages} {119} (\bibinfo {year}
		{1952})}\BibitemShut {NoStop}%
	\bibitem [{\citenamefont {Keller}\ and\ \citenamefont
		{Ting}(1966)}]{keller:periodic}%
	\BibitemOpen
	\bibfield  {author} {\bibinfo {author} {\bibfnamefont {J.~B.}\ \bibnamefont
			{Keller}}\ and\ \bibinfo {author} {\bibfnamefont {L.}~\bibnamefont {Ting}},\
	}\bibfield  {title} {\bibinfo {title} {Periodic vibrations of systems
			governed by nonlinear partial differential equations},\ }\href
	{https://doi.org/10.1002/CPA.3160190404} {\bibfield  {journal} {\bibinfo
			{journal} {Comm. Pure Appl. Math.}\ }\textbf {\bibinfo {volume} {19}},\
		\bibinfo {pages} {371} (\bibinfo {year} {1966})}\BibitemShut {NoStop}%
	\bibitem [{\citenamefont {Lax}(1973)}]{lax:hyperbolic}%
	\BibitemOpen
	\bibfield  {author} {\bibinfo {author} {\bibfnamefont {P.~D.}\ \bibnamefont
			{Lax}},\ }\href@noop {} {\emph {\bibinfo {title} {Hyperbolic Systems of
				Conservation Laws and the Mathematical Theory of Shock Waves}}},\ \bibinfo
	{series} {Regional Conference Series in Applied Mathematics}, Vol.~\bibinfo
	{volume} {11}\ (\bibinfo  {publisher} {SIAM},\ \bibinfo {address}
	{Philadelphia},\ \bibinfo {year} {1973})\BibitemShut {NoStop}%
	\bibitem [{\citenamefont {Virga}(1994)}]{virga:variational}%
	\BibitemOpen
	\bibfield  {author} {\bibinfo {author} {\bibfnamefont {E.~G.}\ \bibnamefont
			{Virga}},\ }\href@noop {} {\emph {\bibinfo {title} {Variational Theories for
				Liquid Crystals}}},\ \bibinfo {series} {Applied {M}athematics and
		{M}athematical {C}omputation}, Vol.~\bibinfo {volume} {8}\ (\bibinfo
	{publisher} {Chapman \& Hall},\ \bibinfo {address} {London},\ \bibinfo {year}
	{1994})\BibitemShut {NoStop}%
\end{thebibliography}
%

\end{document}